\documentclass[a4paper,11pt]{article}
\pdfoutput=1 
\usepackage{jheppub} 
\usepackage[dvipsnames]{xcolor}
\usepackage{booktabs}
\usepackage[T1]{fontenc} 
\usepackage{centernot}
\usepackage{comment}
\usepackage[normalem]{ulem}
\usepackage{environ}
\usepackage{pifont}
\newcommand{\fire}{\protect\includegraphics[height=.8em]{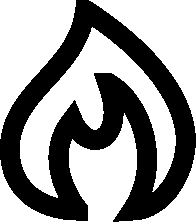}}

\NewEnviron{eq}{%
\begin{align}\begin{split}
  \BODY
\end{split}\end{align}
}
\usepackage{subcaption}
\graphicspath{{./}{./figs/}}
\newcommand{\del}{\partial}
\newcommand{\fsl}[1]{{\centernot{#1}}}

\newcommand{\circled}[1]{\textsf{\textbf{#1}}}

\newcommand{\GeV}{\text{\,GeV}}

\newcommand{\TeV}{\text{\,TeV}}
\newcommand{\MeV}{\text{\,MeV}}
\newcommand{\keV}{\text{\,keV}}
\newcommand{\eV}{\text{~eV}}

\newcommand{\mpl}{m_\text{pl}}
\newcommand{\cm}{\, \text{cm}}

\newcommand{\eps}{\varepsilon}
\renewcommand{\epsilon}{\varepsilon}
\newcommand{\DM}{\text{DM}}

\newcommand{\TSM}{T_{\textsc{sm}}}
\newcommand{\TDM}{T_{\textsc{dm}}}
\newcommand{\TRH}{T_{\text{rh}}}

\newcommand{\ODM}{\Omega_{\textsc{dm}}}

\newcommand{\Tstar}{T_{\star}}
\newcommand{\kstar}{k_\star}

\newcommand{\astar}{a_{\star}}

\newcommand{\qd}{e_\textsc{d}}
\newcommand{\ald}{\alpha_\textsc{d}}
\newcommand{\alem}{\alpha_\text{em}}
\newcommand{\mps}{m_{\psi}}
\newcommand{\mA}{m_{A'}}
\newcommand{\A}{{A'}}
\newcommand{\GamA}{\Gamma_{A'}}
\newcommand{\gamA}{\gamma_{A'}}

\newcommand{\rhoA}{\rho_{A'}}
\newcommand{\nA}{n_{A'}}
\newcommand{\omp}{\omega_\textsc{p}}
\newcommand{\nps}{n_\psi}

\newcommand{\HI}{H_I}
\newcommand{\Ecr}{E'_\text{crit}}
\newcommand{\Hcasc}{H_\text{casc}}
\newcommand{\Wcasc}{\mathcal W_\text{casc}}
\newcommand{\ELvac}{{E'_{L,\text{vac}}}}
\newcommand{\ELvacSq}{{E^{\prime2}_{L,\text{vac}}}}
\newcommand{\omps}{\omega_{\psi}}
\newcommand{\Hplas}{H_\text{plas}}
\newcommand{\aplas}{a_\text{plas}}

\newcommand{\di}{\text{d}}
\newcommand{\pare}[1]{\left(#1\right)}
\newcommand{\parea}[1]{\left[#1\right]}
\newcommand{\avg}[1]{\left<#1\right>}
\newcommand{\op}{\omp}
\newcommand{\kD}{k_\textsc{d}}

\newcommand{\ore}{\omega_\textsc{r}}
\newcommand{\vs}{v_\star}

\newcommand{\ELnop}{{E'_{L}}}

\newcommand{\gstar}{g_{\star}}
\newcommand{\gstarS}{g_{\star,s}}

\newcommand{\Gstar}{\mathcal{G}_{\star}}
\newcommand{\GstarS}{\mathcal{G}_{\star,s}}
\newcommand{\Gstarinf}{\mathcal{G}_{\star,\infty}}
\newcommand{\GstarSinf}{\mathcal{G}_{\star,s,\infty}}

\newcommand{\xfo}{x_\text{f.o.}}

\title{Dark QED from Inflation}

\author[a]{Asimina Arvanitaki}
\author[b]{Savas Dimopoulos}
\author[b]{Marios Galanis}
\author[a]{Davide Racco}
\author[b]{Olivier Simon}
\author[b]{Jedidiah O. Thompson}

\affiliation[a]{Perimeter Institute for Theoretical Physics, Waterloo, Ontario N2L 2Y5, Canada}
\affiliation[b]{Stanford Institute for Theoretical Physics, Stanford University, Stanford, California 94305, USA}

\emailAdd{aarvanitaki@perimeterinstitute.ca}
\emailAdd{savas@stanford.edu}
\emailAdd{mgalanis@stanford.edu}
\emailAdd{dracco@perimeterinstitute.ca}
\emailAdd{osimon@stanford.edu}
\emailAdd{jedidiah@stanford.edu}

\abstract{
One contribution to any dark sector's abundance comes from its gravitational production during inflation.
If the dark sector is weakly coupled to the inflaton and the Standard Model, this can be its only production mechanism.
For non-interacting dark sectors, such as a free massive fermion or a free massive vector field, this mechanism has been studied extensively.  In this paper we show, via the example of dark massive QED, that the presence of interactions can result in a vastly different mass for the dark matter (DM) particle, which may well coincide with the range probed by upcoming experiments.

In the context  of dark QED we study the evolution of the energy density in the dark sector after inflation. Inflation produces a cold vector condensate consisting of an enormous number of bosons, which via interesting processes -- Schwinger pair production, strong field electromagnetic cascades, and plasma dynamics -- transfers its energy to a small number of  ``dark electrons''  and triggers thermalization of the dark sector. 
The resulting dark electron DM mass range is from  50 MeV to 30 TeV, far different from both the $10^{-5}$ eV mass of the massive photon dark matter in the absence of dark electrons, and from the $10^9$ GeV dark electron mass in the absence of dark photons.  
This can significantly impact the search strategies for dark QED and, more generally, theories with a self-interacting DM sector. In the presence of kinetic mixing,
a dark electron in this mass range can be searched for with upcoming direct detection experiments, such as SENSEI-100g and OSCURA.
}

\begin{document} 
\maketitle
\flushbottom

\section{Introduction}
\label{sec:introduction}
The presence of Dark Matter (DM) is the most convincing evidence we have for Beyond the Standard Model (BSM) physics. DM offers an explanation for many observations on galactic and CMB scales, while raising new questions regarding its nature and cosmological origin. Inflation is another leading theory for beyond the Standard Model physics that resolves the flatness, isotropy, and homogeneity problems of the universe, and successfully predicts the observed spectrum of primordial perturbations; if there is a period of de Sitter (dS) dominance in the cosmological history of the universe, then any curvature or inhomogeneities are exponentially suppressed and the primordial curvature power spectrum is naturally approximately scale-invariant.

 A dS period also provides a rapidly varying gravitational background in which particles can be produced. Inflation is thus an irreducible portal to any dark sector, even those that are minimally coupled to the Standard Model and the inflaton. If there are stable particles in these sectors, then they can be at least a component of the dark matter today. 

Particle production in the early universe is an old topic and the basic aspects of this mechanism have long been understood \cite{Parker:1969au, Parker:1971pt}. The key ingredient necessary for particle production is violation of scale invariance. The properties of a Friedmann-Robertson-Walker (FRW) universe are uniquely captured by the scale factor $a(t)$. If the equations of motion of a field in this background (when using conformal time) are independent of $a(t)$, i.e.\ if the fields couple in a scale-invariant way, the effects of the time-dependent metric can be eliminated, since the fields just satisfy the same equations as those in a flat universe. Particle production in a time-dependent gravitational background can only occur when there is breaking of scale invariance. 

Taking this into account for inflation, the energy density of particles produced in a momentum mode $k$, as measured at horizon exit, is captured by this general formula:
\begin{equation}
\frac{d\mathcal{\rho}}{d\ln k}=\mathcal{C} \frac{\HI^4}{(2\pi)^4}
\label{eq:power_at_horizon_exit}
\end{equation}
where $\mathcal{C}$ is a constant that quantifies the breaking of scale invariance.  This can be intuitively understood if one identifies the dS temperature as $T_\text{dS}=\frac{\HI}{2 \pi}$.

For a scalar that has no explicit coupling to the Ricci scalar $R$,\footnote{This setup is usually referred to in the literature as a minimally coupled scalar.} the breaking of conformal invariance is maximal and $\mathcal{C} = \mathcal{O}(1)$. This is the effect that is responsible for sourcing the fluctuations of the inflaton, which  today result in the galaxies and large-scale structures that we observe. However the story is not so simple for a scalar field that is a spectator during inflation.
Unless inflation lasts for a long time \cite{Scherlis2018}, the scalar field retains memory of its initial displacement from the minimum of the potential. This is included in what is usually referred to as misalignment. If inflation lasts for long enough that the memory of the initial field value is lost, then Eq.~\ref{eq:power_at_horizon_exit} gives the energy density per mode at horizon exit \cite{Vilenkin:1982wt,LINDE1982335, STAROBINSKY1982175}. In either case, one can easily see that even when inflation lasts for a very long time, not only is it easy for a scalar of any mass $m_\phi$ to account for the totality of DM, but it is also possible to completely overclose the Universe when $\HI>m_\phi$ unless the initial field displacement is close to the origin or $\phi$ has significant interactions with other sectors. The lower bound of $10^{-22}$~eV on the scalar DM mass is only imposed from structure formation requirements. 

Fermions on the other hand are protected by chiral symmetries, and their coupling to the metric is scale-invariant in the $m \to 0$ limit.  As a consequence, their gravitational production during inflation is suppressed. 
The inflationary production of fermions can only account for the full DM abundance for heavy masses.  Provided reheating completes while the Hubble rate is larger than the fermion mass $\mps$, a present-day fraction $\Omega \approx \left( \frac{\mps}{10^9 \GeV} \right)^{5/2}$ \cite{Lyth:1996yj,Kuzmin:1998kk,Chung:2011ck} is produced.  A massless vector also couples in a scale-invariant way to the metric, but in the presence of a mass the various vector modes actually behave differently.  The transverse modes are protected in the $m \to 0$ limit by a gauge symmetry, and so their behavior is similar to that of fermions. The longitudinal mode, however, behaves like a massless scalar in the relativistic limit, which maximally breaks the scaling symmetry. The longitudinal mode of the vector can thus account for the DM when its mass is much lighter, roughly $10^{-5}\eV$ for $\HI=10^{14}\GeV$ \cite{Graham:2015rva}.

These are the examples that have already been worked out in the literature and they explain why inflationary particle production has been historically associated with extremely heavy or extremely light particles. But as we show in this paper, this picture changes drastically when interactions are added to the dark sector. Interactions affect the DM abundance produced in a couple of ways. First, interactions may break scale-invariance through radiative corrections; in that case, the coefficient $\mathcal{C}$ is determined by the theory's $\beta$-function \cite{Dolgov:1981nw,Dolgov:1993vg,Bautista:2017enk}. Second, interactions in the dark sector can lead to thermalization, which changes its cosmological history. This latter case is what we study in this paper and we leave the former case with its complications to be studied in upcoming work. We will show that thermalization of the energy density produced in a dark sector via inflation populates the universe with a thermal bath of temperature $\TDM$ relative to the SM temperature, $\TSM$:
\begin{equation}
\label{eq:tempratio}
\frac{\TDM}{\TSM}\propto \sqrt{\frac{\HI}{\mpl}}
\end{equation}
We study the cosmology of a dark sector produced through inflation using dark QED as a toy example. This dark QED consists of a dark photon $A'$ with mass $\mA$ and a single fermion $\psi$ with mass $\mps$. The Lagrangian of this dark QED is thus given by:
\begin{equation}
\mathcal{L} = - \frac{1}{4} F^{\mu \nu} F_{\mu \nu} + \frac{1}{2} \mA^2 A'^\mu A'_\mu + \bar{\psi} (i \fsl{D} - \mps ) \psi - \qd \bar{\psi} \gamma^\mu \psi A'_\mu
\label{eq:lagrangian}
\end{equation}
where $F_{\mu \nu} = D_{[\mu} A_{\nu]}$ is the $U(1)$ field strength and $D_\mu$ is the covariant derivative. We assume that the dark photon has a St\"{u}ckelberg mass term and there is thus no radial Higgs mode present in the theory, or at least that the Higgs mass is much heavier than the scale of inflation. The dynamics of the Higgs/radial mode are thus decoupled from our considerations.

\begin{figure}[ht] \centering
\includegraphics[width=.7\textwidth]{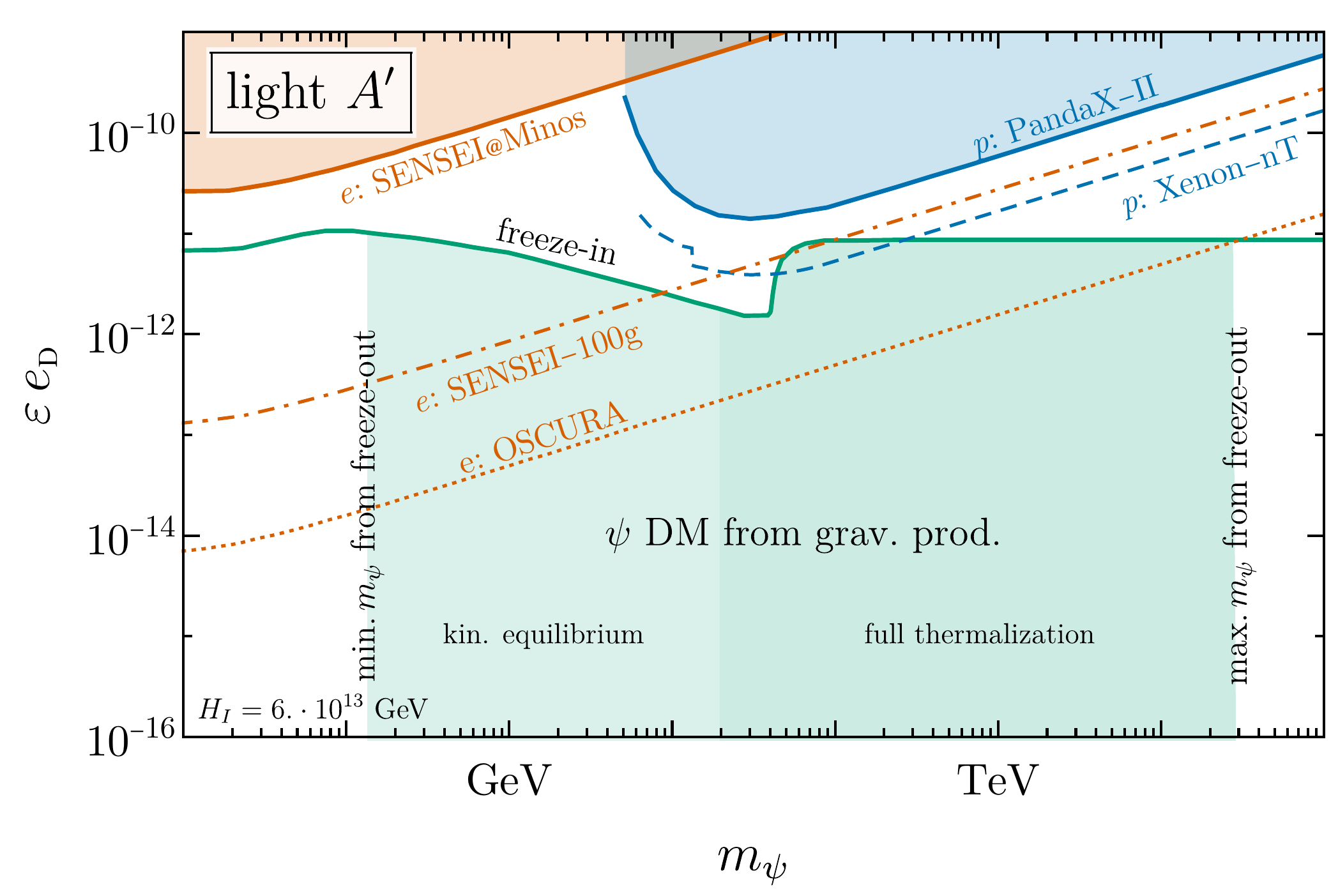}
\caption{Dark fermion parameter space, for $\mps\gg\mA$, that can be probed by the mechanism studied in this paper, assuming a kinetic mixing $\epsilon$ between the dark photon and the SM photon (Sec.~\ref{sec:def_kin_mixing}). Note that production is \emph{independent} of this coupling and relies exclusively on the dynamics of the Dark Sector itself (Eq.~\ref{eq:dmmassrange}). Thus, there is no lower boundary and the green region extends down to arbitrarily small $\epsilon$. The left and right boundaries are set by the mass range for which the dark fermion can be the dark matter (see Fig.~\ref{fig:lightPhotonRelicAbundance} and surrounding discussions for the difference between ``kinetic equilibrium'' and ``full thermalization'') and the upper boundary corresponds to the standard freeze-in production of dark fermions from the Standard Model via kinetic mixing \cite{1911.03389,1112.0493,1902.08623}.
We also outline current bounds (solid) and near-future projections (dashed, dotted) for direct detection experiments (see Sec.~\ref{sec:dark_fermion_light} for details).
}\label{fig:Reach-DD_light-A'}
\end{figure}

In this case, inflation produces the longitudinal component of the massive dark photon, as it maximally breaks scale invariance. As the massive dark photon modes reenter the horizon, they generate a large electric field which has severe implications for cosmology. As it arises from a superposition of soft modes, this initial electric field is far from what a thermal photon state looks like. In addition, depending on the gauge coupling, strong field QED (SFQED) processes like electromagnetic cascades \cite{heisenberg1936,schwinger1951,ritus1985,meuren2015}  become cosmologically relevant and important for producing the fermions in the theory. This primordial ``soup'' of longitudinal gauge modes with large occupation numbers and far fewer fermions can eventually thermalize. When the interactions freeze out in the dark sector,  we find that the dark fermion accounts for the DM of the universe when its mass is:
\begin{equation}
\label{eq:dmmassrange}
\mps\sim 6~\GeV\pare{\frac{\ODM}{0.26}}^{1/2}\left(\frac{\qd}{0.01}\right)^2\left(\frac{\HI}{6\cdot 10^{13}~\GeV}\right)^{-1/4} \quad \text{for }\qd \gtrsim 10^{-3}\,,
\end{equation}
in the case where $\mps > \mA$. When $2 \mps < \mA$, the corresponding fermion mass is:
\begin{equation}
    \mps\sim 46~\text{MeV} \pare{\frac{\ODM}{0.26}}\left(\frac{\HI}{6\cdot 10^{13}~\GeV}\right)^{-3/2} \quad \text{for } \qd \gtrsim 3 \times 10^{-3}
\end{equation}
Assuming a kinetic mixing between the dark photon and electromagnetism, this DM candidate mass coincides with the range probed by upcoming direct detection WIMP experiments, as is shown in Fig.~\ref{fig:Reach-DD_light-A'}, and points to another well-motivated region of parameter space beyond freeze-out and freeze-in. As we will discuss, this contribution to the DM abundance is largely unchanged as long as SM reheating completes before freeze-out happens in the dark sector. Dark sector gravitational production through inflation can thus easily dominate over mechanisms, such as gravitational freeze-in, that require extremely high reheat temperatures.

Our paper is structured as follows: In Sec.~\ref{sec:GPPmassiveVector} we review inflationary production of the massive dark photon, and we  describe how  strong-field QED processes convert some of the energy from the massive photon modes to fermions in Sec.~\ref{sec:thermalization}. In that same section, we describe some sufficient conditions under which the system achieves full thermalization in the two different cases of $\mps > \mA$ and $2 \mps < \mA$. In Sec.~\ref{sec:relicAbundance} we calculate the relic abundance of fermions after interactions freeze-out in the dark sector. Sec.~\ref{sec: signatures} discusses astrophysical signatures and introduces a kinetic mixing between the dark $U(1)$ and the SM photon. After we present the constraints that arise on this model due to this mixing, we show under what conditions this scenario can be probed in future direct detection experiments. We conclude in Sec.~\ref{sec:discussion} by commenting on reheat temperature dependence and how our conclusions can be generalized for an arbitrary dark sector.

In order not to disrupt the flow of the main text, we present  some important results in appendices. In App.~\ref{app:electric_field} we describe how the transition from quantum to classical description happens for the longitudinal modes produced by inflation and how they generate a large electric field as they re-enter the horizon. App.~\ref{app:SFQED} summarizes the relevant strong-field QED processes in a cosmological context. In App.~\ref{app:ProcaPlasmas} and~\ref{app:warmProcaPlasmas}, we derive the properties of Proca plasmas at zero and finite temperature, respectively, and explain how the longitudinal photon modes evolve after the universe is populated with dark electrons. App.~\ref{app:thermalizationAppendix} presents in more detail the processes that lead to thermalization. In App.~\ref{app:freeze-in}  we present calculations for the freeze-in abundance of dark photons and fermions once we introduce a kinetic mixing with the SM photon. Finally, App.~\ref{app:darkPhotonAppendix} contains a detailed discussion of the constraints presented in Sec.~\ref{sec: signatures}, and App.~\ref{app:reheating} closes with a discussion of the reheat temperature dependence.

We work in units where $\hbar = c = k_B = 1$, our convention for the signature of the metric is $(+\,-\,-\,-)$, and we use the unreduced Planck mass $\mpl \equiv G^{-1/2} \approx 1.22 \times 10^{19} \GeV$.

\section{Gravitational production of a massive vector during inflation } \label{sec:GPPmassiveVector}

In this section we briefly recapitulate the inflationary production of a massive dark vector. We consider a massive vector field $A'_\mu = (A'_t,\vec A')_\mu$ with no other interactions other than a minimal gravitational coupling to the FRW background spacetime. 
We will be interested only in the production of longitudinal modes, since it is unsuppressed by the mass $\mA$. In comoving Fourier space therefore, $\vec k \cdot \vec A' = k A'(\vec k)$, with $\vec{k}$ the comoving momentum.
\begin{figure}[ht] \centering
\includegraphics[width=0.8\textwidth]{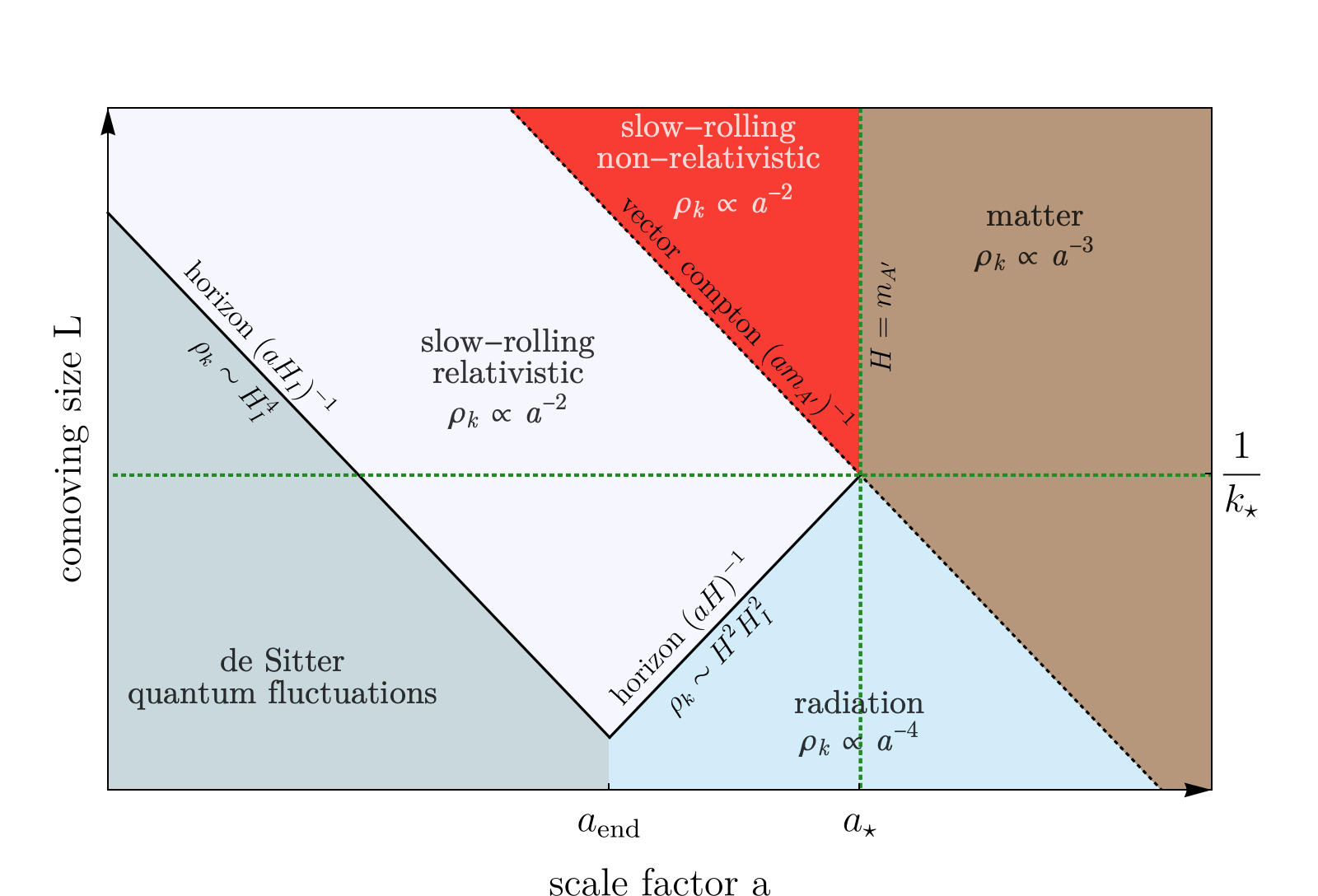}
\caption{Evolution of the energy density in the scenario of \cite{Graham:2015rva}. Modes of the vector field are labeled by a constant wavenumber $k$ and evolve horizontally from left to right. A set of modes around a given $k$ leave the horizon while relativistic during inflation with an energy density $\sim \HI^4.$ As they evolve in time, the modes go through different regimes depending on the hierarchy between the physical wavenumber $k/a$, the size of the cosmological horizon $H^{-1}$, and the Compton wavelength $\mA^{-1}$ of the vector particle.  The energy density contained in the modes redshifts according to those regimes. Mode oscillations are essentially frozen until the mode either re-enters the horizon after the end of inflation, or $H$ decreases below $\mA$, whichever occurs first. Ref.~\cite{Graham:2015rva} studies in detail the scaling of frozen modes. Modes are naturally divided into two sets: those that re-enter the horizon before $H=\mA$ (and therefore while still relativistic), and those that become non-relativistic before $H=\mA$. In the non-interacting theory, the late-time energy density is dominated by the set of modes with comoving wavenumber $\kstar$ that cross the horizon just as $H=\mA$ (dotted green lines). Ref.~\cite{Graham:2015rva} considers the case of instantaneous reheating (shown here), but the results are insensitive to the specifics of reheating provided that it completes before $H>\mA$ \cite{Ahmed:2020fhc,Kolb:2020fwh}.}
\label{fig:massiveVectorEvol}
\end{figure}  

A representation like Fig.~\ref{fig:massiveVectorEvol} is an important guide to the discussion and helps to keep track of the behavior of the comoving Fourier modes as the universe goes through inflation, reheating, and radiation domination. We see that all modes start out highly relativistic and sub-horizon in the asymptotic inflationary past. In accordance with the Goldstone Equivalence theorem, highly relativistic longitudinal vector modes behave as massless scalar modes, the scalar Goldstone boson that has been ``eaten up'' to generate the longitudinal degree of freedom of the massive vector field. 
For such scalars, it is a standard result of quantum field theory in a curved spacetime that during horizon exit (when the scale factor is $a_\text{exit}(k) \equiv k/\HI$), the zero-point quantum fluctuations which originally populate the mode grow in amplitude and become ``locked-in'' as classical, coherent field excitations at some time after the mode has crossed the horizon. This can be described by saying that the modes are initialized to some complex classical amplitude 
$A'_H(\vec k)\equiv A'(\vec k,a_\text{exit}(k))$ at horizon exit, such that the power spectrum of the field is
\begin{equation}
\mathcal P_{A'}(k,a) = \left(\frac{k\HI}{2\pi \mA}\right)^2.
\end{equation}
The power spectrum is defined as in  \cite{Graham:2015rva} such that $\langle X^2 \rangle = \int d\ln k\, \mathcal P_X(k,t)$.
Rather than a field amplitude, this can be phrased in terms of the energy density in the field. If we let
\begin{equation}
  \bar\rho_\A(a) = \int d\ln k \hspace{1mm} \bar \rho_\A(k,a)
  \label{eq:rhoA def int d ln k}
\end{equation}
then a mode that leaves the horizon contributes to the energy density as
\begin{equation}
  \bar\rho_\A\Big(k,a_\text{exit}(k)\Big)\approx\left(\frac{\HI^2}{2\pi}\right)^2
  \label{eq:rho_exit}
\end{equation}
as it exits. 
The energy density is entirely fixed by $\HI$ and is independent of the mass $\mA$.\par
A mode that has exited the horizon is overdamped and slow-rolling due to Hubble friction. Due to the vector nature of the field, the energy density in a mode redshifts approximately as $a^{-2}$ while it is super-horizon, whether the mode is relativistic or not \cite{Graham:2015rva}. Inflation ends at $a_\text{reheat}$ and the Universe enters a period of radiation-domination, after which the size of the comoving horizon increases as $\propto a$. A mode re-enters the horizon at $a_\text{re}(k) = k/H$ (i.e.\ when the physical energy scale of the vector particle is $k/a = H$). 

Depending on the value of $k$, a mode will either re-enter the horizon before $H=\mA$, or after. If we let $a_\star$ be the value of the scale factor when $H = \mA$, then the comoving scale $k_\star = \mA a_\star$ corresponds to the physical scale which enters the horizon just as $H=\mA$. In other words, modes with $k=k_\star$ are the last modes to enter the horizon while relativistic. \par
Modes that re-enter the horizon while relativistic become underdamped only at that time. Their energy density therefore redshifts as $a^{-2}$ all the way up to horizon re-entry. After re-entry, the mode behaves like radiation, its energy density redshifting $\propto a^{-4}$ until the mode becomes non-relativistic at $a_\text{NR}(k) = k/\mA$. Since the universe is still radiation-dominated, $H^2 \propto a^{-4}$, so
\begin{equation}
  \label{eq:rho_enter}
  \bar\rho_A\left(k >k_\star, a_\text{re}\left(k\right) \leq a \leq a_\text{NR}\left(k\right)\right)= \left(\frac{H(a)\HI}{2\pi}\right)^2.
\end{equation}
Note that the right-hand side is independent of $k$. Modes with larger $k$ re-enter the horizon earlier and with more energy density, but they also start redshifting like radiation sooner; the two effects cancel exactly. \par
In the non-interacting scenario, modes with $k<k_\star$ that become non-relativistic before $H=\mA$ become underdamped when $H=\mA$. They start oscillating and behave like matter ($\bar\rho_\A (k,a)\propto a^{-3}$) from that time onward, even while they are still super-horizon in scale.
The outcome of this analysis (see Fig.~\ref{fig:massiveVectorEvol}) is that, if the massive vector modes are non-interacting, then the spectrum of perturbations at late times is dominated by the scale $k_\star$.\par

\subsection{Subhorizon energy density and the parameter \texorpdfstring{$\eta$}{eta}}
We briefly underline an aspect of the non-interacting theory which will play a particularly important role in understanding the cosmological history of the interacting theory.\par
Suppose for the moment that $a<a_\star$. From Eq.~\ref{eq:rho_enter}, one has that every ($e$-fold of) subhorizon modes that redshift like radiation (see Fig.~\ref{fig:massiveVectorEvol}) contributes the \emph{same} amount of energy to the total energy density. This leads us to define a parameter which simply counts the number of such modes:
\begin{equation}
\label{eq:eta}
\eta^2= \int_{aH}^{a_\text{end}H_\text{end}} d\ln k = \ln\left(\frac{a_\text{end}H_\text{end}}{aH}\right),
\end{equation} 
where $H_\text{end}$ is the Hubble scale at the end of reheating ($H_\text{end} = H_I$ for instantaneous reheating) and $a_\text{end}H_\text{end} = k_\text{max}$ is the comoving lengthscale which enters the horizon just as reheating ends. The total contribution of those modes to total energy density is then
\begin{equation}
\label{eq:rho_sub}
\bar \rho_{A'}(a)_\text{sub} = \left(\frac{H(a)H_I\eta}{2\pi}\right)^2.
\end{equation} 
In dark QED, as we will see in Sec.~\ref{sec:thermalization} (also Apps.~\ref{app:ProcaPlasmas} and \ref{app:thermalizationAppendix}), in the presence of a dark plasma, modes with $k<k_\star$ continue redshifting as $a^{-2}$ even after $H=\mA$, and only start oscillating once their wavelength enters the horizon (Fig.~\ref{fig:plasma_effects}). As a result, they too contribute to the $\eta^2$ factor, so that the definitions \ref{eq:eta} and \ref{eq:rho_sub} can be extended past $a_\star.$ The \emph{subhorizon energy density} is important because it is essentially the amount of energy that can interact, and thermalize. Since Eq.~\ref{eq:rho_enter} only depends on the fact that the energy is contained in the form of \emph{radiation}, the relation still holds if we thermalize this energy into a radiation bath. $\eta^2$ then represents the number of modes that have added their energy to the bath. \par
The important feature of $\eta^2$ is its \emph{logarithmic} dependence on the scales. For example, $H_\text{end}$ can be viewed as parametrizing one's ignorance about the history of reheating (see App.~\ref{app:reheating} for details). Parametrically, the energy in the dark sector is characterized by $H$ and $H_I$ only, with only mild logarithmic dependence on other scales.\par

\subsection{Backreaction on the vector field in an interacting theory}
We have now recapitulated the mechanism for the inflationary generation of a minimally coupled vector when no other forms of interactions are present. The work in this paper studies how the vector field created during inflation evolves when interactions are present. A reasonable question however remains: does the presence of interactions alter the inflationary production of the vector perturbations in the first place? For example, the presence of vector perturbations during inflation implies the existence of a physical (as opposed to comoving) electric field $E_\text{phys}$ which will indeed play a crucial role in the evolution of the dark sector after inflation in our scenario. But one might worry that, during inflation, this electric field can act on gravitationally-produced fermions, or produce fermions itself, and thus change the initial conditions given by inflation. \par
We argue that this is in fact not the case. The charged fermion current observable $\langle J^\mu \rangle= \langle \qd \bar \psi \gamma^\mu \psi\rangle $ in de Sitter spacetime in the presence of a constant \emph{classical} electric field has recently been evaluated in various dimensions \cite{vilenkin2014,stahl2015a,stahl2015b,stahl2016,hayashinaka2016}. We can use this to estimate the backreaction on the Proca electric field. In 3+1 dimensions \cite{hayashinaka2016}, the size of the physical current was found to be ($m_\psi^2,\qd E_\text{phys} \ll \HI^2$):
\begin{equation}
|J_\text{phys}|=\sqrt{-g_{ij}J^iJ^j}=\qd^2 \HI |E_\text{phys}|\log\left(\frac{\HI}{m_\psi}\right).
\end{equation}
App.~\ref{app:electric_field} discusses how the size of the electric field during inflation is $E^2_\text{phys}\simeq \mA^2 \HI^2$. We compare the work done by the field over a Hubble time to the energy density $\simeq \HI^4$ contained in the Proca field. One finds
\begin{equation}
\frac{|J_\text{phys}||E_\text{phys}|\HI^{-1}}{\HI^4} \simeq \qd^2 \left(\frac{\mA}{\HI}\right)^2\log\left(\frac{\HI}{m_\psi}\right) \ll 1.
\end{equation}
Note that this succinctly boils down to the fact that $\langle E_\text{phys}^2 \rangle \ll \langle \mA^2 A_\text{phys}^2 \rangle \simeq \HI^4$ during inflation. \par
On the basis of this estimate, we conclude that the initial conditions for the longitudinal vector modes given by inflation are unchanged between the free and interacting theories. The bulk of this paper is devoted to studying how these initial conditions evolve subsequently.

\section{Dynamics in the dark sector and thermalization} \label{sec:thermalization}

In order to understand the present-day dark sector abundance, we must understand not only its production during inflation but also its cosmological evolution.  In this section we focus in particular on how the presence of the $\psi$ field in the Lagrangian of Eq.~\ref{eq:lagrangian} can lead to a redistribution of the dark sector energy density due to interactions.  Such a redistribution can be extremely important because it can substantially affect how the energy density redshifts, which directly determines its late-time value.

In this paper we have limited ourselves primarily to those regions of parameter space where the dark sector comes to thermal equilibrium (with itself, not with the SM) at a relatively early time.  This limitation is for practical reasons: it is difficult to make precise statements about other regions of parameter space without detailed three-dimensional simulations involving SFQED.  Still, the evolution of the system towards thermalization is far from trivial.  It begins after inflation in a highly non-thermal state, and thermalization involves several subtle processes.  This section is dedicated to this topic, beginning just after the end of inflation and ending when the dark sector finally reaches thermal equilibrium.

Throughout, we limit ourselves to studying a conservative and self-consistent path to thermalization.  It is possible that other processes (especially in the strong-field regime) can play a role, but they will always help push the system towards equilibrium, and so they cannot spoil our predictions in the range of parameter space that we outline below. 

We start with a bird's-eye overview of the state of the dark sector after the end of inflation in Sec.~\ref{sec:stateAfterInflation}.  From this point on, the dynamics are somewhat different depending on whether the dark fermion is lighter or heavier than the dark photon, so we segment the discussion.  In Sec.~\ref{sec:lightPhotonThermalization} we deal with the case of a light dark photon ($\mA \ll \mps$), and outline the various processes that influence the system as the universe evolves, referring the reader to the appendices for further discussion.
Finally in Sec.~\ref{sec:heavyPhotonThermalization} we turn to the case of a heavy dark photon ($\mA \gg \mps$), where additional effects are relevant.

In the following, we use $\gstar$ and $\gstarS$ as they are defined in usual cosmology (summing only over the relativistic SM species), and in addition we define functions to count relativistic degrees of freedom in the dark sector:
\begin{subequations}
\begin{equation}
    \Gstar(\TDM) \equiv \sum_{i \in \text{DM bos.}} g_i \left( \frac{T_i}{\TDM} \right)^4
    + \frac{7}{8} \sum_{j \in \text{DM ferm.}} g_j \left( \frac{T_j}{\TDM} \right)^4
    \label{eq:Gstar}
\end{equation}
\begin{equation}
    \GstarS(\TDM) \equiv \sum_{i \in \text{DM bos.}} g_i \left( \frac{T_i}{\TDM} \right)^3
        + \frac{7}{8} \sum_{j \in \text{DM ferm.}} g_j \left( \frac{T_j}{\TDM} \right)^3
        \label{eq:GstarS}
\end{equation}
\end{subequations}
where the above sums run only over relativistic species.  We also define $\Gstarinf$ and $\GstarSinf$ to be $\Gstar(\TDM \to \infty)$ and $\GstarS(\TDM \to \infty)$ respectively.

\subsection{The state of the dark sector after inflation} \label{sec:stateAfterInflation}
Immediately after the end of inflation, the dark sector energy density consists almost entirely of longitudinal photons.  These modes are in a highly non-thermal state, which can be seen as follows.  The number density of photons in a mode of comoving momentum $k$ as they re-enter the horizon (i.e.\ when $k \sim a H$) can be estimated from Eq.~\ref{eq:rho_enter} as:
\begin{equation}
    \nA ( k \sim a H ) \approx \frac{H \HI^2}{(2 \pi)^2} 
    \label{eq:darkPhotonNumberDensity}
\end{equation}
which implies a large occupation number:
\begin{equation}
    \nA ( k \sim a H) \cdot (2 \pi / H)^3 \approx (\HI / H)^2
    \label{eq:darkPhotonOccupationNumberAtEntry}
\end{equation}
A thermal distribution, on the other hand, with the energy density of Eq.~\ref{eq:rho_enter} would have a temperature of:
\begin{equation}
\TDM \sim \sqrt{H \HI} 
\end{equation}
and thus a thermal number density of 
\begin{equation}
    \nA = \frac{3 \zeta(3)}{\pi^2} \TDM^3 \sim (H \HI)^{3/2}
\end{equation}
and a thermal occupation number of $\mathcal{O}(1)$.  Interactions will in general drive the sector towards thermal equilibrium, but it is clear that reaching equilibrium will require the conversion of many low-energy photons into a much smaller number density of high-energy photons.  In addition, a thermal bath will have a thermal population of dark fermions which must be pair-created.  Throughout the rest of this section we outline the most important processes at work.
\begin{figure}
    \centering
    \includegraphics[width=0.7\columnwidth]{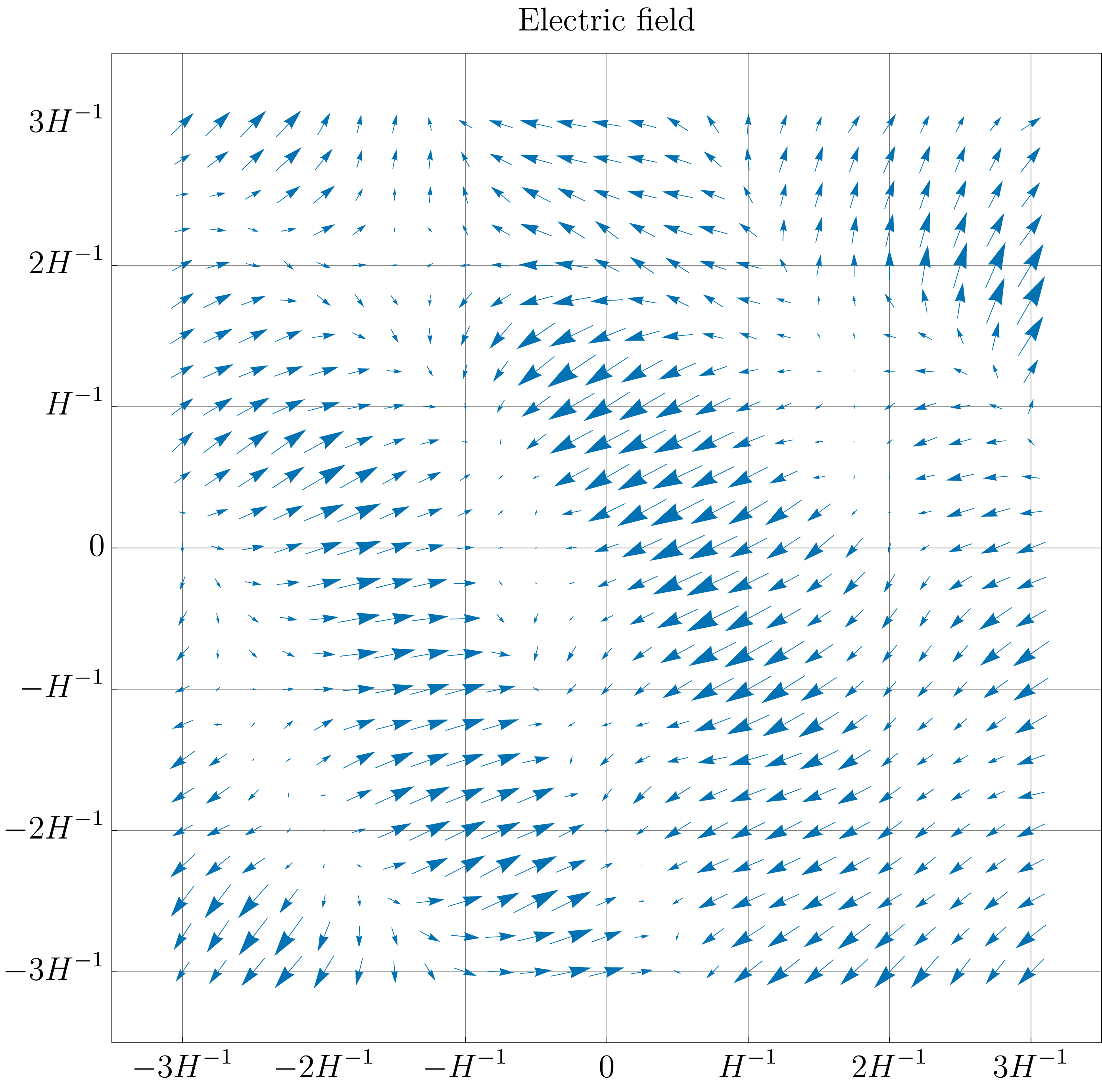}
    \caption{A two-dimensional visualization of the electric field at early times during radiation domination, obtained by randomly drawing from the power spectrum discussed in App.~\ref{app:electric_field}.  Over single Hubble patches (volumes $\sim H^{-3}$), the electric field is approximately constant in both magnitude and direction with typical magnitudes of $\mathcal{O}(\mA \HI)$.}
    \label{fig:EfieldPlot}
\end{figure}

At early times, the most efficient processes are coherent.  The large occupation number of a dark photon mode at horizon entry (Eq.~\ref{eq:darkPhotonOccupationNumberAtEntry}) implies that there is a classical (dark) electric field of size:
\begin{equation}
    \langle E_k^2\rangle= \left(\frac{m_{A'} \HI}{2 \pi}\right)^2 \,,
\end{equation}
whereas the magnetic field is zero because the produced modes are longitudinal. As discussed in App.~\ref{app:electric_field} and shown in Fig.~\ref{fig:EfieldPlot}, this field is coherent over a length $a/k \sim H^{-1}$, meaning that in a region of this size the electric field has a fixed direction and phase with a typical magnitude of $\mathcal{O}(\mA \HI)$.  Such a large classical electric field allows for coherent acceleration of dark fermions as well as strong-field processes such as electromagnetic cascades (App.~\ref{app:SFQED} briefly reviews results of strong-field QED and comments on their application in an expanding universe).  All of these effects extract energy from the electric field, and thereby decrease $\nA$.  Once cascades have produced a large number of fermions in the dark sector, plasma effects also begin to matter, and they significantly modify the behavior of any coherent dark photon modes that have not yet thermalized.

As the universe continues to evolve, all electric fields are damped into the plasma, eventually eliminating the importance of coherent effects.  This leaves only perturbative scatterings of photons and fermions.  These include processes such as pair creation ($A' A' \to \overline{\psi} \psi$) and dark Compton scattering ($A' \psi \to A' \psi$) as well as more complicated $3 \to 2$ processes such as $A' A' \psi \to A' \psi$.  This latter category is higher-order in the coupling $\qd$ but is still important; as discussed above, the overall number density $n_\text{dark} \equiv \nA + \nps + n_{\overline{\psi}}$ must decrease in order to reach thermal equilibrium and the lowest-order processes such as pair creation and Compton scattering all conserve $n_\text{dark}$.  Important for these perturbative processes is the kinematics of the fermion mass.  Dark photons do not scatter efficiently off each other (they can do so only through the one-loop exchange of a dark fermion), but they do scatter efficiently off dark fermions.  The initial coherent processes provide a population of such fermions, and thus play an instrumental role in allowing for thermalization.

Throughout Secs.~\ref{sec:lightPhotonThermalization} and \ref{sec:heavyPhotonThermalization} we carefully track all of these effects to determine the regions of parameter space where thermalization is eventually achieved, equipping us to compute the final dark sector energy density in Sec.~\ref{sec:relicAbundance}. If the reader is willing to assume that thermalization happens for those parts of parameter space that we focus on, they should feel free to skip this section.

\subsection{Thermalization for a light \texorpdfstring{$A'$}{A'}}
\label{sec:lightPhotonThermalization}
\begin{figure}[t] \centering \vspace{-4em}
\includegraphics[width=\textwidth]{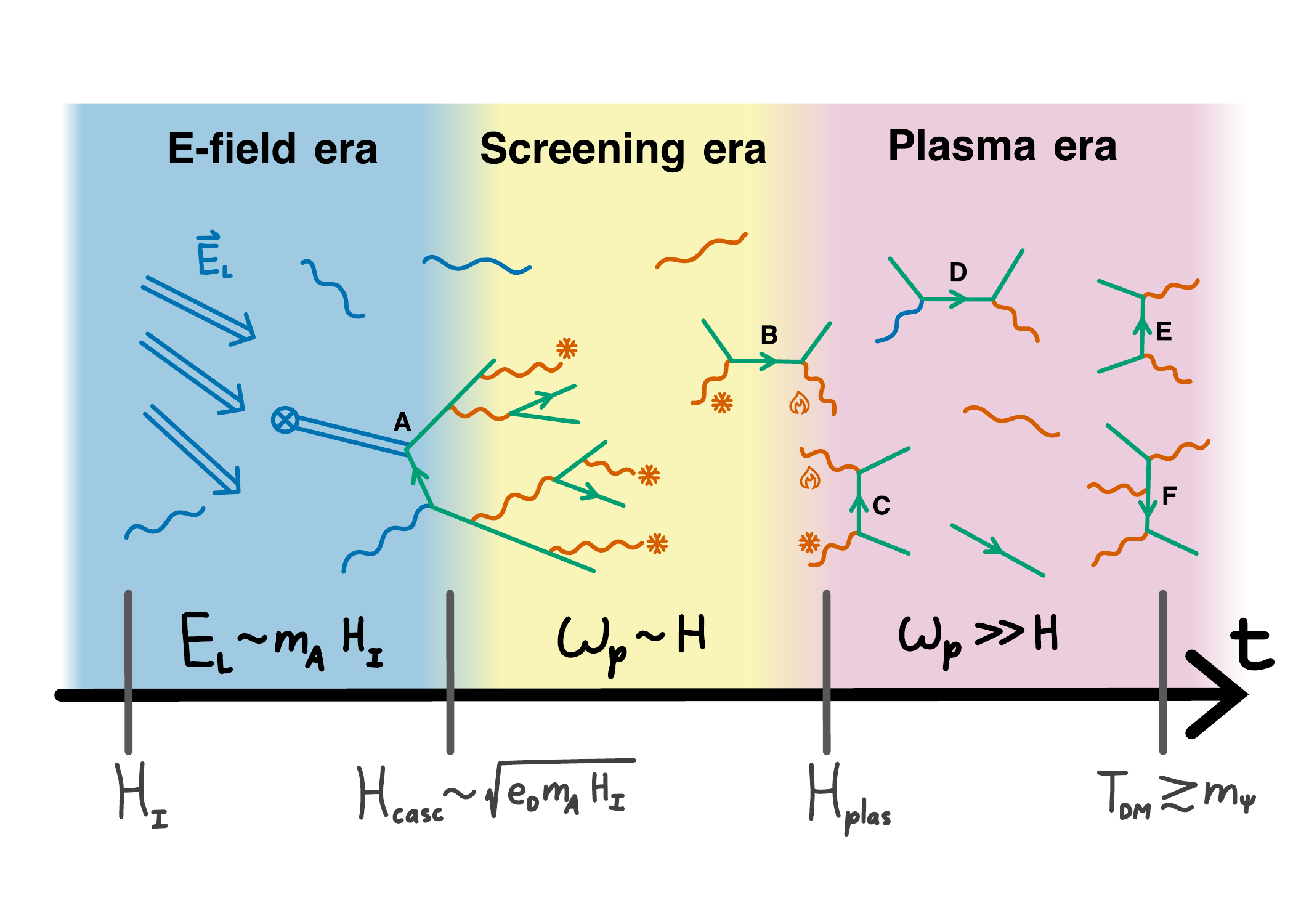}
\vspace{-4em}
\caption{A graphical summary of the thermalization process for $\mA \ll \mps$.
As the $A'_L$ modes (in blue in the figure) that were gravitationally produced during inflation re-enter the horizon, they induce a large dark electric field $E'_L\sim \mA\HI$ (marked by blue double arrows). Strong field processes like cascade (and possibly Schwinger) pair production create $\psi\overline\psi$ pairs (shown in green). As these are accelerated in the $E'$ field, they radiate transverse $A'_T$ photons (in orange) of typically soft momentum $H$ (denoted by \ding{100}).
At early times (``$E$-field era'') the $\nps$ dilution due to the Hubble expansion is strong, and prevents a full electromagnetic cascade (process \circled{A}) until $H<\Hcasc\sim \sqrt{\qd E'_L}$.
Starting from $\Hcasc$, the system enters the ``Screening era'' in which the electric field experienced by the charges starts to be suppressed due to plasma effects (the plasma frequency $\omp$ starts to be $\gtrsim H$).
At $\Hplas\sim \qd\Hcasc$, inverse Compton scattering (process \circled{B}) is efficient, leading to a population of hot photons (denoted by \fire) that can efficiently pair produce as in \circled{C}. These processes increase $\nps$, and an increasing fraction of the $A'_L$ interact via Compton scattering as in \circled{D}, producing $A'_T$ that interact with $\psi$ more efficiently. Eventually, when $2\to2$ (\circled{E}) and $3\to2$ (\circled{F}) processes are efficient, the dark sector thermalizes. More details can be found in Sec.~\ref{sec:thermalization} and in the Appendices, in particular App.~\ref{app:thermalizationAppendix}.
}
\label{fig:cartoon thermalization}
\end{figure}
The case of $\mA \ll \mps$ is the one that most resembles usual QED, so the processes are direct analogues of normal QED processes, albeit with electric field strengths much stronger than those typically probed in laboratory experiments.  In this section, we give a qualitative picture of the evolution of the dark sector in time, as it moves towards thermalization. If the sector thermalizes at some $\TDM \gtrsim \mps$, none of the details of its history actually affect the late-time observables. We thus defer a lengthier, more quantitative picture of this history to the appendices (Apps.~\ref{app:ProcaPlasmas}, \ref{app:warmProcaPlasmas}, and \ref{app:thermalizationAppendix}).  A cartoon picture of the thermalization process we describe is given in Fig.~\ref{fig:cartoon thermalization}.

\subsubsection{Pure electric field dominance era}
Inflation produces predominantly longitudinal vector modes, and, after the universe is reheated and these modes re-enter the horizon, they carry the energy density in the dark sector. A mode just re-entering the horizon sets-up a macroscopic electric field%
\footnote{Note that although $\ELnop$ is quite large, it is only a small component of the energy density contained in the $A'$ field (i.e. $\ELnop^2 \ll \rhoA$ for $H \gg \mA$).  This is because at $H \gg \mA$ all $A'_L$ modes entering the horizon are relativistic, and the electric field of a relativistic longitudinal mode is suppressed relative to its overall energy density by $\mA^2 / p^2$ where $p$ is the physical momentum of the dark photon.  This point is discussed at greater length in App.~\ref{app:electric_field}.}
of average size $\ELnop \sim \mA \HI$, which starts redshifting like $a^{-2}$. The properties of the field inside a Hubble volume are on average determined primarily by the mode entering the horizon at that time. In order for the energy density stored in those modes to thermalize we need to transfer some of that energy to fermions. 

There are two ways these fermion seeds arise: (1) Schwinger pair production, and (2) electromagnetic cascade, sometimes referred to as catalyzed Schwinger pair production.  Of the two processes, the latter is more efficient, and it is depicted as process \circled{A} in Fig.~\ref{fig:cartoon thermalization}. As discussed in detail in App.~\ref{app:SFQED}, the electromagnetic cascade must be seeded by a photon moving not parallel to the electric field. This photon is provided by the longitudinal modes already present inside the universe. This photon is then converted to a pair of fermions, which get accelerated in the background field, producing transverse photons which produce more fermions, and so forth. The rate for this process is given by \cite{greiner_qed,fedotov2013}: 
\begin{equation}
\mathcal W_\text{pair} \sim n_{A'} \frac{\mA^2}{\omega_{A'}^2}\frac{\ald m_\psi^2}{\omega_{A'}} \cdot \begin{cases} \chi\exp\left(-\frac{8}{3}\chi\right), & \chi \ll 1\\
\chi^{2/3}, & \chi \gg 1
\end{cases}
\end{equation}
where $\chi\approx \frac{\qd E' \omega_{A'}}{\mps^3}$, and $n_{A'}$ is the density of longitudinal dark photons with energy $\omega_{A'}$. $\chi \sim 1$ sets the kinematic threshold for this process and can be thought of as the analog of the Schwinger pair production threshold of $\qd E'\sim \mps^2$ in a boosted frame where the effective electric field is given by $E'_\text{boost}\sim E' \gamma=E' \frac{\omega_{A'}}{\mps}$. We add the factor $\frac{\mA^2}{\omega_{A'}^2}$ in order to account for the suppression of the longitudinal mode coupling.
Assuming relatively instantaneous reheating, the highest energy longitudinal photon density is given by $(H \HI)^{3/2}$ with $\omega_{A'}\sim \HI\sqrt{\frac{H}{\HI}}$. As discussed in App.~\ref{app:thermalizationAppendix}, the cascade process becomes exponentially fast when $H\sim \Hcasc \equiv \sqrt{\qd \ELvac}$, where $\ELvac\equiv \mA\HI/(2\pi)$ is the typical amplitude of the electric field at horizon entry. 
The cascading fermions get accelerated by the background field $\ELnop$ for a distance $H^{-1}$ to energy $\omega_{\psi}\sim \frac{\qd \ELnop}{H}$, and via synchrotron radiation they emit transverse photons, whose number density peaks at energies of order $H$. 

The cascade does not continue to extract all of the energy from the electric field. It stops when there is enough charge separation in the produced fermions to screen the electric field. This happens roughly when $\nps \sim n_\text{scr} \equiv \frac{\ELvac H}{\qd}$; this is the time where the universe enters the screening regime. 

\subsubsection{Screening era} 
After this time, roughly set by $\Hcasc^{-1}$, every mode that enters the horizon will be able to set-off a cascade that efficiently screens it very fast. Most of the energy density is still stored in longitudinal modes. We take the density of fermions to be given by the screening density, $n_\text{scr}$ from above.\footnote{This can only be a lower bound on the density of fermions. The reason is that $n_\text{scr}$ represents the charge separation over scales of order Hubble, not the total charge density in the plasma. In fact, the cascade could still be continuing at smaller scales, increasing the overall fermions density well beyond $n_\text{scr}$ and pushing the system further towards thermalization.} This density of fermions creates a non-trivial plasmon mass  $\omp^2=\frac{\qd^2 n_\text{scr}}{\omps}\sim H$, since the fermions are accelerated over a Hubble distance and $\omega_{\psi}\sim\frac{\qd \ELvac}{H}$ as previously shown. 

At screening densities, the plasma mass thus tracks the Hubble rate and is generically much larger than the mass of the dark photon. The universe now gets a non-trivial plasma density and the superhorizon vector photon modes get projected onto plasma modes, which drastically changes their cosmological evolution.

\subsubsection{The evolution of the superhorizon vector modes} 

As we have alluded to above, the evolution of the superhorizon vector modes changes the moment there is plasma inside the horizon.
There are two possible longitudinal eigenmodes in the plasma with a massive photon field, i.e.~a Proca plasma: (1) a gapped ``fast'' mode with dispersion relation $\omega_f^2 \approx \omp^2 + \mA^2 +\left(\frac{k}{a}\right)^2$ and (2) a gapless ``slow''/acoustic mode with dispersion relation $\omega_s^2 \approx \frac{\omp^2}{\omp^2 +\mA^2} \left(\frac{k}{a} \right)^2$ in the limit where $k/a\ll\{\omp,\mA\}$. These relations apply in the regime of interest to us, i.e.~$\frac{k}{a}\ll \kD$, where $\kD$ is the Debye screening wavenumber. As we show in App.~\ref{app:warmProcaPlasmas}, these relations are insensitive to thermal corrections.

\begin{figure}[ht] \centering
    \includegraphics[width=0.9\textwidth]{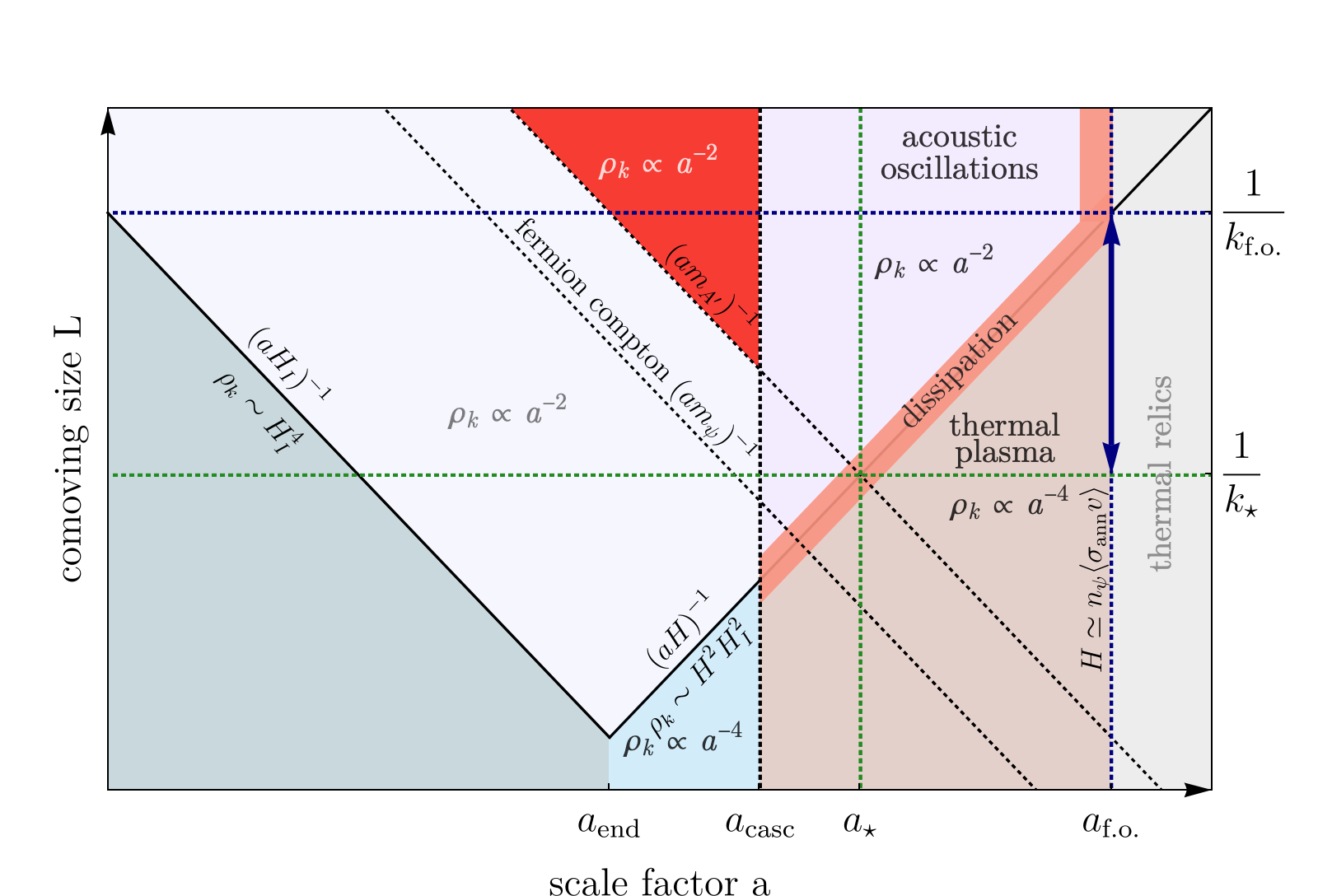}
    \caption{Adaptation of Fig.~\ref{fig:massiveVectorEvol} for the case of massive dark QED ($\mps\gg\mA$): evolution of some key length scales (Hubble radius, Compton wavelengths of $\psi$ and $A'$) together with the scaling of the dark-sector energy density.
    The main differences are the following ones. 
    At $a_{\text{casc}}$ a plasma of $\psi$ and $A'_T$ particles is present (see Sec.~\ref{app:screening regime}), generating a plasma frequency $\op \gtrsim H$ that affects the evolution of super-horizon $A'_L$ modes, preventing them from oscillating at horizon entry (see the discussion around Eq.~\ref{eq:H osc plasma} and in App.~\ref{app:E screening plasma}).
    As these modes cross the horizon, their scattering off the plasma gradually dissipates their energy into the $\psi-A'_T$ system, which for large enough $\qd$ is in thermal equilibrium.
    In this scenario, the time $\astar$ when $H=\mA$ does not play a special role.
    When the $\psi$'s freeze out at $a_\text{f.o.}$, $\nps$ and hence $\op$ drop, leading to a fast redshifting of all super-horizon $A'_L$ modes (see App.~\ref{app:SH_plasma}).
    The total DM energy density at $a_\text{f.o.}$ is of order $\eta^2 \mA^2 \HI^2$ with a mildly different value of $\eta$ relative to the case with no dark fermions (see Eqs.~\ref{eq:TDMa} and \ref{eq:eta dark QED}).}
    \label{fig:plasma_effects}
\end{figure}

When a plasma with $\omp>\mA$ is generated inside Hubble, given that there are no superhorizon correlations for the plasma currents, the superhorizon longitudinal vector modes that are produced by inflation get projected onto the acoustic modes (see App.~\ref{app:ProcaPlasmasGR} and App.~\ref{app:SH_plasma}). Since the plasma mass is so large early on, the dispersion relation of these modes is roughly $\omega_s\approx k/a$. The energy density in those modes still redshifts as $a^{-2}$ while superhorizon ($\frac{k}{a}<H$), and as $a^{-4}$ when subhorizon ($\frac{k}{a}>H$), see Apps.~\ref{app:ProcaPlasmasGR} and~\ref{app:SH_plasma} for details. Those modes are gapless, they no longer oscillate when $H\sim \mA$, and the mode of co-moving wave-number $\kstar$ no longer has any special cosmological significance. This drastic departure from the free massive photon case is illustrated in Fig.~\ref{fig:plasma_effects}, which should be contrasted with Fig.~\ref{fig:massiveVectorEvol}. As we will see in Sec.~\ref{sec:towardsthermalization}, the evolution of these modes further changes when thermalization is achieved.

\subsubsection{Plasma dominated era} 
Assuming that the cascade process cannot produce a fermion density larger than $n_\text{scr}$, we are still stuck in the screening era with a very dilute and hot fluid of fermions and transverse photons, while the energy density is still stored in superhorizon modes. There are several processes which can increase the density of fermions and move the sector away from the screening regime. For example, two high energy longitudinal photons can annihilate to fermions (${A'}_L {A'}_L \to \psi \overline\psi$). To be even more conservative than that, we will present a self-consistent cosmological history based on reprocessing of modes produced by the cascade process in the screening regime. 

One interaction important to this reprocessing is Compton scattering ($\psi A' \to \psi A'$ -- process \circled{B} in Fig.~\ref{fig:cartoon thermalization}). As discussed above, the accelerating fermions mainly radiate transversely-polarized dark photons. This radiation is IR dominated, with most dark photons around $\omega_{A'}\sim H$, but scatters with the hot $\psi$ particles can heat them up. There are far more transverse photons than fermions (bremsstrahlung emits many IR photons), but this process (see App.~\ref{app:first steps thermalization}) manages to heat a subset of the transverse photon population to energies of $\mathcal{O}(\omps)$.

For strong enough coupling (corresponding to the condition $\qd E'_L >\mps^2$, marked by a purple line in Fig.~\ref{fig:lightPhotonRelicAbundance}), 
this hot photon population is then able to pair-create new fermions via scattering off the cold photons ($A' A' \to \psi \bar{\psi}$, process \circled{C} in Fig.~\ref{fig:cartoon thermalization}), which begins significantly increasing the dark fermion number.
Since $\omp^2 \propto \nps$, $\omp$ begins to grow and becomes much larger than $H$.  We denote the time at which this happens by $H = \Hplas$, because plasma effects become the dominant influence on the subsequent evolution of the dark sector. 

\subsubsection{Towards thermalization} 
\label{sec:towardsthermalization}

The evolution of the dark sector beyond $\Hplas$ involves a complicated interplay between the superhorizon longitudinal $A'$ modes still entering the horizon, and the subhorizon dark sector plasma, but in this section we are primarily concerned with how large the dark sector coupling needs to be to ensure that the dark sector thermalizes.  In particular, we want the dark sector to thermalize at a temperature $\TDM \gtrsim \mps$, because if this happens we can trust the thermal freezeout calculations of Sec.~\ref{sec:relicAbundance}.  It turns out that the final criterion is quite simple.  The most stringent requirement to ensure thermalization at some $\TDM \gtrsim \mps$ is to require that some $3 \to 2$ process (such as $A' A' \psi \to A' \psi$, corresponding to the diagram \circled{F} in Fig.~\ref{fig:cartoon thermalization}) be efficient for an almost-thermal bath at $\TDM \sim \mps$.  This is because the initial state is highly non-thermal with a very large total number density $n_\text{dark}$, and interactions become less and less efficient as number densities decrease.  The most difficult step is thus the final step, when the number densities are the smallest and the system is almost completely thermalized already.  We can estimate the rate for such a $3 \to 2$ process in an approximate thermal equilibrium at temperature $T$ as:
\begin{equation}
    \Gamma_{A' A' \psi \to A' \psi}(T) \sim \frac{9\zeta(3)^2}{8\pi^7}\qd^6 T
\end{equation}
and we simply require that $\Gamma_{A' A' \psi \to A' \psi}(\TDM \sim \mps) \gtrsim H|_{\TDM \sim \mps}$.

Assuming this happens, we will be left with a thermalized dark sector with temperature:
\begin{equation}
\TDM(a) =\Tstar \left(\frac{\astar}{a} \right), \quad  
\Tstar \equiv \TDM(\astar)= \left( \frac{15}{2 \pi^4 \Gstarinf} \mA^2 \HI^2 \eta^2 \right)^{1/4} 
\label{eq:TDMa}
\end{equation}
where $\astar$ is the value of scale factor when $H = \mA$, $\Gstarinf$ counts the total number of relativistic degrees of freedom in the dark sector at early times (see Eq.~\ref{eq:Gstar}) and $\eta$ is a logarithmic factor of $\mathcal{O}(3-6)$ that depends on the details of the plasma effects alluded to above.  In particular, $\eta$ is slightly increased relative to the no-fermion case of Eq.~\ref{eq:eta pure A'} because all modes entering the horizon before $a_{\text{f.o.}}$ contribute.

Taking into account that in a radiation-dominated universe the SM temperature is given by $\TSM= \left(\frac{45}{4 \pi^3 \gstar} H^2 \mpl^2 \right)^{1/4}$, we can define the ratio between the dark sector and SM temperatures, $\xi$:
\begin{equation}
\xi \equiv \frac{\TDM}{\TSM}=\left(\frac{2 \gstar}{3 \pi \Gstarinf}\right)^{1/4}\sqrt{\frac{\HI \eta}{\mpl}}
\end{equation}
This estimate for $\TDM$ can be combined with the above discussion about $3 \to 2$ processes (like \circled{F} in Fig.~\ref{fig:cartoon thermalization}) to determine the lower bound on the coupling $\qd$ to ensure thermalization at $\TDM \gtrsim \mps$, which for $\eta \sim 6$ is given by:
\begin{equation}
\label{eq:minQforEquilibrium3to2}
    \qd \gtrsim 0.01 \left( \frac{\mps}{1 \GeV} \right)^{1/6} \left( \frac{6 \times 10^{13} \GeV}{\HI} \frac{6}{\eta} \right)^{1/6}
\end{equation}
This constraint is plotted with a solid red line in Fig.~\ref{fig:lightPhotonRelicAbundance}.

At smaller values of the coupling complete thermalization is not achieved, and the true final abundance in the dark sector will begin to diverge from that predicted in Sec.~\ref{sec:relicAbundance}.  We should note that Fig.~\ref{fig:lightPhotonRelicAbundance} still shows the rate for $2 \to 2$ processes. These processes preserve the total number of particles, i.e.~the sum of dark photons, longitudinal and transverse, as well as fermions and antifermions. Given that we start from a state that has such high occupation numbers, this process alone is not enough to guarantee thermalization. SFQED processes as well as the acceleration of fermions in the background electric field are essentially number changing processes that reduce the total number of particles, so we believe that these, in combination with the $2\to 2$ processes, will drive us close to thermal equilibrium but we have no way to estimate how close that will be. This is the reason why we still show the rate for this process and do not dismiss that part of the parameter space. We leave a more detailed analysis of this regime for future work.

We would next like to comment on the evolution of the superhorizon plasma acoustic modes once the system achieves thermalization and the interactions in the dark sector freeze-out.  While the plasma frequency $\op$ remains large compared to $\mA$, superhorizon modes remain frozen until they re-enter the horizon. Then they start oscillating with frequency $\sim k/a$ and redshift as radiation. When the plasma mass drops below the dark photon mass, the current associated both with the superhorizon modes and with those that entered after freeze-out is forced to scale like $\nps \propto a^{-3}$, and in turn, the field $A'$ has to follow suit.  These observations are backed both by numerical simulations of the full coupled equations for $A'$ and the currents in an FRW background (see App.~\ref{app:ProcaPlasmasGR}) and by analytic approximations (see App.~\ref{app:SH_plasma}). At this point, the energy density in \emph{all} plasma modes starts redshfiting away extremely fast and proportionally to $a^{-8}$. What is left in the  dark sector is a relic abundance of thermal fermions and a subdominant component of photons.

Finally, until the time of freeze-out, each acoustic plasma mode that enters Hubble contains $H^2 \HI^2$ energy density. Comparing this to the energy density stored in the SM degrees of freedom $\sim H^2 \mpl^2$, we find that the ratio is fixed to be $\frac{\HI^2}{\mpl^2}$, i.e. it is proportional to $\xi$. This has a very important implication for the dynamics of the dark sector that we just described: As long as reheating completes before freeze-out occurs in the dark sector, our conclusions for the evolution of the dark sector remain unchanged up to the value of $\eta$, which introduces only a logarithmic dependence of the value of $\TRH$, as explained in App.~\ref{app:reheating}. $\TRH$ needs to be larger than $\frac{T_\text{f.o.}}{\xi}$ for our story to remain unchanged. As we will see in Sec.~\ref{sec:relicAbundance}, this means that $\TRH$ can be as low as $\mathcal{O}(\text{TeV})$.

\subsection{Thermalization for a heavy \texorpdfstring{$A'$}{A'}} \label{sec:heavyPhotonThermalization}
If $\mA>2\mps$, the decay $A' \to \overline{\psi} \psi$ is kinematically allowed, which significantly changes the cosmological history.  The decay rate is given by
\begin{equation}
    \label{eq:heavyPhotonDecayRate}
    \GamA \approx \frac{1}{3} \ald \mA \,,
\end{equation}
so provided $\qd$ isn't vanishingly small (say, $\qd > 10^{-10}$), all photons eventually decay to fermions (and the fermions become non-relativistic) long before the SM nucleosynthesis.  The final step in the cosmological history of the dark sector is then not freeze-out as for the light $A'$ case, but $A'$ decay. 
We can distinguish two opposite limits to streamline the discussion for heavy $A'$.

If the coupling $\qd$ is large enough, then the dark sector successfully thermalizes before $H \sim \GamA$.\footnote{There is a minor subtlety here because in the presence of a thermal background with $\TDM \gtrsim \mA$, the dark fermions pick up a thermal mass of $\mathcal{O}(\TDM)$ and kinematically prevent the direct photon decay.  If the sector has thermalized, then $\TDM \gg \mA$ when $H \sim \GamA$, so the photons will not decay until a much later time.  Still, to be conservative we limit ourselves to the case where the sector has thermalized by $H \sim \GamA$ because if thermalization were significantly delayed then some dark photons could have already begun to decay before thermalization is achieved and a hot thermal background is formed.  Once the sector has thermalized, number densities always evolve as $a^{-3}$ so it makes no difference that the actual photon decay is delayed until $\TDM \lesssim \mA$.} The detailed history of this scenario is quite similar to that illustrated in the light $A'$ case throughout Sec.~\ref{sec:lightPhotonThermalization} and App.~\ref{app:thermalizationAppendix}. The main qualitative difference is that now the cascade production is always kinematically allowed ($\mps$ is now smaller than both the other relevant scales $\mA$ and $\sqrt{E'}$) for the couplings $\qd\gtrsim\mathcal O(10^{-1})$ that apply to this case. The ultimate requirement for thermalization is again the efficiency of $3\to2$ processes.  
The calculation of the present-day dark matter density is thus essentially identical to the case of a fermion that freezes out from the dark sector thermal bath while still relativistic (as a hot relic), and the final $\nps$ after $A'$ decay just consists of $\nps+2\nA$ evaluated before decay. The result is discussed in Sec.~\ref{sec:heavyPhotonRelicAbundance}.

In the opposite limit $\qd\ll 1$, the dark sector never reaches thermalization. In the light $A'$ case, this implies that the cosmological history is mostly identical to the pure $A'$ case of Ref.~\cite{Graham:2015rva}, because the negligible fermion production does not significantly reduce $\nA$. In the heavy $A'$ case though, given that eventually all $A'$'s decay to fermions, the cosmological history can vary significantly with respect to the pure $A'$ case. 
At early times, $\nA$ can be significantly reduced in two ways. Rates for perturbative $n\to 2$ annihilations of many $A'_L$ into $\psi\overline \psi$ are boosted by the very large occupation numbers present for $A'_L$ at horizon crossing, and this boost can beat the suppression due to higher powers of $\qd$, thus overcoming the rates for $2\to2$ processes. Even when $\qd$ is small enough to neglect these, SFQED processes are very effective in the whole parameter space (see App.~\ref{app:thermalization heavy A}) leading to $\psi$ as a viable DM candidate.
These two effects, and the absence of thermalization in the dark sector, preclude an analytical control of the DM abundance for this case. We expect numerical simulations to confirm that fermions with $\mps \sim \keV-\GeV$ and coupling $\qd<\mathcal O(10^{-2})$ can achieve the right relic abundance through this mechanism, where the lower bound of $\mathcal{O}(\text{keV})$ is observational (LSS constraints on warm dark matter) rather than theoretical.  Despite the potential interest of light DM candidates with such a minimal production mechanism, we anticipate from Sec.~\ref{sec:dark_fermion_heavy} that the introduction of interactions (required for experimental direct detection) like kinetic mixing leads to a much larger cosmological production through freeze-in.  This inflationary production is thus subdominant in those regions of parameter space which will be probed by upcoming direct detection experiments, and so the heavy $A'$ case is less interesting than the light $A'$ case from an observational point of view.

\section{Dark matter relic abundance} 
\label{sec:relicAbundance}
Armed with an understanding of how the dark sector is populated and under what conditions it thermalizes, we have all we need to compute the present-day abundance of both the dark fermions and the dark photons.  In this section we do precisely that, splitting our discussion into two cases depending on whether the dark photon is lighter (Sec.~\ref{sec:lightPhotonRelicAbundance}) or heavier (Sec.~\ref{sec:heavyPhotonRelicAbundance}) than the dark fermion.

\subsection{Light photon \texorpdfstring{$(\mA < \mps)$}{(mA' < mPsi)}} 
\label{sec:lightPhotonRelicAbundance}
For the case of a dark photon lighter than the dark fermion, there will in general be surviving populations of both dark fermions and dark photons.  We wish to calculate the abundances of each.  In this paper we have limited ourselves to the regions of parameter space where the dark sector comes to thermal equilibrium, implying that the final abundance of the dark fermion (the heavier particle) will be set by a standard thermal freeze-out calculation (see e.g.\ Chapter 5 of Ref.~\cite{Kolb:1990vq}) with one small but significant adjustment.  In the usual freeze-out computations for a particle coupled to the SM thermal bath, the temperature that sets the Hubble rate is simply the temperature of the relevant thermal bath.  In our case, however, the SM radiation bath dominates the energy density of the universe and thus sets the Hubble rate, but the temperature of the dark sector thermal bath is much lower, as can be seen in Eq.~\ref{eq:tempratio}.

It turns out that we can account for this in a very simple way.  Since the change only affects the Boltzmann equations for freeze-out through the Hubble rate $H$, we can write:
\begin{equation}
\label{eq:m_pl eff}
    H^2 = \frac{8 \pi}{3} \frac{1}{\mpl^2} \frac{\pi^2}{30} \gstar \TSM^4 
    = \frac{8 \pi}{3} \frac{1}{\mpl^2} \frac{\pi^2}{30} \Gstar \TDM^4 \left( \frac{\gstar}{\Gstar} \frac{1}{\xi^4} \right)
    = \frac{8 \pi}{3} \frac{1}{(\mpl^{\mathrm{eff}})^2} \frac{\pi^2}{30} \mathcal{\Gstar} \TDM^4
\end{equation}
where we have defined $\mpl^{\mathrm{eff}} = \mpl \xi^2 \sqrt{\frac{\Gstar}{\gstar}}$.  Since $\mpl$ only enters into the standard freeze-out calculation through the Hubble rate, this allows us to adapt the standard results by first replacing all SM quantities with their DM counterparts, and by then replacing $\mpl$ by $\mpl^{\mathrm{eff}}$.

One last important ingredient we need for this calculation is the fermion annihilation cross-section, which sets the time (and the temperature) at which the fermion freezes out.  For $E_\psi \lesssim \mps$, this is given simply by:
\begin{equation}
    \label{eq:ann_cross_section}
    \langle \sigma_\text{ann}v\rangle \approx \frac{\pi\ald^2}{\mps^2}
\end{equation}

\begin{figure}[ht]\centering
    \includegraphics[width=.75\textwidth]{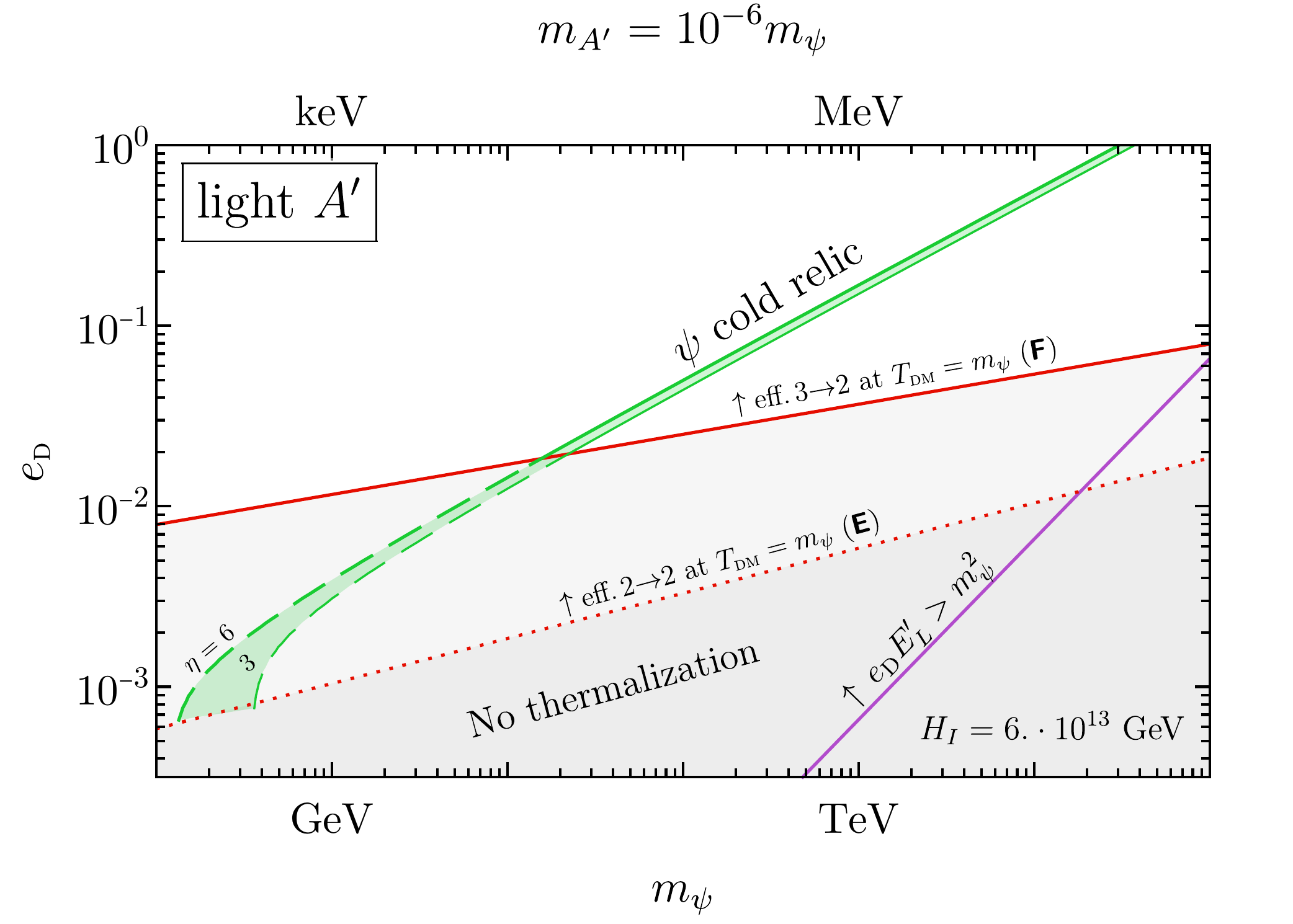}
    \caption{Parameter space where $\psi$ achieves the DM relic abundance via freeze-out for light dark photons ($\mA = 10^{-6} \mps$).
    The dark sector fully thermalizes when $3\to2$ processes like \circled{F} in Fig.~\ref{fig:cartoon thermalization} are efficient at late times, which occurs for large enough $\qd$ as given in Eq.~\ref{eq:minQforEquilibrium3to2}, marked by a red solid line (a more detailed account can be found in Secs.~\ref{sec:thermalization} and \ref{app:thermalizationAppendix}).
    The condition of kinetic equilibrium at late times, when $2\to2$ processes like \circled{E} are efficient, is shown with a red dotted line and marks the region below which the final state of the dark sector appreciably deviates from the thermal one.
    The green line (plotted for $\eta = 3$ and $\eta = 6$) identifies where $\psi$ reaches the DM relic abundance as a cold relic through freeze-out. The line is solid where we can fully trust the thermal relic calculation, and becomes dashed when $\mathcal O(1)$ deviations could appear.
    Finally, the purple line marks where the dark electric field at horizon crossing at early times is larger than the critical field, see the discussion in Apps.~\ref{app:SFQED} and \ref{app:thermalizationAppendix}.
    }
    \label{fig:lightPhotonRelicAbundance}
\end{figure}

One may wonder whether this process is fast or slow compared to the Hubble rate $H$ at a time when $\TDM \sim \mps$, since this determines whether the fermion freezes out while still relativistic (a hot relic) or while nonrelativistic (a cold relic).  
In fact, the requirement that the dark sector reaches thermal equilibrium before $\TDM \sim \mps$ implies that the process \circled{E} in Fig.~\ref{fig:cartoon thermalization} (or other $2 \to 2$ processes which have similar cross-sections) is still active at $\TDM \sim \mps$, implying that we can never truly be in the hot relic regime. 
This requirement for efficient $2\to2$ processes can be computed analogously to Eq.~\ref{eq:minQforEquilibrium3to2} and leads to the red dotted line in Fig.~\ref{fig:lightPhotonRelicAbundance}.
The same line thus also serves as the rough dividing line between the hot relic case (which we do not plot below the dotted red line) and the cold relic case (where the thermal relic calculation is fully trustable).  

With this discussion, we can finally quote the results.  Although we have numerically integrated the Boltzmann equation to obtain the line plotted in Fig.~\ref{fig:lightPhotonRelicAbundance}, a very good analytic approximation can be obtained in both the hot relic and cold relic portions. 

In the cold relic case, we simply have the standard result from e.g.\ Chapter 5 of Ref.~\cite{Kolb:1990vq} with the replacements discussed above, yielding a final answer of:
\begin{equation}
\label{eq:psi_freeze-out}
      n_{\psi,0}^{\mathrm{cold \; relic}} =\frac{x_\text{f.o.}}{\lambda}s_{\textsc{dm},\star}\pare{\frac{\astar}{a_0}}^3,
\end{equation}
where $s_{\textsc{dm},\star} \equiv \frac{2 \pi^2}{45} \GstarS (\Tstar) \Tstar^3$ and
\begin{equation}
\label{eq:lambda_fo}
\lambda = 0.264(\GstarS/\Gstar^{1/2})\mpl^\text{eff}\mps \langle \sigma_\text{ann} v \rangle
      = 0.21 \HI\eta \mps \langle \sigma_\text{ann} v \rangle,
\end{equation}
with $\gstar,\gstarS,\Gstar,\GstarS$ evaluated at $\TDM = \mps/x_\text{f.o}$ and $\eta$ a logarithmic factor discussed in App.~\ref{app:towards thermalization}.  Finally we have:
\begin{equation}
\label{eq:xfo}
x_\text{f.o.} \equiv \frac{\mps}{\TDM}\bigg|_\text{f.o.}\simeq \ln(0.193\lambda) -\frac{1}{2}\ln\big(\ln(0.193\lambda)\big).
\end{equation}
Altogether,
\begin{align}
\label{eq:lightAColdRelicAbundance}
\Omega_{\psi,0}^{\mathrm{cold \; relic}} &\approx 532 \, x_\text{f.o.} \qd^{-4} \frac{\mps^2 T_0^3}{H_0^2 \mpl^3} \sqrt{\frac{\eta \HI}{\mpl}} \\
&\approx 0.47 \left(\frac{\mps}{10 \GeV} \right)^2 \left( \frac{10^{-2}}{\qd} \right)^4 \left( \frac{\eta}{6} \frac{\HI}{6 \times 10^{13} \GeV} \right)^{1/2} \nonumber
\end{align}
where $T_0 \approx 2.725 \, \text{K}$ is the present-day CMB temperature.

For completeness, we also give the result for the hot relic case. As stated before, strictly speaking this is not a physically realizable case, but it  gives the lower end of the $\mps$ range giving the right abundance:
\begin{equation} \label{eq:hotRelicFermionAbundance}
    \Omega_{\psi,0}^{\mathrm{hot \; relic}} \approx 0.29 \, \frac{\mps T_0^3}{\mpl H_0^2} \left( \frac{\eta \HI}{\mpl} \right)^{3/2} \approx 0.19 \left( \frac{\mps}{100 \MeV} \right) \left( \frac{\eta}{6} \frac{\HI}{6 \times 10^{13} \GeV} \right)^{3/2}
\end{equation}

At late times the dark sector energy density will be dominated by dark fermions, but we should also estimate the fraction of this density in the dark photon field.  We note that once $\psi$ is out of equilibrium, the dark photons no longer interact with each other, so their number density simply redshifts as $a^{-3}$.  To find their number density today it is thus sufficient to take their number density at any time after fermion decoupling (we choose the time when the dark sector temperature is $\TDM = \mA$ and the photons are beginning to be nonrelativistic) and redshift it.  We thus obtain:
\begin{equation}
    n_{A',0} = \frac{\zeta(3)}{\pi^2} \mathcal{G}_{A'} \mA^3 \left( \frac{a_\mathrm{NR}}{a_0} \right)^3
\end{equation}
where $a_\mathrm{NR} = a|_{\TDM = \mA}$ and $\mathcal{G}_{A'} = 3$ counts the number of degrees of freedom of the dark photon field.  
Evolved to the present-day this yields:
\begin{equation}
\label{eq:dpabundance}
\Omega_{A',0}^{\text{light $A'$}} \approx
0.29 \frac{\mA T_0^3}{H_0^2 \mpl^2} \left( \frac{\eta \HI}{\mpl} \right)^{3/2}
\approx 0.0019  \left( \frac{\mA}{\MeV} \right) \left(\frac{\HI}{6 \cdot 10^{13} \GeV} \frac{\eta}{6}\right)^{3/2}
\end{equation}

Provided that $\mA\lesssim 1 \MeV$, dark photons form only a small fraction of the present-day dark matter density. Still, we discuss probes of this dark matter subcomponent in Sec.~\ref{sec:darkPhotonConstraints}.

\subsection{Heavy photon \texorpdfstring{$(\mA > 2\mps)$}{(mA' > 2 mPsi)}} \label{sec:heavyPhotonRelicAbundance}
As already discussed in Sec.~\ref{sec:heavyPhotonThermalization}, in the heavy $A'$ case the final $\psi$ DM abundance amounts to sum of $\nps$ and $2\nA$ (due to $A'\to\psi \overline \psi$) evaluated before $A'$ decay.
\begin{figure}[ht] \centering
 \includegraphics[width=0.75\textwidth]{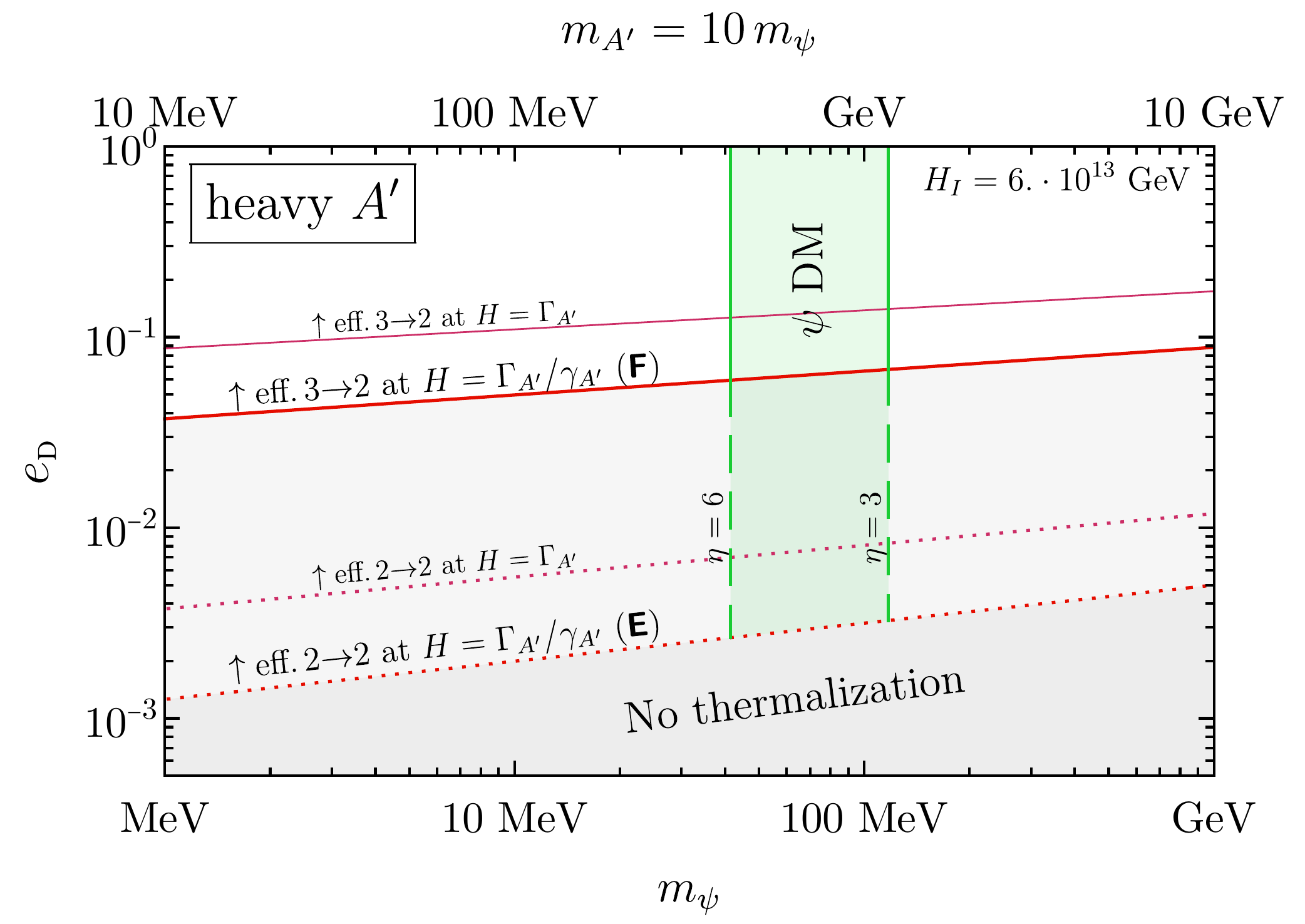} 
 \caption{
 Parameter space where $\psi$ reaches the DM abundance for the case of heavy $A'$ ($\mA=10\mps$), for $\HI=6\cdot 10^{13}\GeV$.
 The red lines mark the regions where the dark sector reaches chemical and kinetic equilibrium (respectively, $3\to2$ and $2\to2$ processes are efficient) before all the $A'$'s decay to $\psi\overline \psi$, accounting for the time dilation $\gamA\sim \TDM/\mA$. For comparison, the thinner violet lines show the rates at $H=\GamA$.
 The green band (corresponding to a variation $\eta=3-6$) shows where $\psi$ achieves the correct DM relic abundance.
 }
 \label{fig:heavyPhotonRelicAbundance}
\end{figure}

We focus on the case in which the dark sector thermalizes before $A'$ decay. As for the light $A'$ case, the condition to impose is that total number-changing $3\to 2$ processes (\circled{F}) are efficient at some time. The latest time at which this can occur is right before $A'$ decays, which corresponds to a time $\GamA$ multiplied by the typical boost factor $\gamma_{A'}$ to account for the relativistic time dilation in the cosmological frame. We evaluate therefore the $3\to 2$ and $2\to 2$ rate (corresponding to kinetic equilibrium, and signalling that the system can reach a state close to thermal and our predictions should be approximately valid) at $H=\GamA/\gamma_{A'}$. The corresponding rates are collected in Eq.~\ref{eq:heavyAthermalizationRates}. 

Just as in the light photon case, the dark sector reaches a thermal bath with temperature $\Tstar$ when $H \sim \mA$ (see Eq.~\ref{eq:TDMa}). All the entropy of this bath will eventually end up as dark fermions, making the final number density of dark fermions in this case essentially identical to that computed in the hot relic case in Sec.~\ref{sec:lightPhotonRelicAbundance}. In fact the only difference is that in this case, \textit{all} entropy ends up in the fermions, whereas in the hot relic case some of it ends up in the residual dark photons (because in that case they are light and thus kinematically forbidden from decaying).  The estimate of the present-day relic abundance is thus modified from Eq.~\ref{eq:hotRelicFermionAbundance} by a ratio of degrees of freedom, with fermions weighted by a factor $3/4$:
\begin{equation}
\label{eq:heavyARelicAbundance}
    \Omega_{\psi,0}^{\text{heavy $A'$, early therm.}}
    \sim \frac{3 \cdot 2+\frac{3}{4}(4)}{\frac{3}{4}(4)} \Omega_{\psi,0}^{\mathrm{hot \; relic}}
    \approx 0.23 \left( \frac{\mps}{40 \MeV} \right) \left( \frac{\HI}{6 \cdot 10^{13} \GeV} \frac{\eta}{6} \right)^{3/2}
\end{equation}
The corresponding band for the right relic density (spanning the range between two representative values $\eta=3$ and $\eta=6$) is shown in green in Fig.~\ref{fig:heavyPhotonRelicAbundance}. 
We note that the abundance is independent of $\mA$ and $\qd$ in this case, which can be understood intuitively because if thermalization happens, neither of these quantities affects the dark sector temperature.  Because the present-day dark sector energy density is determined by its number density, and its number density is determined by its temperature, neither $\mA$ nor $\qd$ can thus affect the abundance.

\section{Signatures and Constraints} \label{sec: signatures}
In this section we discuss signatures and constraints of our mechanism. In Sec.~\ref{sec:halo_signatures} we revisit the effect coming from the ellipticity of observed galaxies, which assumes no interactions between the Dark Sector and the SM other than gravitational, and can be an exciting future probe.  We also check that within our parameter space the fermions in the late-time universe are unable to form bound states.  In Sec.~\ref{sec:def_kin_mixing} we introduce a kinetic mixing term between the Dark Photon and the SM photon. First, in Sec.~\ref{sec:mixing_no_fermions}, we discuss the effects of a SM plasma on the dark photon plasma modes.  We also review constraints on the mechanism of Ref.~\cite{Graham:2015rva} (which we review in Sec.~\ref{sec:GPPmassiveVector}), where the dark photon is the Dark Matter and there are no dark fermions in the theory. Next, in Sec.~\ref{sec:darkPhotonConstraints}, we review existing constraints in the dark photon parameter space and explain how some of them are modified given our production mechanism. Finally, we discuss the dark fermion parameter space in Sec.~\ref{sec:dark_fermion_light} for the ``light dark photon case'' and in Sec.~\ref{sec:dark_fermion_heavy} for the ``heavy photon case.'' 

\subsection{Astrophysical effects: Galactic Halos and Dark Positronium}
\label{sec:halo_signatures}
\underline{Galactic Halos} Dark sectors with internal interactions (what is usually called Self-Interacting Dark Matter (SIDM) in the literature) can lead to observable consequences on astrophysical scales, even if their interaction with the SM is purely gravitational. One of the strongest effects comes from the observed triaxiality of DM halos, whose anisotropies would be smoothed out by Coulomb collisions of the charged fermions \cite{PhysRevD.79.023519,Agrawal_2017}. Essentially the thermalization timescale $\tau_\text{iso}$ must not be too short, namely any $\tau_\text{iso}\lesssim$ few Gyr in galaxies~\cite{Agrawal_2017} would be excluded. This timescale is given by \cite{Agrawal_2017}\footnote{There is a disagreement in the calculation between Refs.~\cite{PhysRevD.79.023519} and~\cite{Agrawal_2017}. We adopt the more conservative result of \cite{Agrawal_2017} here.} 
\begin{equation}
    \tau_\text{iso}=\frac{3}{16\sqrt{\pi}}\frac{\mps^3 v^3}{\ald^2\rho_\psi}\frac{1}{\log\frac{\mps v^2}{\ald\mA}},
\end{equation}
where $v$ is the velocity of DM. Any dark charge with roughly
\begin{equation}
    \qd\gtrsim3\times 10^{-5}\pare{\frac{\mps}{\MeV}}^{3/4}
\end{equation}
is thus potentially excluded. This constraint is weaker than the relic abundance lines of Figs.~\ref{fig:lightPhotonRelicAbundance} and \ref{fig:veryLightPhotonRelicAbundance} and so we do not include it in our plots. In fact, Ref.~\cite{Agrawal_2017} notes that this calculation should not be taken as a strict constraint. For instance, the authors calculate the relaxation of the \emph{velocity} distribution, which overestimates the relaxation of the \emph{number density} distribution in a dynamic galactic halo. Thus, this constraint is possibly even further above our prediction. We refer the interested reader to Ref.~\cite{Agrawal_2017} for a complete discussion of the several uncertainties of this bound. 

Nevertheless, it is worth emphasizing that future observations of galactic halo profiles will be able to probe smaller DM self-couplings \cite{Agrawal_2017} in a more robust way, by using input from possible $N$-body simulations and by improving the statistics of studied galaxies. This would constitute an exciting probe of our scenario, since it does not assume any interaction with the SM, other than gravitational. We leave projections of this to future work.

\underline{Dark Positronium} Another possible effect of the dark sector self-interactions is the formation of $\psi \bar{\psi}$ bound states reminiscent of positronium.  These can only form in the ``light photon'' ($\mA \ll \mps$) case because in order to form a pair of slow-moving fermions they must scatter inelastically, which requires radiating a dark photon.  The formation of such states would be a strong constraint on our parameter space since they would, after some time, annihilate into dark photons and thus deplete the present-day dark matter.  Ref.~\cite{Cyr-Racine:2012tfp} finds that the formation of dark positronium requires the dark charge to be larger than:
\begin{equation}
    \qd \gtrsim 0.09 \left( \frac{\xi}{10^{-2}} \right)^{1/8} \left( \frac{\mps}{1 \GeV} \right)^{1/4}
\end{equation}
where $\xi$ is the ratio of dark to SM temperatures.  A simple comparison of this bound with the expression for the late-time relic abundance (Eq.~\ref{eq:lightAColdRelicAbundance}) demonstrates that it is satisfied for all parameter space that yields the correct dark matter abundance while remaining perturbative (i.e. $\qd \lesssim 1$).  We thus are free from constraints involving the formation of dark bound states.

\underline{Plasma Instabilities} Finally, a dark sector consisting of charged fermions and a light mediator can exhibit certain plasma instabilities that may be of astrophysical relevance.  This possibility has been studied recently by Ref.~\cite{Lasenby:2020rlf}, which finds that for the dark photon masses considered in this paper the effects are negligible.  We thus ignore them.

\subsection{Kinetic Mixing}
\label{sec:def_kin_mixing}
Everything discussed up to this point has assumed a Dark Sector interacting by itself and coupled only gravitationally to the SM. For the remainder of this section, we introduce kinetic mixing between the dark photon and the SM photon. In the low energy theory the mixing manifests itself as a coupling of $U(1)_\textsc{em}$ to $U(1)_D$ characterized by a dimensionless parameter $\epsilon$. The kinetic term of the Lagrangian can be diagonalized, which introduces a direct coupling of the dark photon field with the SM current. To $\mathcal{O}(\epsilon)$ the Lagrangian becomes
\[\mathcal{L}\supset-\frac{1}{4}F^2-\frac{1}{4}F'^2+\frac{1}{2}\mA^2 A'^2+e J_\text{EM}^\mu \pare{A_\mu-\epsilon A_\mu'}+\qd J'^\mu A_\mu'\]
where the primes denote the dark sector operators and the unprimed are the usual QED quantities. We have assumed a St\"{u}ckelberg mass for the dark photon.  We identify the field $A_\mu$ as the photon and the field $A_\mu'$ as the dark photon. 

This new interaction constrains the parameter space of the production mechanism discussed in Secs. \ref{sec:thermalization} \& \ref{sec:relicAbundance}, but also allows for the possibility of detecting both dark states. 

\subsubsection{Cosmological effects of the Standard Model plasma}
\label{sec:mixing_no_fermions}

\underline{The case of dark QED} A kinetic mixing with the SM photon allows the dark sector to be directly probed, but it also forces us to consider how such a sector may be affected by the hot SM plasma.  We recall that the SM temperature $\TSM \gg \TDM$, and so one may be concerned that the effects of the SM plasma will dominate those of the dark plasma\footnote{Since dark matter abundances coming from freeze-in are not in thermal contact with the SM, so is our dark sector, as we are concerned with $\epsilon$ smaller than the ones setting the freeze-in abundance, see Fig.~\ref{fig:Fermion-bounds_light-A'}.}.  In fact this is only a very mild constraint on the kinetic-mixing parameter $\epsilon$.  The leading-order effect of the SM plasma is simply to shift the dark photon's plasma frequency:
\begin{equation}
    \big(\omega_{\textsc{p},\text{dark}}^{\text{(kin. mix.)}}\big)^2 \approx \big(\omega_{\textsc{p},\text{dark}}^{\text{(no kin. mix.)}}\big)^2 + \epsilon^2 \big(\omega_{\textsc{p,sm}}\big)^2
\end{equation}
and since in thermal equilibrium we have $\omega_{\textsc{p,sm}} \sim e \TSM$ and $\omega_{\textsc{P},\text{dark}}^{\text{(no kin. mix.)}} \sim \qd \TDM$, this is a small effect provided $\epsilon \ll \frac{\qd}{e} \frac{\TDM}{\TSM}$.  In the mechanism discussed here, the hierarchy between $\TDM$ and $\TSM$ is roughly $\frac{\TDM}{\TSM} \sim \sqrt{\frac{\eta \HI}{\mpl}}$, which for the high-scale inflation considered here is only $\mathcal{O}(10^{-3})$.  As we discuss in Sec.~\ref{sec:darkPhotonConstraints}, Sec.~\ref{sec:dark_fermion_light}, and App.~\ref{app:freeze-in}, App.~\ref{app:darkPhotonAppendix}, $\epsilon$ is constrained to be significantly lower than this by a combination of direct detection experiments, astrophysical constraints, and the condition that the dark sector isn't overproduced by SM freeze-in processes.  Within the parameter space considered here, the dark sector is thus free of any significant effects from the SM plasma while thermalized.  After the dark fermions freeze-out, the dark sector plasma frequency drops and the SM plasma can in principle dominate the behavior of any further superhorizon modes, but at this point it is no longer a concern because the dark fermion number density has already been fixed.

Even though fermions cannot be affected after freeze-out, a natural question is whether the SM plasma can dominate over the dark one and modify the evolution of superhorizon modes similarly to the dark plasma, as discussed in Sec.~\ref{sec:lightPhotonThermalization} and Apps.~\ref{app:ProcaPlasmasGR} and~\ref{app:SH_plasma}. For SM effects to be dominant after fermion freeze-out, we would need $\epsilon>\qd\sqrt{\frac{n_\psi}{\mps \omega_{\textsc{p,sm}}^2}}$, where $n_\psi$ is determined by Eq.~\ref{eq:psi_freeze-out}. The most stringent requirement would be at the time when electrons freeze-out, as $n_\psi\propto H^{3/2}$ and $\omega_{\textsc{p,sm}}^2\propto H$. This requirement becomes $\epsilon\gtrsim \frac{8.5\xfo^{1/2}}{e \qd}\sqrt{\frac{m_e (\HI\eta)^{1/2}}{\mpl^{3/2}}}$, which is at most $6\times 10^{-11}$ for $\HI=6\times 10^{13}$ GeV and, thus, does not constrain our mechanism (see Fig.~\ref{fig:Fermion-bounds_light-A'}). After electrons freeze out, the bound becomes even weaker. 

\underline{The case without dark fermions} Another SM effect one may be concerned about is the resonant conversion of dark photons into SM plasmons, which is dominated by the time when the SM plasmon frequency is approximately the dark photon mass: $\omega_{\textsc{p,sm}} \sim \mA$.  If the dark sector has fermions and thermalizes, this effect disappears because the dark photons are no longer in the form of a cold bath that can efficiently resonantly convert, but if we had instead considered the case of no dark fermions, with dark matter composed solely of dark photons produced by inflation, this effect leads to a constraint on $\epsilon$.  We briefly review this here.

In a dark sector composed only of a dark photon, the relic abundance is set simply by the Hubble scale at inflation and the dark photon mass.  Ref.~\cite{Graham:2015rva} finds that the proper DM abundance today is achieved when the dark photon has a mass:
\begin{equation} \label{eq:massForProperAbundancePurePhoton}
    \mA^\text{(pure $A'$)} \sim 6 \times 10^{-6} \eV \left( \frac{10^{14} \GeV}{\HI} \right)^4.
\end{equation}
If this photon has a kinetic mixing, however, then dark photons and SM photons can convert between each other.  Because the early-universe SM energy content is in the form of a hot plasma, the SM photon has an effective plasma mass, and when $\omega_\textsc{p,sm} \sim \mA$ this conversion probability can receive a resonant enhancement.  Since the dark photon is in a very-high-number-density state compared to the SM, this resonant conversion generically depletes the dark matter abundance.  This is not a new idea, and the details have been worked out previously (see e.g.\ Refs.~\cite{Jaeckel:2008fi, Arias:2012az, McDermott:2019lch}), but in many cases it cannot be taken as a true constraint on the dark photon model because the initial dark matter density is virtually unconstrained before the CMB.  Even if the vast majority of dark photons are resonantly converted to SM states, perhaps the initial energy density in the dark sector was just much larger and the proper present-day abundance is only achieved after this resonant conversion.  The only constraint then comes from avoiding a large energy dump into the SM, which would affect for example $N_\text{eff}$ (the effective number of light degrees of freedom in the SM) and can lead to spectral distortions in the CMB.  However, for the inflationary dark photon production proposed by Ref.~\cite{Graham:2015rva}, the dark photon mass fixes the initial abundance and thus this is a real constraint.  

From Ref.~\cite{Arias:2012az} we have that the total depletion rate of the dark photon energy density is given by $\exp{(- \tau)}$ with:
\begin{equation}
    \tau \approx \eps^2 \frac{\pi \mA}{r H_\text{res}}
\end{equation}
where $r$ is an $\mathcal{O}(1)$ factor coming from the evolution of the SM photon plasma mass and $H_\text{res}$ is the Hubble constant at the time when resonance occurs ($\omega_\textsc{p,sm} = \mA$).  For dark photon masses in the $10^{-8} \eV \lesssim \mA \lesssim 1 \keV$ mass range (which includes the range most relevant for inflationary production), this resonance occurs after SM $e^+ e^-$ freezeout but before matter-radiation equality.  The SM photon plasma frequency is then given by:
\begin{equation}
    \omega_\textsc{p,sm} \approx \frac{e^2 \eta_B}{m_e} \frac{2 \zeta (3)}{\pi^2} \TSM^3
\end{equation}
where $\eta_B \sim 6 \times 10^{-10}$ is the baryon-to-photon ratio.  With this we can compute $H_\text{res}$ and find that the total depletion factor is given by $\exp{(- \tau)}$ with:
\begin{equation}
    \tau \approx 3 \frac{1}{r} \left( \frac{\eps}{10^{-9}} \right)^2 \left( \frac{10^{-5}}{\mA} \right)^{1/3}
\end{equation}
Requiring that $\tau \lesssim 1$ then yields a constraint:
\begin{equation}
    \eps \lesssim 6 \times 10^{-10} \sqrt{r} \left( \frac{\mA}{10^{-5} \eV} \right)^{1/6}
\end{equation}
For the pure dark photon (i.e.~no fermions) case, this constraint lies slightly below existing constraints from shifts to $N_{\text{eff}}$, but for the dark QED of this paper, the constraint disappears entirely because the photons do not play the role of the late-time DM.

Finally, one could wonder whether the SM plasma in the pure $A'$ scenario can modify the evolution of superhorizon modes, similarly to what we discussed in Sec.~\ref{sec:lightPhotonThermalization} and Apps.~\ref{app:ProcaPlasmasGR} and~\ref{app:SH_plasma}, where the effect would now come from a plasma frequency $\sim \epsilon\omega_{\textsc{p,sm}}$. Notably, the acoustic mode does not exist in this theory \cite{Dubovsky_2015} and the two modes have frequencies $\sim\op$ and $\sim\mA$ (in the case \emph{with} dark fermions the acoustic mode exists and receives small $\propto \epsilon^2$ corrections). Therefore, we do not expect modifications to the evolution of superhorizon modes. Finally, the decay rate of the dark photon mode is $\gamma\propto \epsilon^2\frac{\omega^2_\textsc{p,sm}}{\mA^2}\nu$ \cite{Dubovsky_2015}, where $\nu$ is the collision rate. A simple estimate reveals that dissipation is a very suppressed effect for the relevant parameters of the pure $A'$ theory.

\subsubsection{Dark Photon parameter space} \label{sec:darkPhotonConstraints}
\begin{figure}[ht]\centering
\includegraphics[width=.9\textwidth]{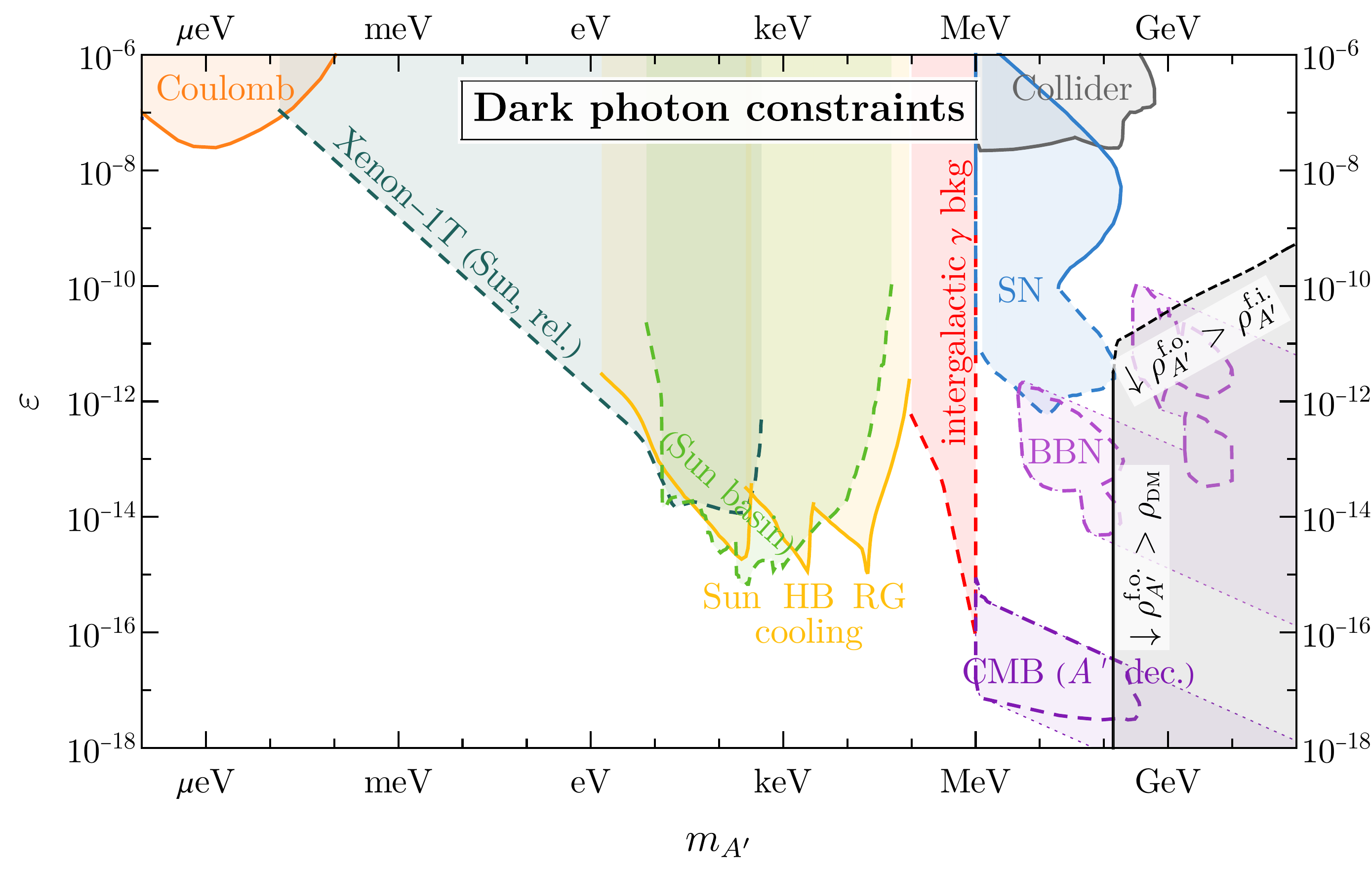}
\caption{Summary plot of the constraints in the plane $(\mA,\varepsilon)$ in the presence of a kinetic mixing.
The colored bounds are experimental and cosmological constraints on dark photons (see the text for their description).
The bounds shown with a dashed line would sensibly shrink or disappear in the case of heavy dark photon ($\mA>2\mps$) if $\qd\gg \epsilon e$, because the dark photon would decay into $\overline \psi \psi$ shortly after production.
The black lines identify the regions where the freeze-out relic density $\rhoA^\text{f.o.}$ would be overabundant with respect to dark matter (solid) or to the freeze-in abundance of $A'$ (dashed). These lines assume $\HI=6 \times 10^{13}$ GeV and $\eta=6$.
}
\label{fig:dark photon constraints}
\end{figure}

The parameter space of the kinetically mixed dark photon, excluding additional BSM states, has been thoroughly constrained by laboratory experiments, astrophysical observations and cosmological measurements. In this section we briefly review the relevant constraints and explain how some of them might change due to the production mechanism considered in this work and in the presence of dark fermions. We refer the reader to App.~\ref{app:darkPhotonAppendix} for a more detailed discussion. More thorough reviews of the Dark Photon parameter space can be found in \cite{2005.01515,1303.1821}.

\uline{Accelerator-based experiments.}
Searches at fixed target experiments and colliders look either for excesses from decays of the dark photon into visible SM particles or for missing momentum/energy from their escape. The former can probe dark photons above an MeV through its decay to $e^+e^-$, whereas the latter can be used if the decay rate to dark states is larger than to SM, which happens if $\mA>2m_\psi$ and $\qd/e>\epsilon$.
Some of these bounds may thus be modified depending on the branching ratio of the dark photon (which depends on the dark coupling and the dark fermion's mass).  We show them as ``Collider'' bounds in Fig.~\ref{fig:dark photon constraints}. 

\uline{Energy injections in Cosmology.} A kinetically mixed dark photon can be produced thermally in the Early Universe. If $\mA >2m_e$, decay channels into SM states are open, leading to possible energy injections during BBN and matter-radiation decoupling. The bounds denoted as ``BBN'' and ``CMB'' in Fig.~\ref{fig:dark photon constraints} were derived in \cite{1407.0993} and assumed a dark photon abundance produced via the freeze-in mechanism.  Their ``island'' shape comes about from the combined requirements of decaying at the right time and of injecting enough energy to alter the standard cosmological evolution at these times. Since the abundance of dark photons in this work is in principle different from the one derived via freeze-in, we find that these constraints are modified. Below the dashed black line labeled ``$\rho_{A'}^\text{f.o.}>\rho_{A'}^\text{f.i.}$'' in Fig.~\ref{fig:dark photon constraints} the dark photon yield of this work is larger than that from freeze-in and independent of both $\epsilon$ and $\mA$. Therefore, we can exclude any dark photon mass and coupling below this line as long as its decay happens during the relevant epoch. For a more detailed calculation we refer the reader to App.~\ref{app:darkPhotonAppendix}.

In addition to these we include the vertical solid line at $\sim 100$ MeV, to the right of which the dark photon abundance of Eq.~\ref{eq:dpabundance} overcloses the Universe if only gravitational interactions are in play. Within this region, we deem robust the extension of the bounds coming from BBN and CMB energy injections as discussed above and in App.~\ref{app:darkPhotonAppendix}. However, we remain uncertain about the rest as it is conceivable that a more proper treatment taking into account our larger final abundance of dark photons rule out part of it. For instance, one might think that even after BBN is over, a decaying dark photon might inject enough charged states to destroy light element nuclei. We do not attempt a complete investigation of these phenomena and leave it to future work.

For a ``heavy dark photon'' these constraints may disappear altogether if it decays to dark fermions faster than it does to SM states. This will be the case for any $\qd/e>\epsilon$. Since these bounds correspond to very small values of $\epsilon$, they do not apply whenever thermalization of the dark sector is efficient.

\uline{Bounds from Supernovae.} Inside the core of SNe the temperature can reach $\sim 30$ MeV, which can thermally produce dark photons below $\sim 100$ MeV \cite{1611.03864, Hardy:2016kme}. If they escape unimpeded, they cool the SNa. Comparison of this energy loss against the well-measured neutrino flux of SN1987a constrains the solid line `SN' region of Fig.~\ref{fig:dark photon constraints}. This argument applies when the theory contains no dark fermions. Their existence can, in principle, modify the bounds \cite{1803.00993} if they can also be produced inside the SNa core. In the ``light dark photon'' case, the large mass hierarchy between $\mA$ and $m_\psi$ of our mechanism means that the dark fermions relevant for these dark photon masses are much heavier than the core temperature and so they cannot be produced efficiently. This renders their existence irrelevant. For the ``heavy dark photon'' case, there could be a modification of the constrained region depending on the mass hierarchy, but we ignore it in anticipation of its uninteresting direct detection prospects (see Sec.~\ref{sec:dark_fermion_heavy}). This is why we are plotting the constraints that correspond to a pure $U(1)_D$ of \cite{1611.03864}. 

In addition to cooling bounds, Ref.~\cite{1901.08596} identified three more ways of using SNe to constrain dark photons, which rely on the decay of the dark photon to positrons. In the ``heavy dark photon'' case, if $\qd/e>\epsilon$ and the dark photon decays primarily to dark fermions, these bounds can be evaded, which is why we draw their boundaries with dashed lines.

\uline{Intergalactic Diffuse Photon Background.} If $\mA<1$ MeV, then the principal decay channel into SM states is to three photons, through an electron loop. The bound labeled ``Thermal'' is derived demanding that the resulting flux does not exceed the observed Intergalactic Diffuse Photon Background \cite{0811.0326}. The original calculation assumed an initial dark photon abundance produced via freeze-in and thus we find that it becomes slightly modified given the dark photon production scenario discussed in this work (see Sec.~\ref{app:darkPhotonAppendix} for details).

\uline{Helioscopes.} Dark Photons produced in the Sun can travel all the way to the Earth and be observed in direct detection experiments \cite{2006.13929}. These bounds do not get modified by the presence of a dark fermion in the ``light photon case''.

The remaining bounds from stellar cooling \cite{An:2013yfc} and modifications of the Coulomb law (from measuring the Coulomb force between charged spheres in laboratory experiments, hence the peak around 1 m) \cite{PhysRevD.2.483} apply without modifications as they do not depend on the specifics of the cosmological production mechanism.

\subsubsection{Dark Fermion Parameter Space}
\label{sec:dark_fermion_light}
\begin{figure}[ht] \centering
\includegraphics[width=.7\textwidth]{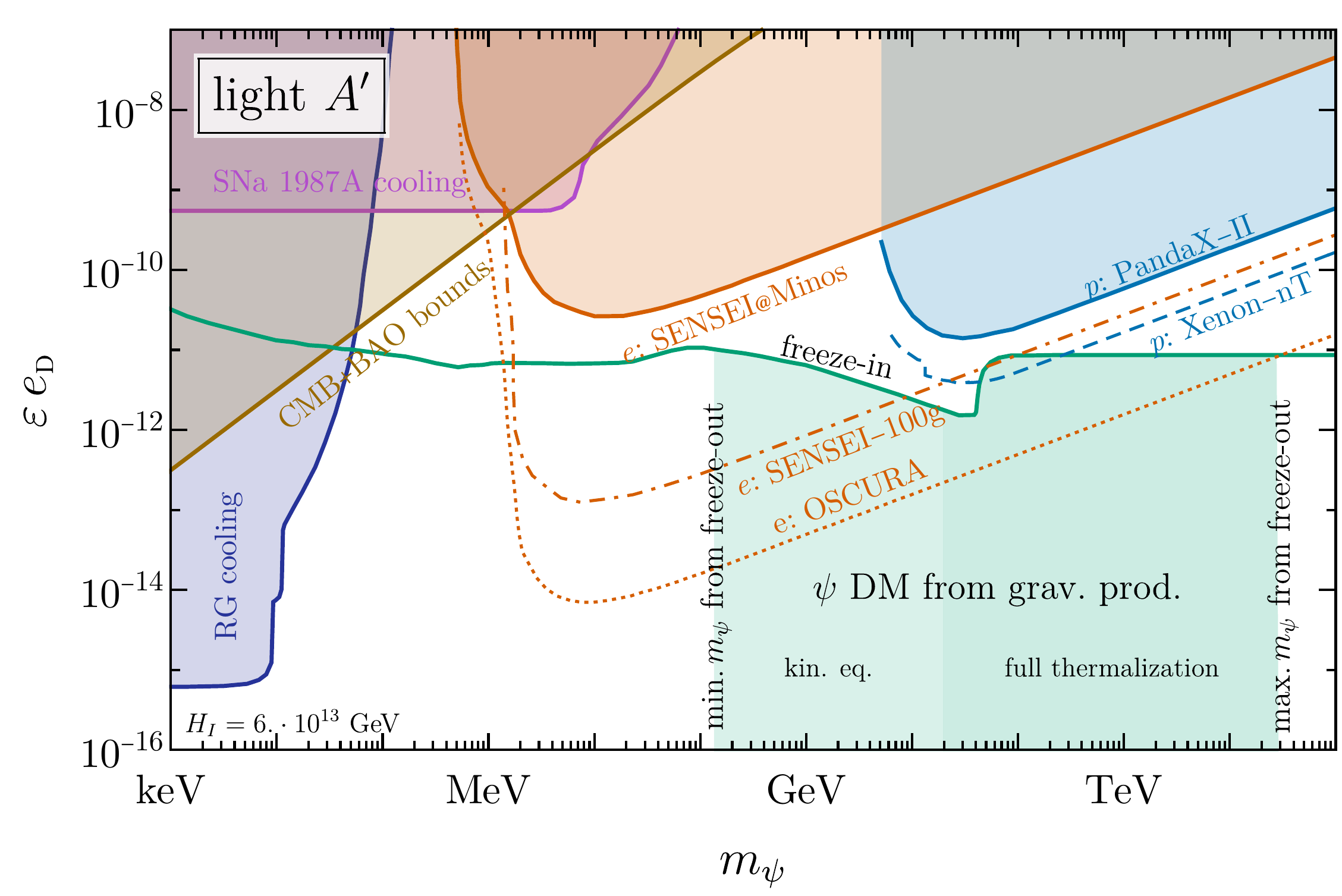}
\caption{The green region shows the dark fermion parameter space, for $\mps\gg\mA$, that can be probed by the mechanism studied in this paper, assuming a kinetic mixing $\epsilon$ between the dark photon and the SM photon. The left and right boundaries are set by the mass range for which the dark fermion can be the dark matter (see Fig.~\ref{fig:lightPhotonRelicAbundance}) and the upper boundary corresponds to the standard freeze-in production of dark fermions from the Standard Model via kinetic mixing \cite{1911.03389,1112.0493,1902.08623} (see App.~\ref{app:freeze-in} for details).
In the darker green region, the dark sector achieves full thermalization, whereas in the lighter green one it only reaches kinetic equilibrium (see Sec.~\ref{sec:thermalization}).
With solid lines we outline the currently constrained regions, choosing representative bounds from electron recoils by SENSEI \cite{1804.00088,1901.10478,2004.11378}, and from nuclear recoils by PandaX-II \cite{1802.06912,2104.14724}. We plot near-future experimental reaches with dashed-dotted (SENSEI-100g \cite{USvisions}) and dashed (Xenon-nT \cite{1907.11485,2007.08796}) lines. The dotted line is the reach for OSCURA \cite{snowmass}, which is more futuristic but relies on existing technology. We also include constraints from cosmological and stellar cooling arguments, as well as from observations of the SNa 1987A, so as to illustrate where our parameter space lies in the broader literature of millicharged particles.
}
\label{fig:Fermion-bounds_light-A'}
\end{figure}

The kinetic mixing gives a small effective \textit{millicharge} to the fermions (at least at momentum transfers larger than the dark photon mass).  Millicharged Dark Matter is, in general, constrained by several cosmological arguments, astrophysical observations and direct detection experiments. For the region of parameter space we are interested in, only the latter are relevant, but we show some of the strongest constraints on Fig.~\ref{fig:Fermion-bounds_light-A'} in order to define the parameter space that is currently being probed. Apart from direct detection prospects, current bounds do not constrain our mechanism.

\uline{Cosmological Constraints} Dark sectors present during the BBN and CMB epochs may contribute to dark radiation, in conflict with measurements of $\Delta N_\text{eff}$ \cite{1311.2600}, or lead to efficient DM-baryon scattering before recombination.  They can also affect the CMB and the linear matter power spectrum on small angular scales \cite{Boddy:2018wzy, 1802.06788,Buen-Abad:2021mvc}. Even though formally these constraints are for millicharged particles, they may in principle apply in scenarios with $m_{A'}\ll m_\psi$. These correspond to the region ``CMB+BAO bounds'' of Fig.~\ref{fig:Fermion-bounds_light-A'}.  The $\Delta N_\text{eff}$ constraints, taken from \cite{1802.06788}, lie inside this region.

\uline{Astrophysical Cooling Arguments} Arguments about cooling of astrophysical objects can be used for new fermionic sectors coupled under a new gauge group. 

\emph{SNa 1987A cooling}: Inside the hot interiors of a SNa proto-neutron star there can be production not only of the dark gauge boson \cite{1611.03864}, as discussed in Sec.~\ref{sec:darkPhotonConstraints}, but also of the dark fermion \cite{1803.00993}. The qualitative shape of the bound differs from that for a dark photon and is flat because in the latter case production decouples as $m_{A'}\epsilon$ due to the effective mixing parameter in the plasma. In contrast the dark fermion may be produced at lower masses without suffering a suppression. In plotting the bound of Fig.~\ref{fig:Fermion-bounds_light-A'} we have taken the millicharged case, which is again a good approximation given the large mass hierarchy needed in the dark sector.

\emph{Stellar Cooling}: Production of DS particles inside stars leads to cooling which can be constrained from astrophysical observations \cite{1911.03389}. These are important for $m_\chi\lesssim 100$ keV, and even though they exclude regions below the freeze-in bound for DM abundance (i.e. may exclude even sub-components of DM), they are irrelevant for the parameter space of interest here. We include them in order to better establish the part of the parameter space that is currently being probed.

\uline{Current Direct Detection bounds} These experiments probe direct interactions of the DM  with atomic electrons or nuclei. In Figs.~\ref{fig:Reach-DD_light-A'} and~\ref{fig:Fermion-bounds_light-A'} (right panel) we plot the bounds from SENSEI \cite{1804.00088,1901.10478,2004.11378}, and from PandaX-II \cite{1802.06912,2104.14724}, which rely on detection of electron and nuclear recoils respectively. For both experiments we recast their cross-section constraints to bounds on $\epsilon \qd$, assuming that the mediator has a mass $\mA$ much smaller than the typical momentum transfer $q_0$ of the experimental process. Note that if $\mA\gtrsim q_0$, which can naturally be the case in our mechanism in both the electron and nuclear recoil probed parameter spaces, the bounds get weaker with a growing mediator mass. The reason is that the propagator becomes $\propto \mA^{-2}$, suppressing the cross-section and thus relaxing the bound on the coupling (i.e. the same observable cross-section demands larger couplings as the mediator mass becomes larger). For SENSEI this translates to no more than a factor of $\sim 3$ around $\mA\sim \alpha m_e\simeq 4\text{ keV}$, but weakens $\propto \mA^2$ at higher masses. The same situation holds for PandaX-II, for mediator masses $\mA\gtrsim 100$ MeV \cite{2104.14724}.
We choose to plot the most restrictive current bounds to highlight the fact that our parameter space has not yet been probed. Several other experiments such as XENON10 \cite{1206.2644,1104.3088}, XENON100 \cite{1605.06262,1703.00910}, DarkSide-50 \cite{1802.06998}, DAMIC \cite{1907.12628}, XENON1T \cite{1907.11485,2006.09721}, SuperCDMS \cite{1804.10697,2005.14067,2007.14289}, EDEL-WEISS \cite{1901.03588,2003.01046}, and CRESST-III \cite{1707.06749,1904.00498,1905.07335} have set similar or weaker limits which we do not include for clarity.

\uline{Near-future projections} We also plot projections of future experiments that rely on existing technologies and are in construction phase. For electron recoils, we choose to plot two upgraded versions of SENSEI, the 100g \cite{USvisions} version (dashed-dotted), and OSCURA \cite{snowmass} (dotted), which is more futuristic as it assumes 30kg-year exposure. For reference, the current bound from SENSEI uses 1g of target material and has more backgrounds than initially expected \cite{USvisions}. However, future versions will be built deeper underground (SNOLAB), significantly improving both backgrounds and kg-year exposure, giving realistic projections several orders of magnitude below the current bound. For nuclear recoils, we plot a rough projection of the Xenon-nT \cite{2007.08796} reach, by rescaling the bounds the XENON collaboration set for self-interacting dark matter \cite{1907.11485} from Xenon-1T. A proper projection line would need a detailed calculation involving the momentum transfer dependence of the cross-section (including the nuclear form-factor), which we do not perform. Other current and future direct detection experiments will probe similar regions of parameter space. For a review we refer the reader to \cite{USvisions}. All plotted projections assume a light mediator (where ``light'' depends on the experiment).

It is worth noting that there are experimental proposals that can probe even lower values of $\epsilon$~\cite{USvisions}, some even covering a sizable part of the green region of Figs.~\ref{fig:Reach-DD_light-A'} and \ref{fig:Fermion-bounds_light-A'}. However, the fact that even the most realistic near-future experiments will start probing our scenario is by itself very exciting. In particular, detection of a dark fermion below the freeze-in line would point towards the production mechanism discussed in this paper and could potentially be an indirect signature of inflation.

\subsubsection{Heavy photon case}
\label{sec:dark_fermion_heavy}
In this case several direct detection bounds and projections can in principle apply to our scenario. These come either from nuclear recoils for $m_\psi\gtrsim 1$ GeV from Xenon-1T \cite{1907.11485} and CRESST-III \cite{1905.06348}, as well as from electron recoils at lower masses, from Xenon-1T \cite{1907.11485} and SENSEI \cite{1804.00088,1901.10478,2004.11378},. The bounds do not differ significantly from those in the light photon case of Fig.~\ref{fig:Fermion-bounds_light-A'} and correspond to the region around $\sigma_p\sim10^{-40}\cm^2$ for nuclear recoil cross-sections. However the freeze-in line for the Heavy DP case lies around $\epsilon \qd\simeq 10^{-11}$ (see App.~\ref{app:freeze-in} and in particular Eq.~\ref{eq:hdp_freezein_bound} for an estimate), which we can recast into a value for $\sigma_p$ using the definition
\begin{equation}
\label{eq:nuclear_cs}
    \sigma_p=\frac{\alem (\epsilon \qd)^2 \mu_{\psi p}^2}{\mA^4},
\end{equation}
where $\mu_{\psi p}$ is the nucleon-DM reduced mass. A mediator mass of $\mA\simeq 1$ GeV corresponds to $\sigma_p\simeq 10^{-56} \text{ cm}^2$ around $\mps\sim 1$ MeV. Any cross-section above this value (for the same $\mA$ and $\mps$) is already excluded from freeze-in overproduction of dark fermions. Because of the large denominator in Eq.~\ref{eq:nuclear_cs}, the cross-sections corresponding to our production mechanism lie so many orders of magnitude below current constraints and projections that we deem direct detection in the Heavy Dark Photon case presently uninteresting.

\section{Discussion}
\label{sec:discussion}

Particle production in a time-varying gravitational background has been an active topic of research for many years, and it is well-appreciated that this is a minimal mechanism to produce the totality of dark matter.  However nearly all the work on this front has focused on single particle sectors, either a free scalar, fermion, or massive vector.  While these cases are more theoretically tractable, there is no reason to believe that the dark sector is so simple, especially given the complexity of the observed particles and interactions in the Standard Model. In this paper, we show that even minimal dark sector complexity can be enough to significantly change the parameter space that yields the correct DM abundance.

The crucial point is that inflation produces bosonic fields (massive vectors or scalars not sourced by misalignment) in a non-thermal state, with a distribution function weighted deeply towards the IR.   
When their modes re-enter the horizon with a characteristic energy $E$, their number density is $n \gg E^3$. Introducing any interactions into this system inevitably pushes it in the direction of thermalization: characteristic energies rise and number densities fall. In particular, if thermalization completes, a secluded dark radiation bath forms at a temperature suppressed by $\sqrt{\frac{\HI}{\mpl}}$ compared to that of the SM. The energy of the dark bath rapidly redshifts away, raising the mass of a DM candidate required to achieve the proper late-time abundance.

Quantitatively, a non-interacting massive vector produced by inflation  comprises the totality of DM for $m_{A'}^\text{(pure $A'$)}\sim 10^{-5}\eV$ for $\HI \simeq 10^{14} \GeV$. At the other extreme, a free fermion produced gravitationally by inflation must be very heavy, of the order of $10^9$ GeV, since Pauli exclusion essentially restricts the occupation number of each mode to unity.
In striking contrast to the non-interacting theories, dark QED from inflation yields the correct abundance of fermionic DM for $\mps^\text{(dark QED)} \sim \mathcal{O}(50 \MeV - 30 \TeV)$, if the dark sector thermalizes. This is a difference of more than ten orders of magnitude. This vast shift from extremely light or heavy DM to the more experimentally accessible GeV--TeV range for the case of dark QED, is one of the main results of this paper.

The pathway to thermalization can be very non-trivial because inflation sets up highly non-thermal classical states on large scales.
In the case of dark QED, this results in an era when non-perturbative strong-field effects become important. Furthermore, the resulting dark plasma affects the evolution of superhorizon vector modes in unexpected ways, a point that has not been touched on previously in the literature.
For instance, non-relativistic modes of a free massive vector redshift as matter after $H=\mA$, but in the presence of a dark bath where the plasma frequency is the largest scale in the system, we find that these superhorizon modes remain ``frozen'' (i.e.\ redshift slower than matter) until either their wavelength $k^{-1}>\mA^{-1}$ enters the horizon, or the plasma frequency becomes too small.\par

The mechanism described here is insensitive to the details of reheating. $A'$ modes are unaffected by reheating while superhorizon, and any set of modes that enters the horizon after the end of reheating contributes the same amount $\sim H^2 \HI^2$ to the late-time energy density. In contrast, modes that enter before the end of reheating redshift away, thus reducing the final relic abundance only logarithmically, provided reheating ends before the $\psi$ freeze out. For the lowest possible value of the DM mass, the SM reheat temperature could be as low as $\mathcal{O}(\text{TeV})$ without any significant change to our results.  This should be contrasted with another commonly-studied dark sector production mechanism involving gravity---gravitational freeze-in---which has a steep dependence on the reheat temperature, and quickly becomes subdominant to the mechanism of this paper unless the reheat temperature is extremely high.

The observed gravitational coupling of DM, and the well-motivated hypothesis of an epoch of primordial inflation, provide a natural production mechanism that is unavoidable for any field that couples to the background metric in a non-scale-invariant way. 
As we have shown here, interactions and the possibility of a thermal history for the dark sector can reshape the favored parameter space.
Within the context of dark QED, we predict a dark fermion mass in the 50 MeV--30 TeV range. 
In the presence of a kinetic mixing between dark and SM photons, upcoming direct detection experiments will probe part of this parameter space, which extends beyond the freeze-in prediction and can be an indirect probe of inflation.

\acknowledgments
We thank Junwu Huang and Robert Lasenby for very valuable insights.
We also thank Sergei Dubovsky, Daniel Ega\~na-Ugrinovi\v{c}, Anson Hook, Nickolas Kokron, Sebastian Meuren, Cristina Mondino, and David Reis for useful discussions.\\
SD is grateful for support from the National Science Foundation under Grant No.\ PHYS-2014215, and from the Gordon and Betty Moore Foundation Grant GBMF7946.
Research at Perimeter Institute is supported in part by the Government of Canada through the Department of Innovation, Science and Economic Development Canada and by the Province of Ontario through the Ministry of Colleges and Universities. 

\addcontentsline{toc}{section}{References}
\bibliographystyle{JHEP}
\bibliography{refs.bib}

\appendix

\section{Dark electric fields and energy densities in the non-interacting theory}
\label{app:electric_field} 
In this appendix we elaborate on aspects implicitly contained in Ref.~\cite{Graham:2015rva}, but not central to the discussion there.  Whereas \cite{Graham:2015rva} is mostly concerned with the evolution of the Proca field amplitude $\vec A'$ and its energy density, we are also interested in the associated \emph{classical dark electric and magnetic fields}, because those are the quantities that couple to a matter sector. The description is \emph{classical} because the superhorizon excursion of a given Fourier mode of the field rapidly increases the phase space occupation number of that mode to large values (Sec.~\ref{sec:qtc_transition}). The electric and magnetic fields are the appropriate objects of description for a classical interacting theory of electrodynamics (or a semiclassical theory of quantum electrodynamics in a classical electromagnetic background (Sec.~\ref{app:SFQED})). We therefore elaborate on the evolution of the electromagnetic fields in the scenario of \cite{Graham:2015rva} (Sec.~\ref{sec:classical_evolution}), with an eye on implications for an interacting theory.

\subsection{Heuristic for the quantum-to-classical transition}
\label{sec:qtc_transition}
In this subsection, we recast some of the results in \cite{Graham:2015rva} in terms of phase space distributions. This provides a heuristic\footnote{We stress that arguments based on the production of large occupation numbers during inflation like the one reviewed here is only a commonly accepted -- though admittedly incomplete -- \emph{heuristic}. The topic of the quantum-to-classical transition of states during inflation remains an object of study. For rigorous treatments, see \cite{kiefer1998,kiefer2007}.} for the quantum-to-classical transition of the vector perturbations and starts us towards the path of understanding thermalization.\par
Eqs.~(18) through (24) in \cite{Graham:2015rva} can be summarized into an equation for the evolution of the total energy density contained in \emph{superhorizon} modes of $A'$:
\begin{equation}
\bar\rho_{A'}(a) \supset{} \int_{k/a<H} d\ln k \left(\frac{\HI^2}{2\pi}\right)^2\left(\frac{k}{\HI a}\right)^2.
\end{equation}
This can be rearranged as
\begin{subequations}
\begin{equation}
\bar\rho_{A'}(a) \supset{} \int_{p<H}\frac{d^3 \vec p}{(2\pi)^3} ( p^2 +m_{A'}^2)^{1/2}f_\text{sup}(p,a)
\end{equation}
where $\vec p = \vec k /a$ is the physical momentum and we have identified the phase space occupation number
\begin{equation}
\label{eq:occupation_number}
f_\text{sup}(p<H,a) \equiv \frac{1}{2}\frac{\HI^2}{p(p^2 + m_{A'}^2)^{1/2}}.
\end{equation}
\end{subequations}
The phase space occupation \ref{eq:occupation_number} provides a useful heuristic for understanding inflationary particle production. Modes exit the horizon during inflation when their physical momentum is $p = \HI\gg m_{A'}$. At this time,  the occupation number of that mode is  $f(\HI,a) = 1/2.$ In other words, the occupation number at the time of horizon exit is merely the zero-point occupation number one expects of a quantum field.\par
The phase space occupation number of decoupled matter or radiation that is neither being created or destroyed is \emph{only a function of the comoving momentum number k}; in other words $f(k/a,a) = f(k,1)= \text{constant}$ for a species that is neither being created or destroyed \cite{Kolb:1990vq}. This is clearly not the case for \ref{eq:occupation_number}. If we follow a  mode after it exits the horizon, the physical momentum decreases like $p \propto a^{-1}$. From \ref{eq:occupation_number} we see that the occupation number of the mode increases as $f \propto a^2$; in other words, particles are being created.\par
A mode re-enters the horizon when $a = a_k$, such that $p=k/a_k = H(a_k)\equiv H_k$. The occupation number of that mode at that time is
\begin{equation}
\label{eq:occupation_at_re_entry}
f(H_k,a) \simeq \frac{1}{2}\frac{\HI^2}{H_k^2} \gg \frac{1}{2},
\end{equation}
for modes that re-enter while still relativistic (i.e $k>k_\star$). Subsequently, in the scenario of \cite{Graham:2015rva}, the vector particles evolve as decoupled radiation or matter. Their phase space occupation number therefore remains constant
\begin{equation}
\label{eq:occupation_after_re_entry}
f_\text{sub}(p > H,a) = f(H_k,a) \simeq \frac{1}{2}\frac{\HI^2 H^2}{p^4},
\end{equation}
where the last equality assumes radiation-domination.\par
Looking at the phase space occupation serves two purposes. Firstly, we see explicitly that excitations of a given mode are being created from the time that mode exits the horizon during inflation, to the time it re-enters the horizon after inflation ends. As \ref{eq:occupation_at_re_entry} indicates, the occupation number is extremely large by the time of re-entry, justifying a classical field theory description. In fact, if we take $f\gg1$ as our criterion for classicality, we see that a classical description is justified even while the mode is superhorizon, starting shortly after horizon exit.\par
Secondly, one sees that the distribution of sub-horizon modes is \emph{very non-thermal} in the absence of interactions. If the energy density of Eq.~\ref{eq:rho_enter} contributed by subhorizon modes were to thermalize, it would reach a temperature $T\sim \bar \rho_{A'}(k,a)^{1/4} \sim (\HI^2H^2)^{1/4}$. Thus, Eq.~\ref{eq:occupation_after_re_entry} gives $f_\text{sub} \sim (T/p)^4$, with $p\lesssim T$. Comparing to the corresponding limit of the Planck distribution, $f_\text{Planck} \sim T/p$, one sees that \ref{eq:occupation_after_re_entry} is more IR weighted than a thermal distribution. There are too many low-energy particles and thermalization will require many-to-few number-changing processes that decrease the number of particles. \par

\subsection{\label{sec:classical_evolution}Evolution of the dark electromagnetic fields in the non-interacting theory}
We now know that inflation produces a classical electric field. This classical electric field is stochastic in nature and characterized by its power spectrum. In this appendix, we look at the evolution of the power spectrum of the electric field, establish its coherence over a Hubble volume, and note that the electric field constitutes a subdominant fraction of the spectral energy density at a given $k$ until $k/a \sim H$.\par
The vector modes produced by inflation via the mechanism in \cite{Graham:2015rva} are longitudinal in the cosmological frame. Therefore the only non-zero elements of the dark electromagnetic strength tensor are $F'^{0i} = - F'^{i0}$. That is, the cosmological frame has zero dark magnetic fields. From the covariant Proca equation, one can derive an equation for the non-zero electric components of $F'^{\mu\nu}$:
\begin{equation}
\label{eq:electric_field_for_A}
F'_{0i} = \frac{m_{A'}^2 a^2}{k^2+m_{A'}^2a^2}\partial_0 A_i'.
\end{equation}
The size of the physical dark electric field measured by a cosmological observer is (see also the discussion around Eq.~\ref{eq:invariant})
\begin{equation}
|\vec E'_\text{phys}| = \frac{\sqrt{{F'_{0i}}^2}}{a}.
\end{equation}
The evolution of the electric field as a mode exits the horizon during inflation and re-enters after the end of inflation is therefore set by the time-derivative of the $A'$ field. The amplitude $A'(\vec k,a)$ is almost frozen to the value $A'_H(\vec k)$ imprinted at horizon exit all the way up to horizon re-entry. The derivative of the field during inflation is largely suppressed while the mode is super-horizon: $\partial_0 A'(\vec k,a) \simeq (k^2/a^2 \HI)A'_H(\vec k)$. On the other hand, we know that modes with $k>k_\star$ behave as radiation when they re-enter the horizon. So when $H=k/a$, we expect the time-derivative to correspond to the physical momentum: $\partial_0 A'(\vec k,a) \simeq (k/a)A'_H(\vec k) = HA'_H(\vec k) $. These two regimes must connect continuously, and one finds that 
\begin{equation}
\partial_0 A'(\vec k,a) \simeq \frac{k^2}{a^2 H}A'_H(\vec k)
\end{equation} 
from inflation, through reheating, and all while the mode is superhorizon. \par
After re-entry, the field behaves as radiation ($k>k_\star$), so that $A'(\vec k)\propto a^{-1}$ and $\partial_0 A'(\vec k) \propto (k/a)A'(\vec k) \propto a^{-2}$, meaning
\begin{equation}
\label{eq:subhorizon_derivative}
\partial_0 A'(\vec k, a) \simeq H A'_H(\vec k).
\end{equation}
\par
Like the $A'$ field, the primordial electric field is a classical Gaussian random field. Its power spectrum is related to that of the $A'$ via Eqs.~\ref{eq:electric_field_for_A} through \ref{eq:subhorizon_derivative}, giving
\begin{equation}
\label{eq:E_power_spectrum}
\mathcal P_{E'_\text{phys}}(k,a) \approx \frac{\mA'^2 \HI^2}{(2\pi)^2}\times
\begin{cases}
k^2/a^2 H^2, & k/a < H,\\
a^2 H^2/k^2,  & k/a > H.
\end{cases}
\end{equation}
where we have used the power spectrum for $A'_H$
\begin{equation}
\mathcal P_{A_H'}(k) = \left(\frac{k\HI}{2\pi \mA}\right)^2
\end{equation}
imprinted at horizon exit \cite{Graham:2015rva}.\par
Interestingly, the power contribution from a mode to the electric field at the time a mode re-enters the horizon is 1) independent of $k$, and 2) the same as it was at horizon exit. All modes leave the horizon with the same energy in the electric field, and they all re-enter with roughly that same amount later. The cross-horizon excursion only serves to make that field classical.\par
The power spectrum \ref{eq:E_power_spectrum} is peaked around modes $k\sim a H$. That is, it is always dominated by the mode which just entered the horizon. Consequently, the amplitude of the electric field is almost constant:
\begin{equation}
\langle E'^2_\text{phys}(a)\rangle \approx\frac{\mA^2 \HI^2}{(2\pi)^2}\left( \int_{-\infty}^{\ln(a H)} d\ln k \frac{k^2}{a^2H^2} +\int_{\ln(aH)}^\infty d\ln k \frac{a^2H^2}{k^2}\right) = \frac{\mA^2 \HI^2}{(2\pi)^2}.
\end{equation}
In the scenario of \cite{Graham:2015rva}, this is true until $a=a_\star$, when the last relativistic mode enters the horizon.\par
The power spectrum \ref{eq:E_power_spectrum} has a width $\Delta k \sim a H$, meaning that the coherence length of the field is $ k_0 \sim a H$. In other words, the physical coherence length of the field $k_0/a$ is about a Hubble length $H^{-1}$. Roughly, one can think that in each Hubble volume, the electric field has amplitude $\sim \mA \HI/(2\pi)$ and a random direction.  For a two-dimensional visualization of this electric field drawn from the above power spectrum, see Fig.~\ref{fig:EfieldPlot}. \par 
Finally, we observe that only a subdominant fraction of the energy density of superhorizon relativistic modes is stored in the electric field. The energy of the longitudinal Proca field in \cite{Graham:2015rva} is partitioned between the energy density $\rho_{\vec{A'}}$ of the $\vec A'$ field, and the energy density $\rho_{\vec{E'}}$ of electric field\footnote{The energy density of the electric field of a Proca field is not only the usual ${E'}^2_\text{phys}/2$ term, but rather ${E'}^2_\text{phys}/2+(1/\mA^2)(\vec \nabla \cdot{\vec E'_\text{phys}})^2$ \cite{greiner_qed}.} (or equivalently, the time-derivative $\partial_0 \vec A'$). For relativistic modes,
\begin{equation}
\frac{\rho_{\vec E}(k,a)}{\rho_{\vec A}(k,a)} \simeq \frac{a^2}{k^2}\frac{\mathcal P_{\partial_0 A_L'}(k,a)}{\mathcal P _{A'}(k,a)}.
\end{equation} 
For superhorizon modes, $\mathcal P _{A'}(k,a) \approx \mathcal P _{A_H'}(k)$ and $\mathcal P_{\partial_0 A_L'}(k,a) \approx (k^4/a^4 H^2)\mathcal P _{A_H'}(k)$ (Eq.~\ref{eq:subhorizon_derivative}). Thus, for superhorizon relativistic modes
\begin{equation}
\frac{\rho_{\vec E}(k,a)}{\rho_{\vec A}(k,a)} \simeq \frac{k^2}{a^2H^2}.
\end{equation} 
Even though this result is strictly speaking only true in the non-interacting theory, it nevertheless provides important intuition about the interacting theory. The Proca field interacts with a matter sector only through the electric field $\vec E'$, while most of the energy density generated at inflation is stored in the $\vec A'$ degree of freedom. As a result, \emph{it is difficult to extract a significant fraction of the energy from (i.e.\ ``discharge'') modes while they are superhorizon.}  For example, even if the electric field of superhorizon modes were to be totally screened (i.e. $ \langle \vec E'(k)^2 \rangle \approx 0$) by medium effects in the interacting theory, this would remove only a tiny fraction of the total energy density. \par 
The energy density in a mode is therefore ``protected'' while the mode is superhorizon. Once a relativistic mode is subhorizon, $\mathcal P_{\partial_0 A'} \simeq (k/a) \mathcal P_{A'}$, the energy is equipartitioned between $\vec A'$ and $\vec E'$, and the mode can dissipate. A simplified picture therefore is that a mode behaves as in \cite{Graham:2015rva} until it re-enters the horizon, at which time it starts interacting. This basic picture is confirmed by a more detailed analysis of plasma effects (or rather, the lack thereof) on relativistic modes (see Apps. \ref{app:ProcaPlasmas}, \ref{app:warmProcaPlasmas}, and \ref{app:thermalizationAppendix}).

\section{Review of Quantum Electrodynamics of Strong Fields}
\label{app:SFQED}

The onset of our thermalization history relies on strong field processes like quantum electrodynamic cascades in the dark sector.  QED cascades play an important role in astrophysical systems \cite{blanford1977,daugherty1982}, and could realistically be observed in the laboratory with the next generation of high-power lasers \cite{Narozhny2004,Narozhny2006,bell2008,fedotov2010,fedotov2011,fedotov2013}. Such phenomena fall within the purview of strong field quantum electrodynamics (SFQED), which studies the behavior of quantized particles (photons, as well as charged fermions and scalars) in the presence of large \emph{classical} electromagnetic backgrounds \cite{ritus1985,greiner_sfqed,meuren2015}. As such, $\hbar$ is involved in the equations only as it pertains to the quantized particles incident on or produced by the strong classical electromagnetic background fields. The classical backgrounds themselves are characterized by the classical (dark) electric and magnetic fields $\vec E'$ and $\vec B'$. For this reason, 
strong field processes are somewhat different in nature from the usual quanta-to-quanta perturbative scattering processes. They involve the coherent action of many photons from the background field, and are often non-perturbative in nature.  \par
A few important semi-classical parameters characterize SFQED processes. The first and most important one dates back to the foundational result of SFQED: Schwinger's study\footnote{Preceded by earlier works on relativistic quantum mechanics by Klein \cite{Klein1929}, Sauter \cite{sauter1931}, Heisenberg and Euler \cite{heisenberg1936}, among others.} of the spontaneous production of particle-antiparticle pairs in strong homogeneous electric fields, the process now known as the \emph{Schwinger effect} \cite{schwinger1951}. In a semiclassical theory of fermions of mass $m_\psi$ and charge $\qd$, once $\hbar$ becomes available to construct physical quantities, one can recognize that the combination $m_\psi^2 c^3/\hbar$ has the units of electric force, and therefore provides a natural scale for the electric field: 
\begin{equation}
\label{eq:critical_field}
\Ecr \equiv \frac{m_\psi^2 c^3}{\qd \hbar} = \frac{m_\psi c^2}{\qd \lambda_\text{C}},
\end{equation}
where $\lambda_\text{C} = \hbar/m_\psi c$ is the Compton wavelength of the fermion. Eq.~\ref{eq:critical_field} is the \emph{Schwinger critical field}; it intuitively corresponds to a classical electric field whose work over the fermion's Compton wavelength is of the order of the fermion mass. An electric field for which $|\vec E'|> \Ecr$ is unstable, and quantum mechanically decays spontaneously into particle-antiparticle pairs\footnote{This is true of an electric field without a magnetic field. The proper covariant generalization for the Schwinger process gives the requirement $F_{\mu\nu} F^{\mu\nu} = \vec E^2-\vec B^2 > E^2_\text{crit}$ for decay to occur.} at a rate per unit volume:
\begin{equation}
\mathcal W_\text{Schwinger} \sim \qd^2 |\vec E'|^2 \exp\left(-\frac{\pi \Ecr}{|\vec E'|}\right).
\end{equation}
In modern language, this can be viewed as a tunneling, or bubble nucleation, process out of an unstable QED vacuum; the study of this process is amenable to instanton methods \cite{page2002, page2003,dunne2005, brown2015}. The exponential form $\sim e^{-S/\hbar}$ does not have a perturbative expansion in powers of the small coupling $\qd$, as is characteristic of non-perturbative effects. \par
We have reintroduced $\hbar$ and $c$ in the notation of Eq.~\ref{eq:critical_field} to emphasize the semiclassical nature of the description. The quantum nature of the phenomenon is entirely contained in the single scale $\lambda_\text{C}$ pertaining to the particles being produced, while the electric field is viewed classically. In what follows, we move back to $\hbar = c =1$.  \par 
The scale \ref{eq:critical_field} is ubiquitous in SFQED results; it essentially defines the strong field regime where non-perturbative effects happen. However, it is not necessary to reach critical values of the electric field \emph{in the lab} for strong field effects to matter.
Given a worldline with 4-momentum $p^\mu = (p^0, \vec p)$ and the dark electromagnetic strength tensor ${F'}^{\mu\nu}$, consider the 4-vector associated with the Lorentz force law on a fermion:
\begin{subequations}
\begin{equation}
f^\mu = \frac{1}{m_\psi}{F'}^\mu_{\;\;\;\nu} p^\nu = \frac{1}{m_\psi}\left(\vec E' \cdot \vec p, \frac{p^0 \vec E' + \vec p \times \vec B'}{a^2}\right),
\end{equation}
Then consider the coordinate-invariant
\begin{equation}
\label{eq:invariant}
\sqrt{-f^\mu f_\mu} = \frac{1}{m_\psi}\sqrt{\frac{1}{a^2}(p^0 \vec E' + \vec p \times \vec B')^2-(\vec E' \cdot \vec p)^2} = |\vec E'_\text{rest}|,
\end{equation}
\end{subequations}
Here $\vec E'$ and $\vec B'$ are general coordinate frame quantities, and $|\vec E'_\text{rest}|$ is the magnitude of the electric field in the local rest frame ($p^{\hat\mu}=(m_\psi,0,0,0)$ and $a=1$ in the local rest frame). \par
\begin{figure}
 \centering
 \begin{subfigure}[b]{0.25\textwidth}
     \centering
     \includegraphics[width=\textwidth]{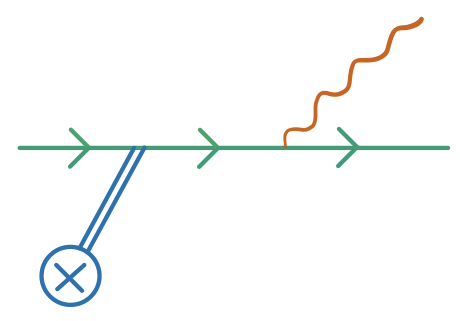}
     \caption{}
     \label{fig:nonlinear_compton}
 \end{subfigure} \hfill
 \begin{subfigure}[b]{0.25\textwidth}
     \centering
     \includegraphics[width=\textwidth]{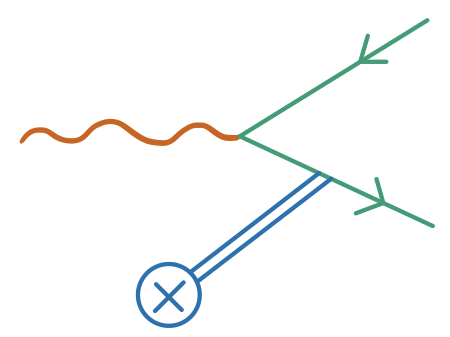}
     \caption{}
     \label{fig:pair_creation}
 \end{subfigure} \hfill
 \begin{subfigure}[b]{0.4\textwidth}
     \centering
     \includegraphics[width=\textwidth]{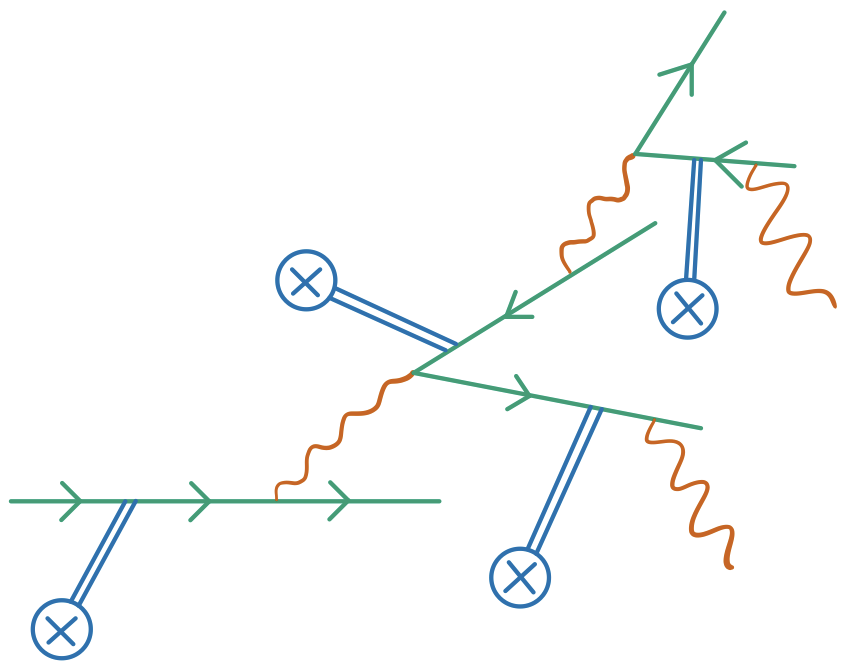}
     \caption{}
     \label{fig:qed_cascade}
 \end{subfigure}
 \caption{Some SFQED interactions of quantized particles in classical background fields. (a) Nonlinear Compton scattering, (b) pair creation in a background field, and (c) a fermion incident on a background field causes the field to decay into a QED cascade.}
\end{figure}

Take $p^{\mu} = (\omega_\psi,0,0,k_\psi)$ to describe a fermion incident on a classical electromagnetic configuration. It turns out that strong field effects are achieved when \begin{equation}
\label{eq:SFQED_condition}
\chi = \frac{\sqrt{-f^\mu f_\mu}}{\Ecr} \gtrsim 1\,.
\end{equation}
$\chi$ is  the \emph{quantum non-linearity parameter} \cite{ritus1985,meuren2015} and
$\chi \gtrsim 1$ can be obtained even if the ``lab'' field $|\vec E'| \ll \Ecr$.
Heuristically, this is because the Schwinger threshold is in fact reached in the highly boosted frame of the fast-moving incoming particle.  A high-energy fermion propagating through a homogeneous, time-independent classical electromagnetic field may therefore undergo \emph{nonlinear Compton scattering} (Fig.~\ref{fig:nonlinear_compton}), many-to-one photon emissions at the rate \cite{greiner_qed,fedotov2013}  
\begin{equation}
\label{eq:single_photon_rad}
W_\text{rad}\sim  \frac{\ald m^2_\psi }{\omega_\psi}\cdot\begin{cases}
\chi,&\chi \ll 1,\\
\chi^{2/3},& \chi \gg 1.
\end{cases}
\end{equation}

The incident particle may also be a high-energy photon (for a light-like worldline, \ref{eq:invariant} loses its interpretation as a rest frame quantity but can still be non-zero, for example if $\vec B = 0$ and $\vec p \cdot \vec E = 0$). The photon traveling through the field can then cause a sort of ``stimulated'' Schwinger process and ``decays'' in the background field into charged pairs (Fig.~\ref{fig:pair_creation}) at the rate \cite{Dunne2009,greiner_qed,fedotov2013},
\begin{equation}
\label{eq:pairs}
W_\text{pair} \sim \frac{\ald m_\psi^2}{\omega_{A'}} \cdot \begin{cases} \chi\exp\left(-\frac{8}{3\chi}\right), & \chi \ll 1,\\
\chi^{2/3}, & \chi \gg 1.
\end{cases}
\end{equation}
If the background electric field is large enough, photons emitted by a seed fermion through \ref{eq:single_photon_rad} can go on to trigger fermion pair production with the rate \ref{eq:pairs}; subsequently, the emitted fermions can go on to emit photons at the rate \ref{eq:single_photon_rad}. This results in a runaway process, a \emph{QED cascade,} wherein the classical electromagnetic field undergoes a ``stimulated decay'' to a shower of fermions and photons (Fig.~\ref{fig:qed_cascade}).\par

\subsection{Spatiotemporal variations of the background fields and SFQED in an expanding universe}
\begin{figure}[t]
    \centering
    \includegraphics[width=3.5in]{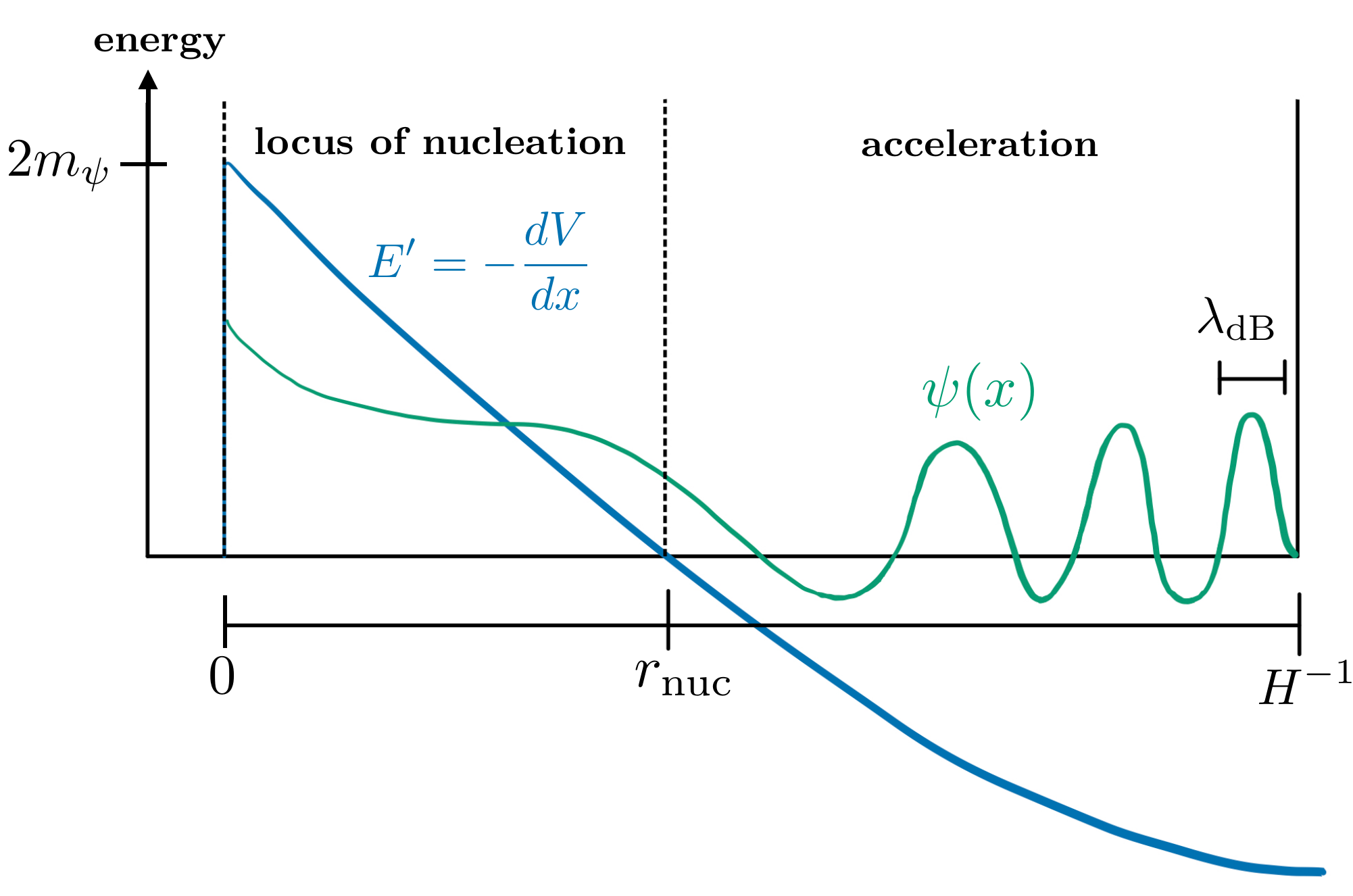}
    \caption{Particle-antiparticle pair nucleation (Schwinger process) in the semiclassical (WKB) approximation as an example of SFQED process in an expanding universe. The nucleation of the vacuum is analogous to the ionization of a bound state by a strong field. The presence of a cosmological scale $H$ can be thought of as placing the system in a box of finite size $H^{-1}$. The fermion tunnels through a distance $r_\text{nuc}$, such that the work done over that distance is equal to the rest mass of the nucleated pairs (i.e. $\qd E' r_\text{nuc} \simeq 2m_\psi$), and is subsequently accelerated. As long as the size of the nucleation volume $r_\text{nuc}$ and the wavelength of the WKB wavefunction $\lambda_\text{dB}$ are much smaller than $H^{-1}$ and other intrinsic scales of variation of the external field, constant-field, flat spacetime results can be invoked, even if $m_\psi \ll H $.}
    \label{fig:WKB_nucleation}
 \end{figure}

Strictly speaking, the results quoted above apply only to idealized homogeneous and time-independent external background electromagnetic fields. Treating rapid temporal and spatial variations of the background electromagnetic fields exactly renders the computation challenging and reveals interesting finite-size effects (e.g.\ \cite{page2002,page2003}). A saving grace, however, is that as long as the formation length of the nucleated particle-antiparticle complex 
\begin{equation}
\label{eq:tunneling_length}
r_\text{nuc} \simeq \frac{m_\psi}{\qd|\vec E'|}
\end{equation}
is short relative to the characteristic scales of spatial and temporal variations of the background field, a constant field approximation is generally appropriate \cite{fedotov2010}.  In the case of the Schwinger process, \ref{eq:tunneling_length} is roughly the size of a pair as it is being ``born'' (Fig.~\ref{fig:WKB_nucleation}), and roughly the amount of time it takes for the pair to tunnel out of the vacuum \cite{brown2015}. In the language of vacuum bubble nucleation, \ref{eq:tunneling_length} is the thickness of the bubble's wall. \par
SFQED effects are almost all born out of vacuum nucleation, whether or not the nucleated pair is in the final state (as in the process \ref{eq:single_photon_rad}).
Thus, it is customary to take Eq.~\ref{eq:tunneling_length} to characterize roughly the volume and duration of the all SFQED processes discussed so far, included the ``stimulated'' ones. As long as Eq.~\ref{eq:tunneling_length} is small compared to external spatial and temporal variations of the background fields, the process can be considered point-like and instantaneous.%
\footnote{In addition to Eq.~\ref{eq:tunneling_length}, a number of other lengthscales are used in the literature to determine the validity of the locally constant field approximation for studying stimulated strong-field processes. In the high-intensity laser community, Eq.~\ref{eq:tunneling_length} is commonly used (e.g.\ in \cite{fedotov2010}), but sometimes also the Compton wavelength $\lambda_\text{C}$ is used instead (e.g. in \cite{Narozhny2004,Narozhny2006}). The rate \eqref{eq:pairs} can be obtained via the semiclassical methods of instantons \cite{Monin2009,Monin2010}; this suggests that the size of the largest instanton $r_\text{inst} = \omega_{A'}/\qd |\vec E'|$ should be used instead. As we will see below, we will ultimately be using a lengthscale -- the semiclassical de Broglie wavelength -- which is more constraining than Eq.~\ref{eq:tunneling_length} and about as constraining as $r_\text{inst}$.} 

The way the cosmological scale $H$ factors into strong field processes is an active area of research. Studies have been done for both bosonic and fermionic particles, and for cosmological spacetimes with various dimensions \cite{vilenkin2014,kobayashi2014,stahl2015a,stahl2015b,stahl2016,hayashinaka2016}. To an extent, the same rationale used for general spatio-temporal variations of the background fields holds in this scenario: as long as the horizon scale $H^{-1}$ is large relative to the nucleation scale \ref{eq:tunneling_length}, an adiabatic version of flat space results can be used. There is however an additional caveat. Intuitively, not only must the nucleation happen within a Hubble horizon, but the nucleated pairs, once created, must be able to ``fit'' inside the horizon. This leads to the definition of another semiclassical parameter for SFQED in an expanding spacetime:
\begin{equation} \label{eq:lambdadB}
\lambda_\text{dB} = \frac{H}{\qd|\vec E'|},
\end{equation}
corresponding to the semiclassical de Broglie wavelength of a particle accelerated by the electric force $\qd \vec E'$ over a distance of order the cosmological horizon. In the semiclassical picture of vacuum breakdown, the fermions are nucleated at rest and subsequently accelerated by the electric field. If $\lambda_\text{dB} \ll H^{-1}$, this subsequent acceleration is large enough to give the fermion large enough momentum within a Hubble volume that the corresponding wavefunction ``fits'' inside the universe (Fig.~\ref{fig:WKB_nucleation}). \par
Additionally, the spatial coherence length (and coherence time) of the background classical stochastic electric field is also $\sim H$ (Sec. \ref{sec:classical_evolution}). \par
\begin{figure}[ht] \centering
\includegraphics[width=.5\textwidth]{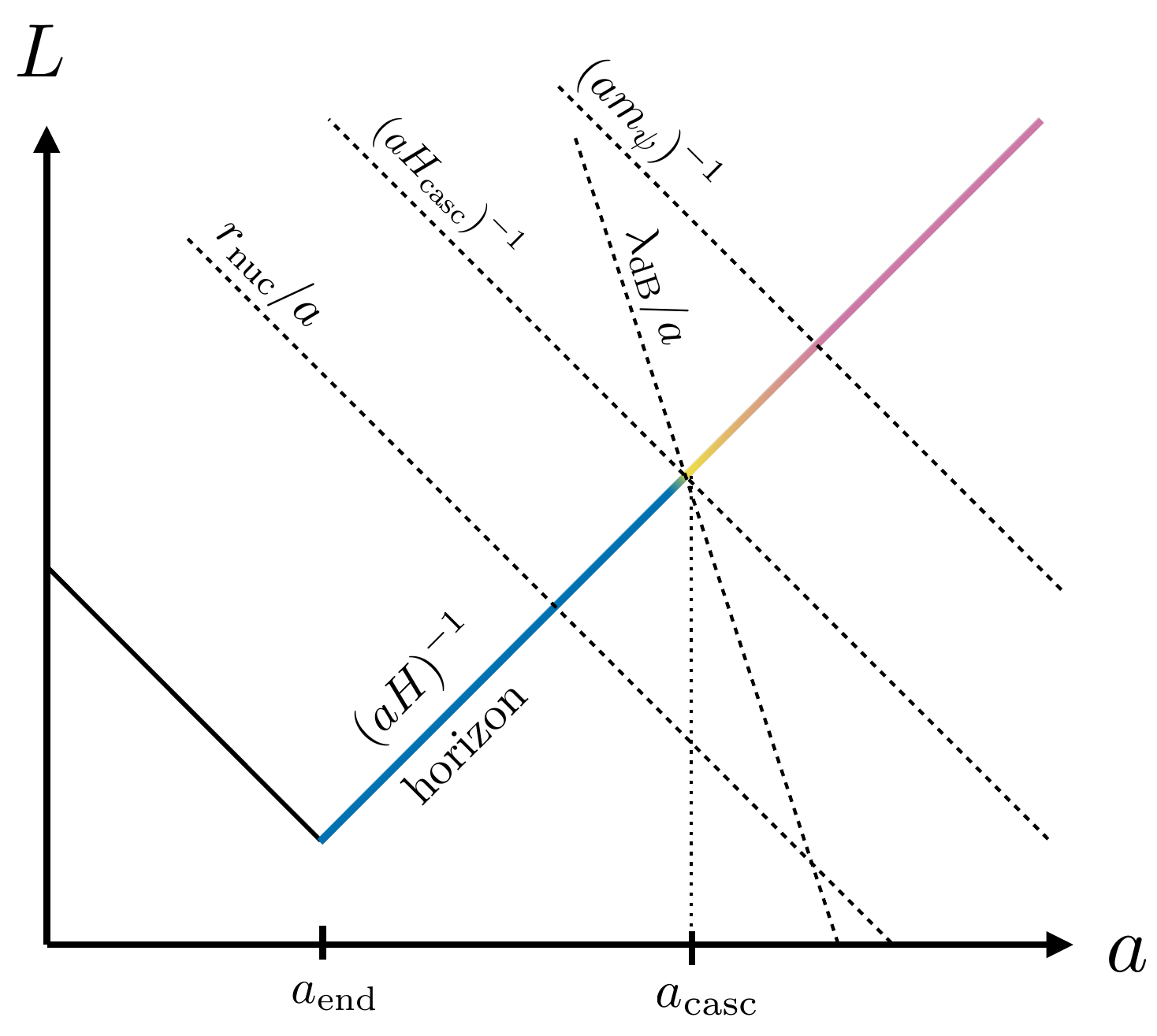}\hfill
\caption{The important lengthscales to consider for SFQED processes in the background primordial electric field from inflation. We construct the diagram by observing that, for $|\vec E'|>E'_\text{crit}$, $r_\text{nuc} < \lambda_\text{C} = m_\psi^{-1}$ and  $r_\text{nuc} = \lambda_\text{dB}$ when $H=m_\psi$.  $H_\text{casc} = (\qd E')^{1/2}$ is the rate of expansion of the universe that is slow enough that a QED cascade can build up. As long as $r_\text{nuc}$ and $\lambda_\text{dB} \ll r_\text{nuc}$ are both less then $H^{-1}$ (which is both the size of the universe \emph{and} the coherence length of the stochastic background field), the approximation of a locally constant background field is appropriate and the production rates of SFQED in time-independent, homogeneous fields can be used. As a visual aid to comparison, the horizon after inflation has been color-coded according to the era defined on Fig.~\ref{fig:cartoon thermalization}.}
\label{fig:SFQED _lengthscales}
\end{figure}
Altogether, in order to use the approximation of locally constant background fields in flat spacetime, we must have $r_\text{nuc} \lesssim H^{-1}$ \emph{and} $\lambda_\text{dB} \lesssim H^{-1}$. While $H>m_\psi$, the latter is more constraining. Coincidentally, one has that the condition $\lambda_\text{dB}H\lesssim 1$ is also the condition for the QED cascade to be efficiently \emph{sustained} (see the definition of $H_\text{casc}$ in Sec.~\ref{sec:thermalization} and App.~\ref{app:before Hcasc}). That is, one can view the value $H_\text{casc}$ at which SFQED begins to be efficient in the cosmological history, as also being the value where the approximation of locally constant fields is appropriate. \par Fig.~\ref{fig:SFQED _lengthscales} tracks the evolution of the important lengthscales for SFQED processes through the cosmological history. The main take-away is that after $H=H_\text{casc}$, all lengthscales of relevance are short compared to the cosmological horizon, and therefore also short compared to the scales of spatial and temporal variations of the background electric fields. 

\subsection{SFQED with a massive $U(1)$}
The studies of SFQED are generally done with Maxwell (massless) electromagnetic fields in mind. In practice, freely propagating laser backgrounds are often considered. One might therefore ask: do those results apply to electric fields produced by \emph{longitudinal modes} supported by \emph{massive} Proca fields?\par
The answer is essentially \emph{yes}. The electromagnetic fields of SFQED are non-dynamical \emph{background} fields. The theory is agnostic as to \emph{how} those fields are produced, and what the on-shell properties of the electromagnetic states are in the free theory. Indeed, the original context envisioned by Schwinger -- the volume inside large capacitor plates -- is in fact a \emph{longitudinal} mode that exists in the electrostatic theory,\footnote{The electric field of a capacitor varies in the direction in which the field points, from $\approx 0$ to the left of the capacitor, to non-zero inside the capacitor, to $\approx 0$ to the right of the capacitor. A field that varies along the direction towards which it points is a longitudinal field.} even though no such longitudinal modes exist in the free Maxwell theory.
It is so because SFQED results are formulated in terms of the gauge-invariant electric and magnetic fields. As long as the background electric and magnetic fields are given, the theory does not care about \emph{how} those fields arise, whether it be by a clever external charge-current distribution, or because they are freely supported in Proca theory. \par 
A more technical way of stating this point is that the Schwinger vacuum decay rate is related, via the optical theorem, to the imaginary part of the Euler-Heisenberg one-loop effective action of QED \cite{schwinger1951,martin1989,page2002,schwartz2014quantum}. The photon propagator plays no role in computing the effective action, which therefore does not know about the massive or massless nature of the photon.\par
\begin{figure}[t]
    \centering
    \includegraphics[width=3.5in]{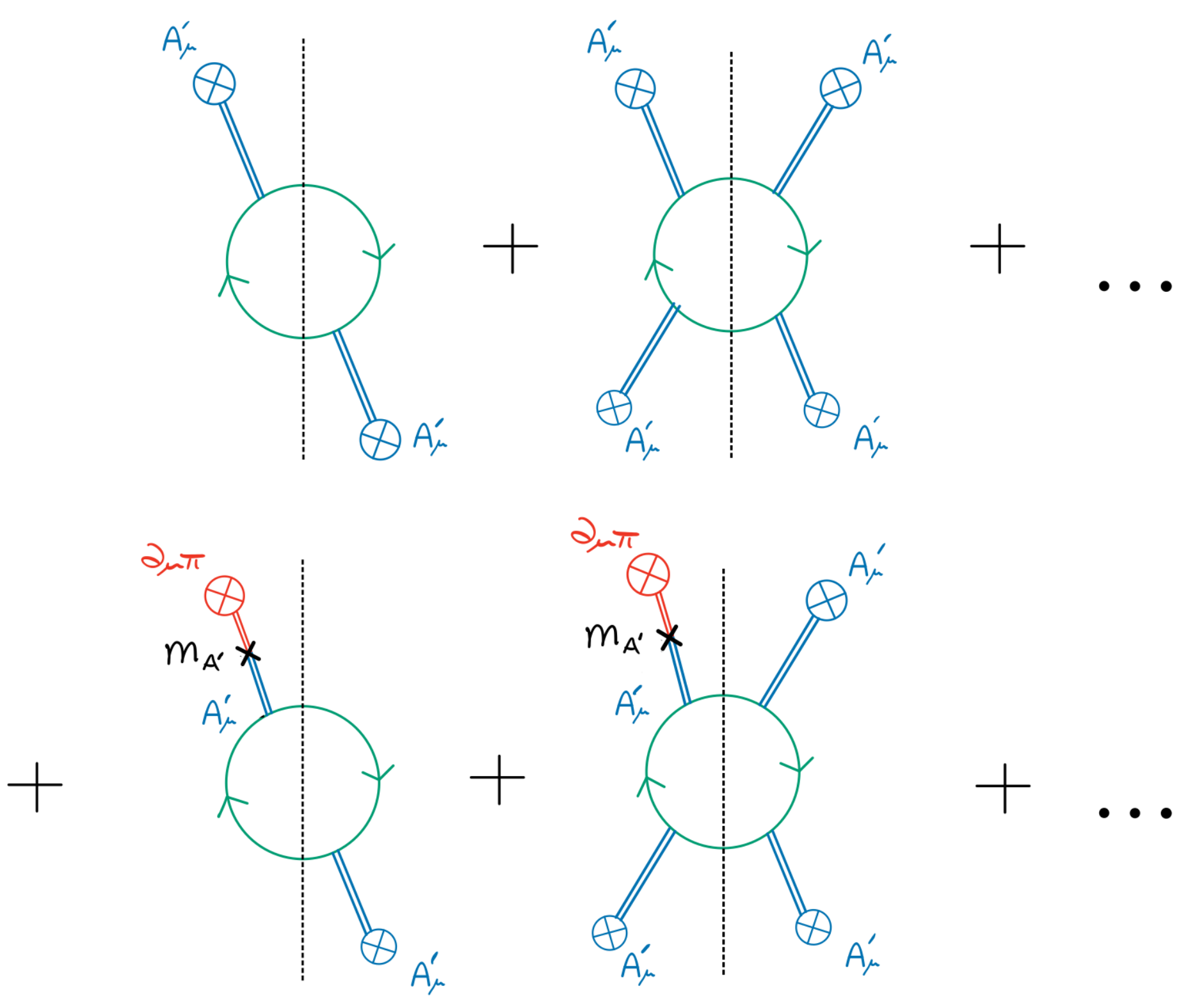}
    \caption{Diagrammatic expansion of the one-loop effective action $\mathcal L_\text{eff}$ of a theory of QED with a massive photon. The rate of vacuum decay is related to the imaginary part of $\mathcal L_\text{eff}$ via the optical theorem (represented as usual by a vertical cut in the diagrams). The main difference from a theory with a massless photon is the scalar degree of freedom $\pi$ which must in principle be taken into account as an external field and can oscillate into $A'_\mu$ via a mass vertex $\mA$. However, because of the Ward identity (i.e.\ because the current $J^\mu = \overline \psi \gamma^\mu \psi$ is conserved), such diagrams evaluate to zero.}
    \label{fig:HE_effective_action}
 \end{figure}

One can make an even better argument by working in St\"uckelberg theory, which restores $U(1)$ gauge invariance to the field theory of a massive vector by introducing a massless scalar field $\pi$:
\begin{equation}
    \mathcal L_\text{St\"uck} = -\frac{1}{4}F'^2 + \frac{1}{2}\mA^2\left(A'_\mu + \frac{1}{\mA}
    \partial_\mu\pi\right)^2+i\overline\psi\partial_\mu \gamma^\mu \psi-\mps\overline\psi \psi +\qd A'_\mu \overline\psi\gamma^\mu\psi.
\end{equation}
This has a gauge symmetry under which $A_\mu' \mapsto \widetilde A_\mu'= A_\mu' - \partial_\mu \alpha$, $\pi \mapsto \widetilde \pi = \pi + \mA\alpha$ and $\psi \mapsto \widetilde \psi = e^{-i\qd \alpha}\psi$, where tildes denote quantities in the new gauge. We can sketch the computation of the one-loop effective action in this theory, which amounts to the resummation of the diagrams in Fig.~\ref{fig:HE_effective_action} with background external legs given by solutions to the free Proca field pointing in the $z$-direction. The Proca gauge is $\pi=0$. In this gauge
\begin{subequations}
\begin{gather}
A'_\mu = \frac{a}{\mA}(p,0,0,-\omega_{A'})e^{-i(\omega_{A'} t-p z)}, \qquad \omega_{A'}^2 = \mA^2 +p^2,\\
\pi = 0,
\end{gather}
\end{subequations}
where $a$ is the amplitude of the massive photon field.\par
A convenient and common gauge to use for usual SFQED processes with Maxwell fields is the temporal gauge $A_0 =0$ (i.e. $F_{0i}=\partial_0 A_i$) \cite{padmanabhan1991quantum,srinivasan1999,itzykson2012quantum,ott2016}. We can exploit gauge freedom in the St\"uckelberg Lagrangian to move to the temporal gauge at the cost of introducing $\pi \neq 0$ in the background field. The relevant choice of $\alpha$ is
\begin{equation}
\alpha = \frac{v}{\mA}iae^{-i(\omega_{A'} t-pz)}, \qquad v=p/\omega_{A'}.
\end{equation}
In this new gauge,
\begin{subequations}
\begin{gather}
\label{eq:new_gauge}
\widetilde A'_\mu =\left(0,0,0,-\frac{\omega_{A'}^2-p^2}{\omega_{A'}^2}\right)\frac{a\omega_{A'}}{\mA}e^{-i(\omega_{A'} t-pz)} = \left(0,0,0,\frac{\mA^2}{\mA^2+p^2}A_3'\right) \\
\widetilde \pi =\mA\alpha \neq 0.
\end{gather}
\end{subequations}
Now that $\pi \neq 0$, we must in principle consider oscillations $\partial_\mu \pi\to A'_\mu$ mediated by the vertex $\mA A'^\mu \partial_\mu \pi$. However, because of Lorentz invariance, the polarization 4-vector is conserved at the vertex and through propagation. Therefore, the photons in the diagram have polarization 4-vector $\varepsilon_\mu \propto \partial_\mu = p_\mu$. By the Ward identity $p_\mu \mathcal M^\mu_\text{loop} = 0$ always. Thus, all oscillations from $\pi$ to $A'_\mu$ can be ignored. \par 
The background is now reduced to a space and time dependent external vector field in the temporal gauge. It should be clear then that the usual results of SFQED of Maxwell fields apply to Proca fields, as long as the appropriate gauge-transformed value $\widetilde A_3' = \frac{\mA^2}{\omega_{A'}^2} A_3'$ is used. In particular, we see that, in moving to the temporal gauge, the $A_\mu'$ field has been suppressed by $\mA^2/(\mA^2+p^2)$. That is, $\widetilde F_{03}' =\partial_0 \widetilde A_3' = \mA^2(\mA^2 + p^2)^{-1} \partial_0 A_3'$. This is exactly the same suppression as the one in Eq.~\ref{eq:electric_field_for_A} for the electric field. Again, we see that one gets the correct answer if one works in terms of the gauge-invariant $\chi$, defined in Eq.~\ref{eq:SFQED_condition}. \par
This settles the issue of how to treat the \emph{background} Proca fields. For \emph{stimulated} Schwinger decay however, the high-energy photon seed is not part of the background strong fields. The rate of stimulated decay can be connected to the imaginary part of the  propagator of the seed photon in the strong electromagnetic background fields (i.e.\ Fig.~\ref{fig:HE_effective_action} with one additional photon leg on either side of each loops) \cite{Dunne2009}. We expect the result to depend on the polarization of the high-energy seed photon.  In particular, because the electric field of a longitudinal photon is suppressed relative to that of a transverse photon, we expect to find a corresponding suppression for the rate of stimulated vacuum decay when the seed photon is longitudinally polarized. Consider a longitudinal seed photon with polarization $\varepsilon^\mu_L =\mA^{-1}(p,0,0,\omega_{A'})$ incident in the $z$-direction with $p^\mu = (\omega_{A'}, 0, 0,p)$. We can get the parametric suppression by observing that $\varepsilon^\mu_L \mathcal M_{\text{loop,}\mu} = (\mA \omega_{A'})^{-1}\left(pp^\mu + (0,0,0,m_A^2)^\mu \right)\mathcal M_{\text{loop,}\mu} = \frac{\mA}{\omega_{A'}} \mathcal M_{\text{loop,}3},$ where the part of the polarization proportional to $p^\mu$ gives zero when contracted with $\mathcal M_{\text{loop,}\mu}$. Thus, when seeds photons are longitudinally polarized, we expect a suppression of $\mathcal O(\mA^2/\omega_{A'}^2)$ relative to the result of Eq.~\ref{eq:pairs}, which is given for seed photons with transverse polarization. The emission of massive longitudinally polarized radiation in bremsstrahlung suffers the same suppression \cite{Lechner:2018}.

\section{Proca Plasmas} \label{app:ProcaPlasmas}
In this appendix we outline the general behavior of massive vector fields in a plasma within the boundaries of a cold fluid approximation.  By its nature this treatment is incomplete, failing to capture certain thermal effects which may play a role in the dynamics (some of which are discussed in App.~\ref{app:warmProcaPlasmas}), but we believe it to capture the broad behavior of such plasmas for length scales larger than the Debye screening length (which are the modes that carry the bulk of the energy density in the model of this paper), and we leave a more thorough treatment to future work.  We focus here on a number of interesting physical effects which already demonstrate themselves within this approximation.  Our discussion here takes heavy inspiration from Refs.~\cite{Hardy:2016kme, Lasenby:2020rlf}.

We begin this discussion in Sec.~\ref{app:ProcaPlasmasMinkowski} with an overview of Proca plasmas in Minkowski space.  In this case we may perform a mode decomposition, taking a Fourier transform in both space and time to yield a simple equation that may be solved for $\omega(k)$.  This allows us to gain some intuition for the system.  We follow this in Sec.~\ref{app:ProcaPlasmasGR} with a treatment of Proca plasmas in an FRW background, a topic which to our knowledge is undiscussed in the current literature.  It turns out that our intuition from the Minkowski-space case carries over well to this regime, and the consequences include the presence of dark photon plasma modes that do not begin oscillating until long after $H \sim \mA$,  a significant delay compared to the expectation in a vacuum.

Throughout this section we refer to the Proca field, the dark currents, and the corresponding electric field without the primed labels used in the rest of this paper.

\subsection{Minkowski space} \label{app:ProcaPlasmasMinkowski}
In Minkowski space we may study modes of the Proca plasma by moving to Fourier space via $A^\mu(\vec{x}, t) \equiv \sum_k A_k^\mu \exp{(- i \omega t + i \vec{k} \cdot \vec{x})}$.  In this case, the Maxwell and Lorentz laws become:
\begin{equation}
\label{eq:dispersion}
    (\omega^2 - k^2 - \mA^2) \vec{A} = - \vec{J}
\end{equation}
\begin{equation}
    \mps \del_t \vec{u} = - \qd \vec{E} - 2 \mps \nu \vec{u}\label{eq:cold-lorentz}
\end{equation}
where $\nu$ is a phenomenological collision term whose value is given by the collision rate of the $\psi$ particles.  The cold plasma approximation amounts to taking $\vec{J} = \qd \nps \vec{u}$ where $\nps$ is the total number of $\psi$ and $\bar{\psi}$, and using the definition of the electric field $\vec{E} \equiv E_i = F_{0 i} = i ( -\omega \vec{A} + i \vec{k} A_0 )$ we may combine the above two equations into:
\begin{equation}
    (\omega^2 - k^2 - \mA^2) \vec{A} = \omp^2 \frac{\omega}{\omega + 2 i \nu} \left( \vec{A} - \frac{1}{\omega^2} \vec{k} (\vec{k} \cdot \vec{A} ) \right)
\end{equation}
where $\omp^2 \equiv \frac{\qd^2 \nps}{E_\psi}$ and $E_\psi$ is the fermion energy (this reduces to $\mps$ in the nonrelativistic limit).  Because the coherent modes produced by inflation are longitudinal modes ($\vec{A} \,||\, \vec{k}$), we simplify to that case, producing the longitudinal dispersion relation:
\begin{equation} \label{eq:dispRelMinkowski}
    \omega^2 - k^2 - \mA^2 = \omp^2 \frac{\omega}{\omega + 2 i \nu} \left( 1 - \frac{k^2}{\omega^2} \right)
\end{equation}
For completeness we also note that the electric field of a longitudinal mode is given by:
\begin{equation}
    \vec{E} = - \frac{i}{\omega} ( \omega^2 - k^2 ) \vec{A}
\end{equation}
For generic values of $m, k, \nu, \omp$, Eq.~\ref{eq:dispRelMinkowski} admits two pairs of solutions for $\omega$.  For $\nu > 0$, each of these solutions is damped (i.e. $\Im{(\omega)} < 0$) corresponding to the fact that collisions of the $\psi$ particles will tend to thermalize any coherent modes of the $\psi$--$A$ system.  For our purposes here it will suffice to quote a few specific results from solving this equation assuming various hierarchies of the relevant parameters.
\begin{figure}[t] \centering
\includegraphics[height=.36\textwidth]{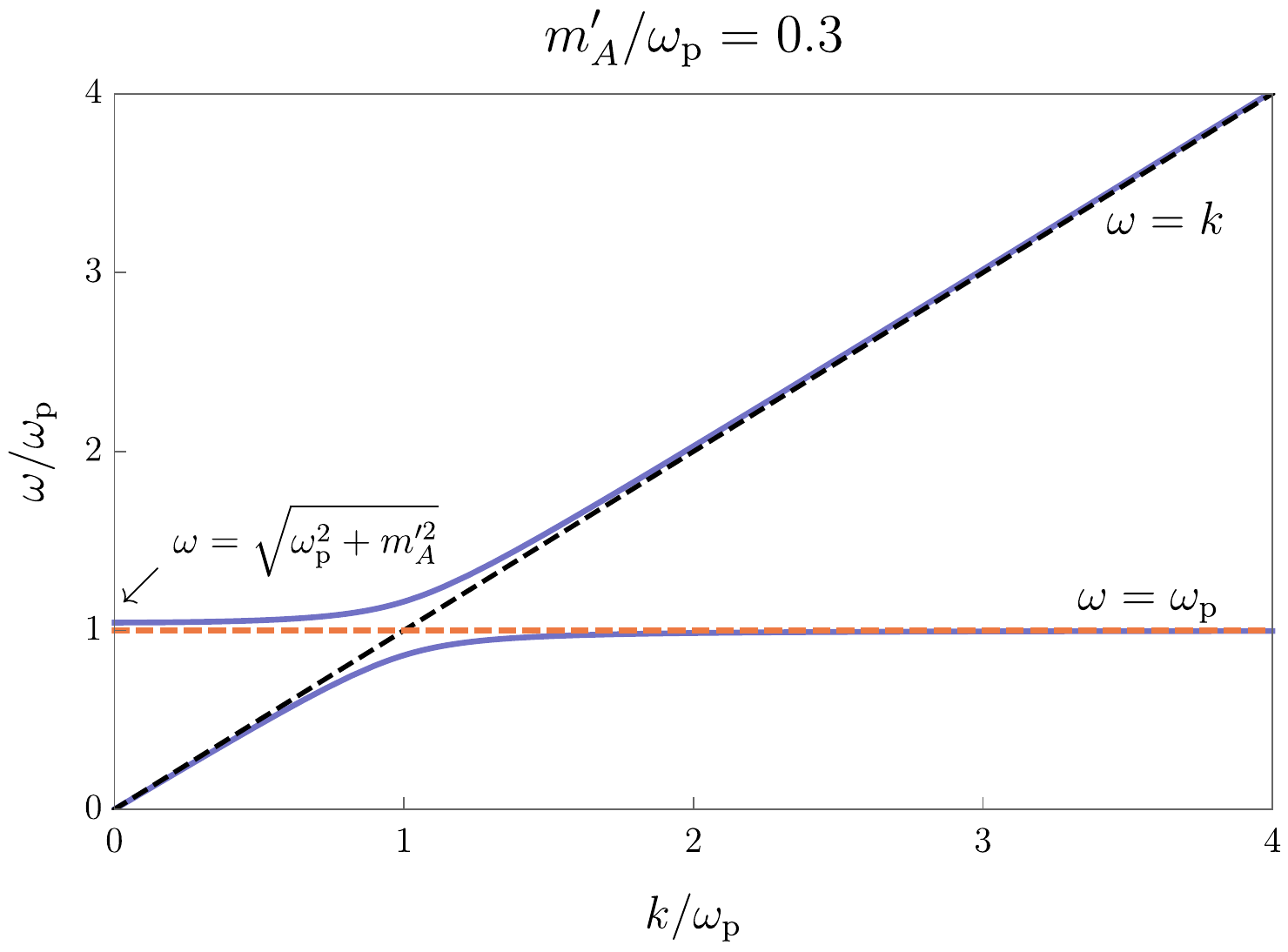}\hfill
\includegraphics[height=.36\textwidth]{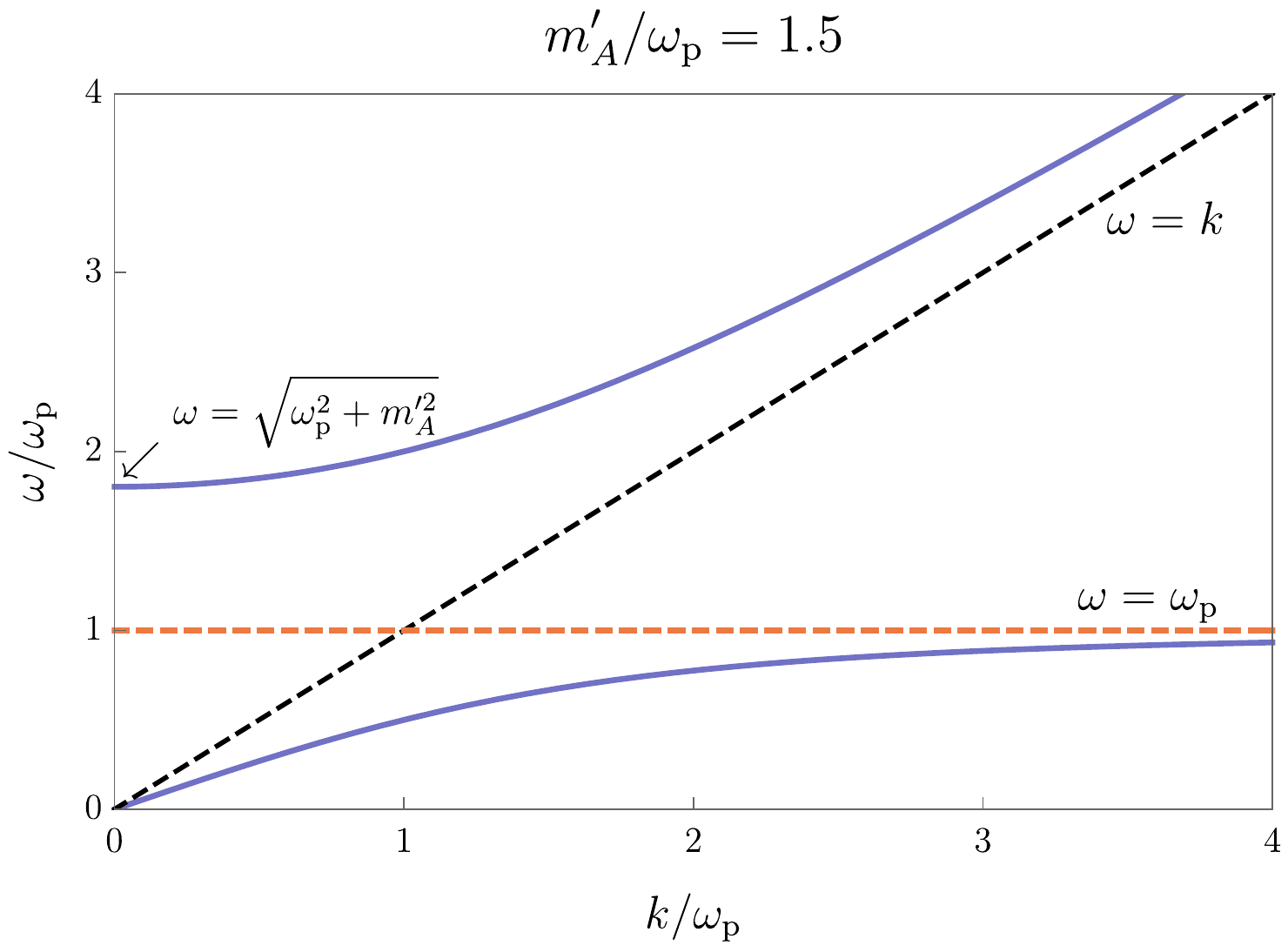}
\caption{Plot of the two branches of the dispersion relation in a cold collisionless plasma for $\mA=0.3\op$ and $\mA=1.5\op$, using Eq.~\ref{eq:dispRelMinkowski} with $\nu=0$. The slow (lower) branch asymptotes to $\omega=\op$ (dashed, orange) at large $k$, in both cases, whereas the fast (upper) branch asymptotes to the light cone $\omega=k$ (dashed, black). As $\mA$ increases, the branches move further away from each other. For any $\mA$, the fast branch has $\omega=(\op^2+\mA^2)^{1/2}$ at $k=0$.}
\label{fig:cold_dispersions}
\end{figure}

\subsubsection*{Modes with $\omp \gg k \gg \mA$}
For modes with $\omp \gg k \gg \mA$, the dispersion relation admits two branches, which we refer to as the ``fast'' and ``slow'' modes.  These can be seen in the left-hand side of the left panel of Fig.~\ref{fig:cold_dispersions}.  To leading order these are given by $\omega_\text{fast} \sim \omp$ and $\omega_\text{slow} \sim k$ in a collisionless plasma ($\nu = 0$).  Solving Eq.~\ref{eq:dispRelMinkowski} for generic $\nu$ shows that as the plasma becomes more and more collisional, the fast mode becomes significantly damped and eventually overdamped for $\nu \gtrsim \omp$ while the slow mode remains underdamped for all values of $\nu$, achieving a maximal damping rate of $\Im{(\omega_\text{slow})} = - \frac{1}{4} \frac{m^2}{k} \frac{\omp^2}{\omp^2 - k^2}$ when $\nu = \frac{1}{2} \frac{\omp^2}{k} - \frac{k}{2}$.  This tells us that the presence of a collisional plasma does not significantly affect the behavior of the slow mode (up to a small damping rate), and as we will see in Sec.~\ref{app:ProcaPlasmasGR}, the initial conditions sourced by inflation tend to project primarily onto a mode that in the Minkowski limit corresponds to this slow mode.

\subsubsection*{Modes with $\omp \gg \mA \gg k$}
For modes with $\omp \gg \mA \gg k$ the effects of the plasma are more noticeable.  In vacuum, all modes with $k \ll \mA$ would obey the dispersion relation $\omega^2 = k^2 + \mA^2$ and so would oscillate with a natural frequency of approximately $\mA$, however in a plasma this is no longer the case.  Eq.~\ref{eq:dispRelMinkowski} again yields two solutions, which we refer to as the ``fast'' and ``slow'' modes and display in the left panel of Fig.~\ref{fig:cold_dispersions}, and we again have for $\nu = 0$ that $\omega_\text{fast} \sim \omp$ and $\omega_\text{slow} \sim k$.  In particular, note that the slow mode now oscillates much more slowly than it would in a vacuum.

Now we may ask what happens in a collisional plasma.  For $\nu \lesssim \omp$, both the slow and fast modes become damped with damping rates proportional to $\nu$, but they are bounded in the complex plane by $|\omega_\text{fast}| \sim \omp$ and $|\omega_\text{slow}| \sim k$.  This means that in a weakly collisional plasma, the timescale to damp a longitudinal mode (fast or slow) away is never smaller than the timescale for the analogous noncollisional mode to oscillate.

In a thermalized plasma, $\nu$ is generically smaller than $\omp$ because it is suppressed by additional powers of the coupling, but if the plasma is out of thermal equilibrium $\nu$ can in principle be much larger than $\omp$.  In this case, the long-wavelength modes reorganize into a pair of underdamped vacuum-like modes with $|\omega | \sim \sqrt{\mA^2 + k^2}$ and a pair of overdamped modes (i.e.\ $\Re{(\omega)} = 0$) with damping rates $\Im{(\omega)} \sim - 2 \nu, - \frac{\omp^2 k^2}{2 \nu \mA^2}$.

\subsubsection*{Modes with $\mA \gg \omp \gg k$}
In our model, the plasma frequency drops below the dark photon mass at late times, and for $\mA \gg \omp \gg k$, the ``slow'' and ``fast'' modes reorganize, as can be seen in the right panel of Fig.~\ref{fig:cold_dispersions}.  In a vacuum there would only be one mode, and it would have $\omega^2 = \mA^2 + k^2 \sim \mA^2$.  In a collisionless plasma this mode persists but we also gain an additional slow mode with $\omega \approx \frac{\omp}{\mA} k$.  As the plasma becomes more collisional, the fast mode remains underdamped with a damping rate given by $\Im{(\omega_\text{fast})} \approx - \omp^2 \frac{\nu}{\mA^2 + 4 \nu^2}$.  The slow mode on the other hand is well-modeled as a damped harmonic oscillator, and for $\nu \gtrsim k \frac{\omp}{\mA}$ it becomes overdamped with a damping rate for the slowly-damped mode of $\Im{(\omega_\text{slow})} \approx - \frac{k^2 \omp^2}{2 \mA^2 \nu}$. 

We provide simple closed-form expressions that encapsulate all limits discussed above in Table~\ref{tab:longWavelengthModes}.

\begin{table} 
\begin{center}
\begin{tabular}{lcc}
\toprule
Quantity &Slow mode & Fast mode \\
\midrule
Frequency $\Re{(\omega)}$ &$\frac{k\omp}{\sqrt{\mA^2+\omp^2}}$& $\sqrt{\mA^2 +\omp^2}$\smallskip\\
E-Field $\vec{E}$& $+i\frac{k}{\omp} \frac{\mA^2}{\sqrt{\omp^2+\mA^2}} \vec{A_L} $& $- i \sqrt{\omp^2 +\mA^2} \vec{A_L} $\smallskip\\
Current density $\vec{J}$& $-\mA^2 \vec{A_L}$ & $- \omp^2 \vec{A_L}$\smallskip\\
Charge density $J_0$& $\frac{\sqrt{\mA^2+\omp^2}}{\omp}\mA^2 \vec{A_L}$ & $\frac{k}{\sqrt{\mA^2+\omp^2}}\omp^2 \vec{A_L}$\\
Energy in $A'$ mass term $\rho_A$& $\mA^2 | \vec{A_L} |^2$ & $\mA^2 | \vec{A_L} |^2$\smallskip\\
Energy in  $\tfrac 12 E^2/\rho_A$&$\frac{k^2}{\omp^2} \frac{\mA^2}{\mA^2+\omp^2}$ & $\frac{\omp^2 + \mA^2}{\mA^2} $\smallskip\\
Fermion kinetic energy $\rho_u/\rho_A$& $\frac{\mA^2}{\omp^2}$ & $\frac{\omp^2}{\mA^2}$\smallskip\\
Damping rate $\Im{(\omega)}$ &$-\frac{\nu \mA^2}{\mA^2+\omp^2}$ & $-\frac{\nu \omp^2}{\mA^2+\omp^2}$\\
\bottomrule
\end{tabular}
\caption{Summary table for the long-wavelength ($k \ll \max{(\mA, \omp)}$) ``slow'' and ``fast'' longitudinal modes of a Proca plasma in the cold fluid approximation in Minkowski space.  These results are only valid for collision rates $\nu \lesssim \omp$.  In our case, when plasma effects are important, the modes of $A'$ produced by inflation have a behavior (in curved space) that is quite analogous to the slow mode discussed here.
\label{tab:longWavelengthModes}}
\end{center}
\end{table}

\subsection{FRW background} \label{app:ProcaPlasmasGR}

In the early universe, we cannot trust the mode decomposition of Sec.~\ref{app:ProcaPlasmasMinkowski} for the simple reason that the background changes on timescales that are fast compared to the oscillation times of some modes.  We thus need to develop a more sophisticated cold fluid approximation.  This can be done by writing out a coupled system of equations for the $A$ field and the dark current $J^\mu$ in the presence of a radiation-dominated FRW background.  Without loss of generality we consider the longitudinal mode to point in the $z$-direction, and we recall that the Lorenz gauge condition allows us to write everything in terms of $A_3$.  We may then write the equation of motion for $A_3$ as:
\begin{equation} \label{eq:A3eomGR}
    \left[ \del_t^2 + H \left( \frac{3 k^2 + \mA^2 a^2}{k^2 + \mA^2 a^2} \right) \del_t + \frac{k^2}{a^2} + \mA^2 \right] A_3 = J_3 - \frac{2 H i k a^2 J_0}{k^2 + \mA^2 a^2}
\end{equation}
In addition we have the Lorentz force law for the velocity $u_3$ of a charged particle in an FRW background:
\begin{equation} \label{eq:u3eomGR}
    \del_t u_3 = - \frac{\qd}{\mps} \left( \frac{\mA^2 a^2}{k^2 + \mA^2 a^2} \del_t A_3 + \frac{i a^2}{k} \frac{k^2}{k^2 + \mA^2 a^2} J_0 \right) - 2 \nu u_3
\end{equation}
and the conservation equation for $J^\mu$:
\begin{equation} \label{eq:J0eomGR}
    \del_t J_0 + 3 H J_0 - \frac{i k}{a^2} J_3 = 0
\end{equation}
Recalling the definition of $J_3 \equiv \qd n_{\psi} u_{3, \psi} - \qd n_{\bar{\psi}} u_{3, \bar{\psi}}$ we may rewrite Eq.~\ref{eq:u3eomGR} as an equation for $J_3$:
\begin{equation} \label{eq:J3eomGR}
    \del_t J_3 = - \omp^2 \frac{\mA^2 a^2 \del_t A_3 - i k a^2 J_0}{k^2 + \mA^2 a^2} - 2 \nu J_3 + \frac{\del_t (\omp^2)}{\omp^2} J_3
\end{equation}
where $\nu$ and $\omp$ may be functions of the scale factor $a$ (for example due to the changing background number densities).  This system of equations (Eqs.~\ref{eq:A3eomGR}, \ref{eq:J0eomGR}, and \ref{eq:J3eomGR}) may now be integrated numerically to find the behavior of a given longitudinal mode of the dark plasma (see App.~\ref{app:E screening plasma}).

From our discussion above (Sec.~\ref{app:ProcaPlasmasMinkowski}), we expect that the above set of equations supports multiple types of oscillations at different frequencies.  The precise behavior of the system will thus depend on initial conditions.  In our case, the initial conditions are given by the behavior of $A_3$ at the end of inflation.  Immediately after inflation the number density of $\psi$ particles has been diluted to effectively zero, implying that $J_\text{0,init} = J_\text{3,init} = 0$.  The $A_3$ field, on the other hand, has been initialized to a large value of order $A_\text{3,init} \sim \frac{k \HI}{2 \pi \mA}$, and as discussed in App.~\ref{app:electric_field} its initial time derivative is given by $\del_t A_\text{3,init} \sim - \frac{k^2}{3 a^2 H} \frac{k \HI}{2 \pi \mA}$.  The initial time in the above equations may be chosen to occur at any time when $H(t) < \HI$ but before the first fermions are created.

The principal question is whether the presence of a plasma can cause a mode produced by inflation to begin oscillating before its oscillation time in vacuum. One may suspect the answer is yes, because oscillations in a plasma can be supported at a frequency $\omp$ which can be much higher than $\mA$ and $k$ (the quantities that set the oscillation time in vacuum).  However this is not the case: both numerical integration and analytic approximations (see App.~\ref{app:SH_plasma}) of the above system demonstrates that in a weakly-collisional plasma ($\nu \lesssim \omp$), whenever $\omp$ is larger than $k/a$, the mode produced by inflation begins oscillating at a time when 
\begin{equation}
\label{eq:H osc plasma}
H \sim \frac{\omp}{\sqrt{\mA^2 + \omp^2}} \frac{k}{a}    
\end{equation}
and subsequently oscillates with a frequency $\omega \sim \frac{\omp}{\sqrt{\mA^2 + \omp^2}} \frac{k}{a}$, a fact that can be intuitively understood by looking at Table~\ref{tab:longWavelengthModes}.  Because the initial conditions selected by the cosmology are $J_\text{0,init} = J_\text{3,init} = 0$, the behavior of the full FRW-background equations should be most similar to the behavior of the Minkowski mode with the smallest value of $J_3$ and $J_0$.  This mode is the slow mode, oscillating at precisely this frequency, which for large $\omp$ reduces to $\sim k/a$ (in the Minkowski case the physical momentum $k/a$ is replaced simply by $k$). 

Our numerical integration confirms that the Minkowski intuition works quite well even in an FRW background, and the behavior of the inflation-produced field $A_3$ and the currents $J_0$ and $J_3$ are well-modeled by the behavior of the slow mode discussed in Sec.~\ref{app:ProcaPlasmasMinkowski} (we refer the reader also to App.~\ref{app:SH_plasma} for an analytic derivation of this fact).  This implies that in the presence of a plasma with $\omp \gtrsim \mA$, modes with $k < \kstar$ will not begin oscillating until $H \sim k/a$ rather than the vacuum expectation of $H \sim \mA$.  The net impact of this (as discussed in Sec.~\ref{sec:thermalization}) is a shift to the parameter $\eta$ that measures the number of $e$-folds of comoving wavenumber that contribute to the final energy density of the dark thermal bath.  In our case, there are specific forms of $\omp(a)$ and $\nu(a)$ that are predicted by the processes that eventually lead to thermalization, and we discuss these in App.~\ref{app:thermalizationAppendix}.

\section{Warm Proca Plasmas} \label{app:warmProcaPlasmas}

In this appendix we outline how thermal effects modify the dispersion relations of Proca plasmas in a Minkowski spacetime.  We do not attempt a complete treatment of this complex subject, but we point out that we do not expect them to alter any of our main calculations (i.e.\ those of App.~\ref{app:thermalizationAppendix}), where they are not included explicitly.  This is because the plasma effects that matter for us are those which affect superhorizon modes (for which $k/a < H$), and as we show here, significant thermal corrections do not begin to set in until the Debye wavenumber $\kD$.  When the dark sector forms a thermalized plasma, its plasma frequency is $\omp \gg H$ and $\kD \gtrsim \omp$, so $k/a < H \ll \omp \lesssim \kD$ and superhorizon modes are not significantly affected.  We leave further work on this front to future work.

We start from the relativistic Vlasov equation for a collisionless, unmagnetized plasma
\begin{equation}
\label{eq:vlasov_flat}
\frac{\partial f}{\partial t}+\frac{p^i}{E}\frac{\partial f}{\partial x^i}-qF^{i0}\frac{\partial f}{\partial p^i}=0,
\end{equation}
where $f=f(\vec{x},\vec{p},t)$ is the phase-space density. Collisions can, in principle, be taken into account in this formalism, if we include the usual collision integral in the right-hand-side \cite{Dodelson2003}. A simpler approach would be to write a covariant phenomenological term that drives the phase space distribution towards equilibrium. For our purposes, we expect that collisions are well modelled by the phenomenological $\nu$ term in the cold-fluid formalism, Eq.~\ref{eq:cold-lorentz}. Here, we are interested only in thermal effects that do not arise in the cold plasma approximation, so we will neglect collisions. 

Treating the external electric field as a perturbation, we may write $f=f_\text{eq}(|\vec{p}|,t)+\delta f(\vec{x},\vec{p},t)$, where $f_\text{eq}$ is the distribution in thermal equilibrium, which we take to be the Maxwell-J\"{u}ttner distribution $f_\text{eq}=\exp\pare{-\sqrt{p^2 + m_\psi^2}/T}$.

Linearizing the equations in the flat spacetime limit and Fourier-transforming in both time and space yields 
\begin{equation}
    J^\mu\equiv2q\int\frac{\di^3 p}{(2\pi)^3}\frac{p^\mu}{E} f=2q^2\int\frac{\di^3 p}{(2\pi)^3}\frac{p^\mu}{E}\frac{F^{i0}}{\omega-\frac{k_jp^j}{E}}\frac{\partial f_\text{eq}}{\partial p^i},
\end{equation}
where the integral over $f_\text{eq}$ yielded exactly zero due to the integrand being an odd function of the momenta. We decompose $A^i$ into a longitudinal and transverse component, writing $A^i=\frac{k^i}{k}A_\text{L}+A_{\text{T}}^i$, such that $k_i A_\text{T}^i=0$. From Eq.~\ref{eq:dispersion}, the dispersion equation for the longitudinal mode from is
\begin{equation}
\label{eq:proca_disp}
    \pare{-\omega^2+k^2+\mA^2}A_\text{L}=\frac{k_i}{k}J^i \,.
\end{equation}
Using the relation $\partial_\mu A^\mu=0$, we may write $F^{i0}=i\pare{k^iA^0-A^i k^0}=i\frac{k^ik}{\omega}\pare{1-\frac{\omega^2}{k^2}}A_\text{L}$. We choose $k^i$ to lie along the $z$-axis, so that $k^ip_i=kp\cos\theta$. The longitudinal current then becomes
\begin{equation}
\label{eq:rel_current}
\frac{k_j}{k}J^j=- \frac{2q^2}{4\pi^2}\frac{k^2}{\omega}\pare{1-\frac{\omega^2}{k^2}}A_L\int_0^{+\infty}\di p\,\frac{p^3}{E}\frac{\partial f_\text{eq}}{\partial p}\int_{-1}^{1}\di x\,\frac{x^2}{\omega-\frac{k p x}{E}},
\end{equation}
where we defined $x\equiv\cos\theta$ and carried out the azimuthal integration. One can see clearly that the denominator of Eq.~\ref{eq:rel_current} is singular when $\omega=kpx/E$, which is the mathematical manifestation of Landau damping \cite{1981}. As such, we write $\omega=\ore+i\omega_\textsc{i}$, before carrying out the integration over $x$. Upon integrating, we essentially reproduce a known result \cite{Buti1962}, 
\begin{multline}
\label{eq:angle_int}
I_\theta\equiv\int_{-1}^1\frac{x^2\,\di x} {\omega-\frac{kpx}{E}}= -\frac{\ore}{k^2v^2}\parea{2-2\frac{\omega_\textsc{i}}{kv}\pare{\tan^{-1}\frac{kv-\ore}{\omega_\textsc{i}}+\tan^{-1}\frac{kv+\ore}{\omega_\textsc{i}}}+\frac{\ore^2-\omega_\textsc{i}^2}{kv\ore}\ell}\\
-\frac{i\omega_\textsc{i}}{k^2v^2}\parea{2-\frac{\ore^2-\omega_\textsc{i}^2}{kv\omega_\textsc{i}}\pare{\tan^{-1}\frac{kv-\ore}{\omega_\textsc{i}}+\tan^{-1}\frac{kv+\ore}{\omega_\textsc{i}}}-\frac{2\ore}{kv}
\ell},
\end{multline}
where $v\equiv p/E$, $\ell=\log\frac{\ore-kv}{\ore+kv}$. This formula is exact to all orders in $v$ and $\omega_\textsc{i}$. We expand in powers of $\omega_\textsc{i}<0$, as is usually done \cite{1981}, which yields
\begin{equation}
\label{eq:angle_int_small}
    I_\theta\simeq -\frac{\ore}{k^2v^2}\parea{2+\frac{\ore}{2kv} \ell}-\frac{i\pi\ore^2}{k^3v^3}\Theta(kv-\ore)-\frac{2i\omega_\textsc{i}}{k^2v^2}\parea{1-\frac{\ore^2}{k^2v^2-\ore^2}+\frac{\ore}{kv} \ell},
\end{equation}
First we will look at the non-relativistic limit of this equation. Here, we may expand in powers of $v$, in which case the dispersion relation Eq.~\ref{eq:proca_disp} becomes
\begin{multline}
\label{eq:proca_landau}
1+\frac{\mA^2}{k^2-\ore^2}\pare{1+\frac{2i\omega_\textsc{i}\ore}{k^2-\ore^2}}=   
  \frac{\op^2}{\ore^2}\pare{1-i\frac{\omega_\textsc{i}}{\ore}}\left(1+\frac{3k^2\op^2}{\ore^2\kD^2}-\frac{i\omega_\textsc{i}}{\ore}-\frac{9i\omega_\textsc{i} k^2\op^2}{\ore^3\kD^2}\right)+\\
  -i\frac{\kD^3}{k^3}\sqrt{\frac{\pi}{2}}\frac{\ore^3}{\op^3}e^{-\frac{\mA\ore^2}{2k^2T}}.
\end{multline}
where $\kD\equiv\sqrt{4\pi\alpha n_\psi/T}$ is the Debye length, where $n_\psi$ is the fermion number density. For $\mA=0$, this reduces to the usual dispersion relation for warm plasmas, where the term in the second line was first noted by Landau \cite{landau} and corresponds to a collisionless damping mechanism. The real part of Eq.~\ref{eq:proca_landau} can be solved exactly to yield the real part of the frequency, but the answer is not particularly illuminating. In the small $k$ limit, we find the two solutions
\begin{subequations}
\label{eq:branches_nonrel}
\begin{align}
    \ore^2=&\op^2\pare{1+\frac{3k^2}{\kD^2}\frac{\op^2}{
    \op^2+\mA^2}}+\mA^2\pare{1+\frac{k^2}{\mA^2+\op^2}},\\ 
    &\omega_\textsc{i}=0,\quad \text{ since }\ore>k\\
    \ore^2=&\frac{k^2\op^2}{\mA^2+\op^2}\pare{1+\frac{3\mA^2}{\kD^2}},\\ &\omega_\textsc{i}\simeq-\frac{k^3\kD^3 \mA^2\op} {\pare{\mA^2+\op^2}^4}\sqrt{\frac{\pi}{8}}e^{-\frac{\kD^2}{2\pare{\mA^2+\op^2}}},\quad \kD\to\infty
    \end{align}
\end{subequations}
For large $k$, the slow branch has dispersion
\begin{subequations}
\begin{align}
    \ore^2&=\op^2\pare{1-\frac{\mA^2}{k^2}+\frac{3k^2}{\kD^2}}\\
    \omega_\textsc{i}&\simeq -\op\sqrt{\frac{\pi}{8}}\pare{\frac{\kD}{k}}^3e^{-\frac{\kD^2}{2k^2}\pare{1-\frac{\mA^2}{k^2}}},\quad \mA\ll k,\op
\end{align}
\end{subequations}
where the Landau damping rate of the slow branch is the usual non-relativistic one \cite{1981} with an $\mA$ correction. This demonstrates that the slow branch inherits the usual QED dispersion at large $k$, as long as $\mA$ is small.

From these relations, it becomes clear that thermal corrections to non-relativistic modes of the slow branch induce only small perturbations on the frequencies and that Landau damping is exponentially suppressed for $k\ll \kD$. As in the usual case of Standard Model plasmas \cite{1981}, we cannot trust our calculation past that point at which the damping becomes significant. However, this result illustrates that for small $k$, which is the region we are interested in, the slow branch oscillations will not be affected by thermal effects.

We now move to relativistic plasmas, for which $v\simeq 1$. In the regime $\op\gg \mA,k$ we are interested in, we expect, as in the non-relativistic case, $\ore\simeq k$. Since all quantities in $I_\theta$ of Eq.~\ref{eq:angle_int_small} are integrated against $f_\text{eq}$, we expect that the dominant contribution will come from the region near the typical velocities of the particles in the plasma. For QED plasmas, it was first proved in \cite{Braaten1993} that substituting $v$ for the typical velocity $\vs$ is a good analytic approximation to the dispersion relation, and exact in the cases of a non-relativistic, degenerate and ultra-relativistic plasma. The typical velocity $\vs$ is defined as
\begin{equation}
\vs \equiv\frac{\omega_1}{\op}\,,\text{ where } 
\left\lbrace \begin{aligned} 
\op^2&\equiv\frac{4\alpha}{\pi}\int_0^{+\infty}\di p\, f_\text{eq} p\pare{v-\frac{1}{3}v^3}\\
    \omega_1^2&\equiv\frac{4\alpha}{\pi}\int_0^{+\infty}\di p\, f_\text{eq} p\pare{\frac{5}{3}v^3-v^5}\end{aligned} \right.
\label{eq:vstar}
\end{equation}
It is interesting to understand how $\vs$ depends on the ratio $a \equiv m_\psi/T$ . Using its definition, Eq.~\ref{eq:vstar}, and expanding for large $E/m_\psi$, we find $\vs\simeq 1-11a^2/16$.

\begin{figure}[ht] \centering
\includegraphics[height=.36\textwidth]{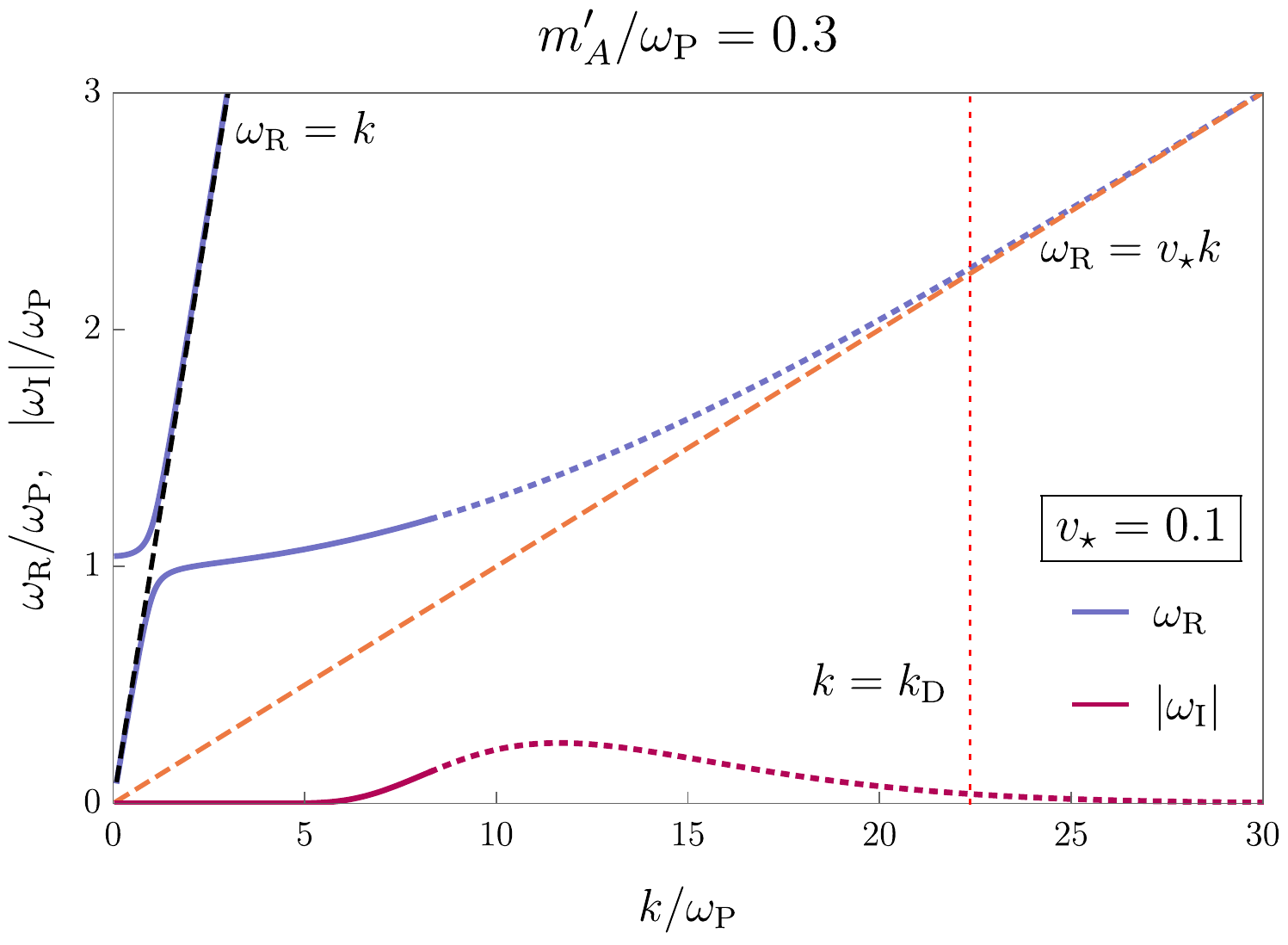}\hfill
\includegraphics[height=.36\textwidth]{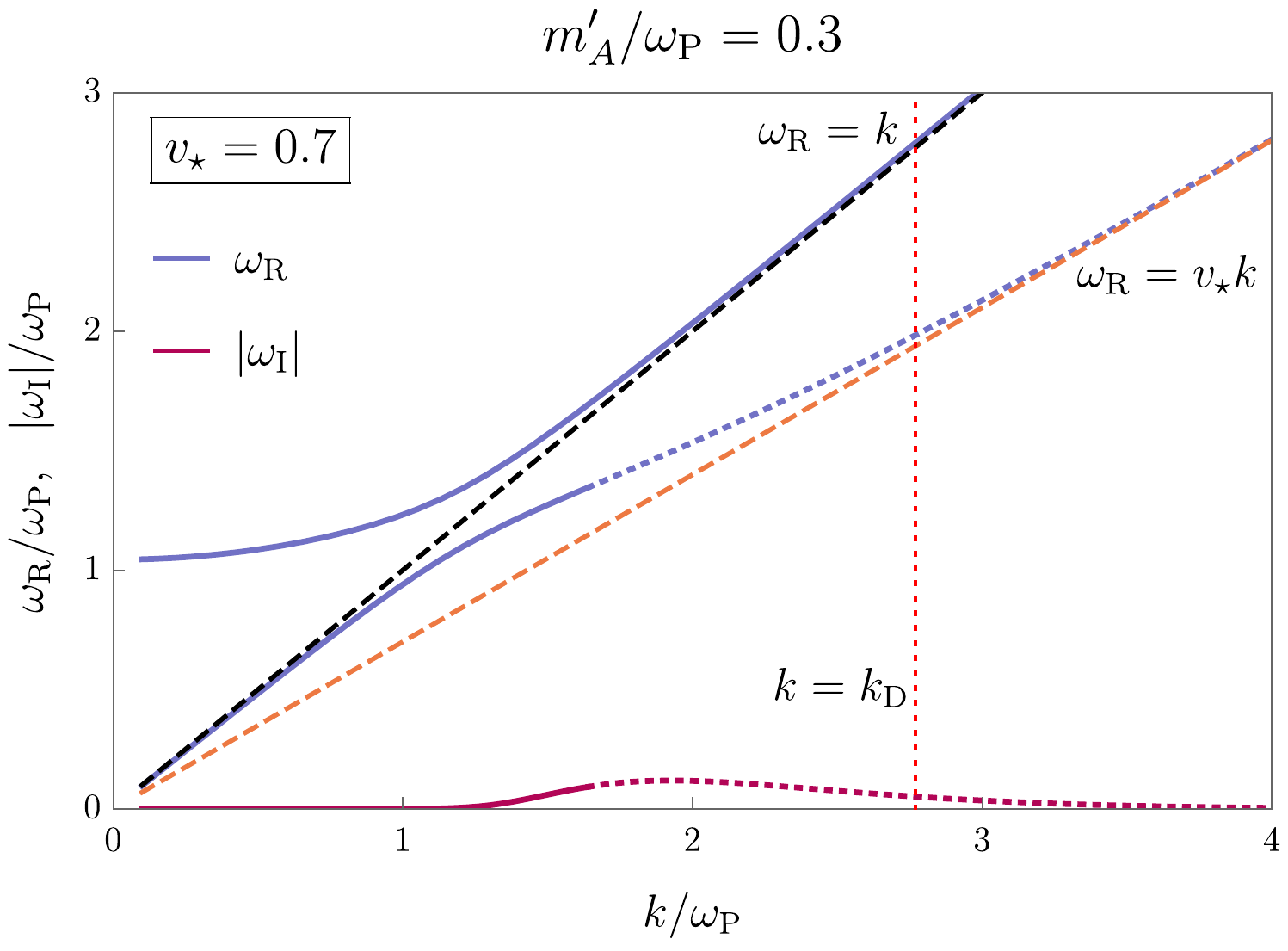}
\caption{Plot of the two branches of the dispersion relations in a warm collisionless plasma for $\mA=0.3\,\op$, using Eqs.~\ref{eq:branches_nonrel} and \ref{eq:branches_rel}. We plot the dispersion relations for two typical electron gas velocities $\vs=0.1$ and $\vs=0.7$. The slow (lower) branch asymptotes to the lines $\ore=\vs k$ (dashed, orange) at large $k$. As the velocity of the gas increases, the slow branch moves closer to the light cone (dashed, black). In the opposite limit of a small velocity, the lower branch matches the cold plasma dispersion of App.~\ref{app:ProcaPlasmasMinkowski} for $k\ll\kD$, namely it flattens out. Landau damping (red) becomes important for the slow branch around $\kD$, as in the usual non-relativistic QED case \cite{1981} (the fast branch is not Landau damped as it lives only above the light cone). Past this point, we cannot trust our calculations and proceed in plotting the real part with a dotted line, as in \cite{Raffelt:1996wa}. For $k\ll \mA,\op$, thermal corrections do not affect the acoustic part of the slow branch that we are mostly interested in.}
\label{fig:warm_dispersions}
\end{figure}

Note that we are still interested in $\ore>k\vs$, even for the slow branch, but the $\Theta$-function term receives contributions entirely outside that region, namely from the ``tail'' of the distribution function $v>\ore/k$. We adopt the approach of Ref.~\cite{Braaten1993} for both the real and the imaginary parts (i.e.\ the bracket terms in Eq.~\ref{eq:angle_int_small}), except for the $\Theta$-function term, which we integrate exactly yielding\footnote{The integration can be carried out by setting $t\equiv\sqrt{1+p^2/m_\psi^2}$, for which the lower limit set by the $\Theta$-function translates into $t>\parea{1-\pare{\ore/k}^2}^{-1/2}$.}
\begin{equation}
\int_0^{+\infty}\di p\, E^2\frac{\partial f_\text{eq}}{\partial p}\Theta(kp-\ore E)=-m_\psi^2e^{-\frac{a k}{\sqrt{k^2-\ore^2}}}\frac{(2+a^2)k^2-2\ore^2+2 a k\sqrt{k^2-\ore^2}}{a^2\pare{k^2-\ore^2}},
\end{equation}
where as above, $a \equiv \mps/T$. The dispersion relations for small $\omega_\textsc{i}$ (after separating the real and the imaginary parts) are
\begin{subequations}
\label{eq:branches_rel}
\begin{align}
1+\frac{\mA^2}{k^2-\ore^2}=&-\frac{3\op^2}{k^2\vs^2}\parea{1+\frac{\ore}{2\vs k}\log\frac{\ore-\vs k}{\ore +\vs k}}\label{eq:rel_proca_real2}\\
\frac{2\omega_\textsc{i} \mA^2\ore}{\pare{k^2-\ore^2}^2}=&-\frac{3\op^2}{k^2\vs^2}\left[\omega_\textsc{i}\pare{\frac{1}{2k\vs}\log\frac{\ore-\vs k}{\ore +\vs k}+\frac{\ore}{\ore^2-k^2\vs^2}}+\right.\nonumber\\
&\left.+\frac{\pi\ore\vs^2}{k}\frac{2\alpha \mps^2}{3\pi\op^2} 
e^{-\frac{a k}{\sqrt{k^2-\ore^2}}} \cdot
\frac{(2+a^2)k^2-2\ore^2+2 a k\sqrt{k^2-\ore^2}}{a^2\pare{k^2-\ore^2}}\right]\label{eq:re_proca_im2}
    \end{align}
\end{subequations}
Eq.~\ref{eq:rel_proca_real2} can be solved numerically to give $\ore$ as a function of $\vs$, whereas Eq.~\ref{eq:re_proca_im2} is linear in $\omega_\textsc{i}$ and has $\ore$ as input. We plot the numerical solution to these equations in Fig.~\ref{fig:warm_dispersions} for two values of the typical gas velocity $\vs$. One can clearly see that we get two branches, one entirely above the light cone (which suffers no Landau damping) and one below.

Before solving for $\omega_\textsc{i}$, let us understand a little the structure of the two modes analytically. An interesting point in the $\ore-k$ plane is the one where the two branches of the dispersion relation avoid crossing (at least for small $\mA$). 
This is the point at which the longitudinal mode of a massless photon would cross the light-cone, i.e. when $\ore=k_\text{cross}$ in Eq.~\ref{eq:rel_proca_real2}, for $\mA=0$. Solving for $k_\text{cross}$ yields
\begin{equation}
\label{eq:kcross}
k_\text{cross} =\frac{\op}{\vs^{3/2}}\pare{\frac{3}{2}}^{1/2}\parea{-\log\pare{\frac{1-\vs}{1+\vs}}-2\vs}
\simeq\op\pare{\frac{3}{2}}^{1/2}\log\frac{1}{1-\vs}
\end{equation}
It is evident that the crossing point goes  (logarithmically) to infinity as $a\to 0$ or as $\vs\to1$. In particular, for an exactly massless fermion the longitudinal mode of a massless photon does not cross the light cone, which means that, for a massive photon, the slow mode reduces to the pure gauge mode $\ore=k$, essentially becoming decoupled.

For small $k$, we expect that the slow branch will behave as $\ore\simeq k$ when $\op$ is the largest scale. We set $\ore=k(1-\delta)$ and linearize Eq.~\ref{eq:rel_proca_real2} in $\delta$. Note that such an expansion implicitly assumes that $\mA$ is also small. We find
\begin{equation}
    \delta\simeq-\frac{\mA^2}{\op^2}\frac{\vs^3}{\log(1-\vs)},\quad \text{ for $1-\vs\ll 1$}
\end{equation}
which is a positive quantity. 

We then return to the calculation of the Landau damping rate for the slow mode. Solving Eq.~\ref{eq:re_proca_im2} for $\omega_\textsc{i}$ and expanding for small $\delta$ gives
\begin{equation}
\omega_\textsc{i}\simeq-\op\exp\pare{-\frac{a}{\sqrt{2\delta}}}\frac{3a^2k\pi}{4\mA^2}\delta.
\end{equation}
Again, we see that thermal effects introduce only small corrections to the cold dispersion relation of App.~\ref{app:ProcaPlasmasMinkowski} in the region of interest (namely for small $k$), and so they can be neglected to a good approximation. 

In conclusion, the cold fluid approximation used in the previous sections does not miss any qualitative phenomena at long enough wavelengths above $\kD^{-1}$. In particular, when the plasma thermalizes, $\op\gg H$ (see App.~\ref{app:first steps thermalization}), so superhorizon modes with $k<H\ll\op\lesssim\kD$ are unaffected by thermal effects.

\section{A more detailed description of thermalization in the dark sector} \label{app:thermalizationAppendix}
In this section, we review in more detail the thermalization process outlined in Sec.~\ref{sec:lightPhotonThermalization}--for the sake of clarity, in the Apps.~\ref{app:before Hcasc}-\ref{app:towards thermalization} we mirror the titles of the corresponding sections in  Sec.~\ref{sec:lightPhotonThermalization}.
In order to support the picture summarized by Fig.~\ref{fig:cartoon thermalization}, we need to describe in further detail the evolution of the dark sector in the early stages with strong $E'$ fields, from the initial phase of large $H$ to the stage of dark electromagnetic cascades (App.~\ref{app:before Hcasc}).  This builds on the discussion of Apps.~\ref{app:electric_field} and \ref{app:SFQED}. 
The dark sector then enters what we call ``screening regime'' (App.~\ref{app:screening regime}) when the presence of the fermion plasma starts affecting the evolution of the $A'_L$ modes. Armed with the results of Apps.~\ref{app:ProcaPlasmas} and \ref{app:warmProcaPlasmas}, we discuss in Apps.~\ref{app:first steps thermalization},  and \ref{app:towards thermalization} and \ref{app:E screening plasma} the subsequent steps towards thermalization. Finally, in App.~\ref{app:SH_plasma} we study the evolution of $A_3'$ modes that re-enter after the fermions have frozen out, and provide analytic arguments as to why they cannot overclose the universe.
The discussion applies not only to the case of light $A'$ (i.e.~$\mps\gg \mA$), but also to that of heavy $A'$ ($\mA\gg 2\mps$). We illustrate in App.~\ref{app:thermalization heavy A} the few relevant differences.

\subsection{Pure electric field dominance era -- expanded}
\label{app:before Hcasc}
At early times after the end of inflation ($H\gg \sqrt{\mA\HI}, \mps$), most of the energy density per each mode at horizon entry in the dark sector (gravitationally produced during inflation) is stored in frozen longitudinal modes $A'_L$ (with $\rho_{A'_L}\sim H^2 \HI^2$, see Eq.~\ref{eq:rho_enter}), with subleading contributions from transverse modes (we expect $\rho_{A'_T}\sim \qd^4 H^4$ on the basis of \cite{Dolgov:1981nw,Dolgov:1993vg}) and fermions ($\rho_{\psi}\sim \mps^4$ when $H \sim \mps$ \cite{Lyth:1996yj,Kuzmin:1998kk,Chung:2011ck,Ema:2019yrd}).
As discussed in Sec.~\ref{sec:thermalization} and especially in App.~\ref{app:electric_field}, as the $A'_L$ modes enter the horizon and start oscillating with frequency $\sim H$, they generate a dark electric field of size $E'\sim \ELvac \equiv \mA \HI/(2\pi)$ with a coherence length of order $H^{-1}$.

For the values of $\mps,\,\qd$ we are interested in to achieve the right DM abundance for $\psi$ through freeze-out, and for $\mA$ not too much smaller than $\mps$ (or, more precisely, $\mA\gtrsim 2\pi\mps^2/(\qd \HI)$), this electric field is larger than the critical field $\Ecr$ defined in Eq.~\ref{eq:critical_field}, meaning that the SFQED processes described in Sec.~\ref{app:SFQED} are relevant.
In particular, the main process for the production of $\psi$ before $H\sim \mps$ is the dark electromagnetic cascade.
We can rewrite the rate in Eq.~\ref{eq:pairs} as the rate of $\psi\overline\psi$ pairs produced per unit volume out of a number density $\nA$ of transverse photons of energy $\omega_{A'}$, and approximate $\chi \sim \frac{\qd E' \omega_{A'}}{\mps^3}$ (i.e.\ considering $\vec E'$ roughly orthogonal to $\vec k_{A'}$): 
\begin{equation}
\Wcasc = \frac{\di N_{\psi\overline\psi}}{\di t\di V} \sim \nA \cdot \begin{cases}
\label{eq:cascade rate}
\qd^3 \dfrac{E'}{\mps} e^{-8/(3\chi)} & \chi \lesssim 1\smallskip\\
\qd^{8/3} \dfrac{E'^{2/3}}{\omega_{A'}^{1/3}} & \chi \gg 1
\end{cases}
\end{equation}
For the case of a longitudinal photon in the external leg of Fig.~\ref{fig:pair_creation}, we must include an additional suppression factor of $(\mA/\omega_{A'})^2$ from the polarisation of the external $A'_L$.
Dark photon modes entering the horizon (and hence experiencing large electric fields on their length scales) at early times satisfy the condition $\chi > 1$, and can produce $\psi \overline \psi$ pairs whenever $\vec E' \nparallel \vec k_{A'}$.

At early times, when the Hubble rate is large, the fermion production rate of Eq.~\ref{eq:cascade rate} is not phenomenologically relevant because of the severe Hubble dilution of the produced fermions ($\Wcasc\ll H^4$). 
In particular, the acceleration of the produced fermions and subsequent cascade process is not efficient as long as the dark charges are not accelerated for a long enough time to compensate for their redshift. 
The typical fermion energy after being accelerated over a Hubble radius ($H^{-1}$ determines both the coherence length and the oscillation frequency of the $E'$ field) is $\qd \ELvac/H$.
The maximum energy injected by the electric field in a Hubble time is thus $\qd \ELvac \nps /H$, which is smaller than the Hubble dilution $3H\nps$ until $H\sim \Hcasc \equiv\sqrt{\qd \ELvac}=\sqrt{\qd \mA\HI/(2\pi)}$. 
Cascade production of $\psi$ can then only occur for $H \lesssim \Hcasc$.
One can see that at any given time (not much before $\Hcasc$) most of the pair production comes from $A'_L$ modes with $k\sim aH$, rather than from $A'_T$ or $\psi$ modes generated during inflation, or $A'_L$ modes that entered the horizon earlier.

Pair-produced dark fermions radiate $A'$ photons while they get linearly accelerated (over a typical distance $H^{-1}$) in the electric field $E'$, and especially when they significantly change their direction. 
The radiated $A'$ are mostly transversely polarized (given that the emission of $A'_L$ is suppressed by $\mA^2/\omega_\psi^2$ \cite{Lechner:2018}), and if they satisfy $\chi \gtrsim 1$ they can in turn pair produce. 
A detailed study of the spectrum and angular distribution of the radiated photons would require a numerical simulation that goes beyond the scope of this work. 
In order to provide a qualitative picture of the dynamics, it is enough for us to understand the typical energy of the radiated $A'_T$. 
The spectrum emitted by a charge undergoing acceleration over a short timescale is roughly flat in energy density, and the corresponding number density peaks in the IR as $1/\omega_{A'}$ (the so-called IR catastrophe) down to an IR cut-off determined by the problem at hand \cite{Jackson:1999,Lechner:2018}.
We expect this simplified description to capture the physics of a charge moving in approximately constant fields for distances $\sim H^{-1}$ and bending over regions of similar size. 
Given that the IR cut-off for our problem is the Hubble radius, we expect most radiated photons to have $\omega_{A'}\sim H$.

In order to assess whether and when an exponential cascade (as depicted in Fig.~\ref{fig:qed_cascade}) can take place, the condition that $\ELvac =\mA \HI /(2\pi) > \Ecr$ (see the discussion around Eq.~\ref{eq:critical_field}) turns out to play a relevant role, even without requiring that Schwinger pair production is accessible.
In the parameter space where $\qd \ELvac >\mps^2$ (as above the purple line in Fig.~\ref{fig:lightPhotonRelicAbundance}), as soon as $H<\Hcasc$ (so that the electric field can inject energy into the fermions faster than the Hubble dilution) we have $\omega_{A'}\sim H > \mps$ for the $A'_T$ radiated by the accelerated $\psi$'s, hence $\chi>1$ and the $A'_T$ can in turn pair produce.
\footnote{One could worry about the fact that the radiation emitted from an ultra-relativistic charge $\psi$ is localised in a narrow cone of opening angle $\mps/\omps$, potentially leading to $\vec k_{A'}|| \vec E'$ and reducing $\chi$. This is clearly not a problem for $\qd E'>\mps^2$ because the $\psi$ do not need to reach ultra-relativistic velocities for the softest radiated $A'_T$ around $\Hcasc$ to satisfy $\chi >1$. It turns out not to be an issue also for the case $\qd E' <\mps^2$ where we rely on harder radiated $A'$ to reach $\chi\sim 1$, because the $\psi$ deflection angle (when $\vec E' \nparallel \vec k_\psi$) is larger than the radiation emission angle.}
Further production channels for the fermions are present: in particular, Schwinger pair production can occur also without initial seeds in this case.
These considerations ensure that, for $\qd \ELvac>\mps^2$, $\nps$ grows exponentially around $\Hcasc$, quickly reaching the screening regime described in the next section. 

\begin{figure}[ht] \centering
    \includegraphics[width=.7\textwidth]{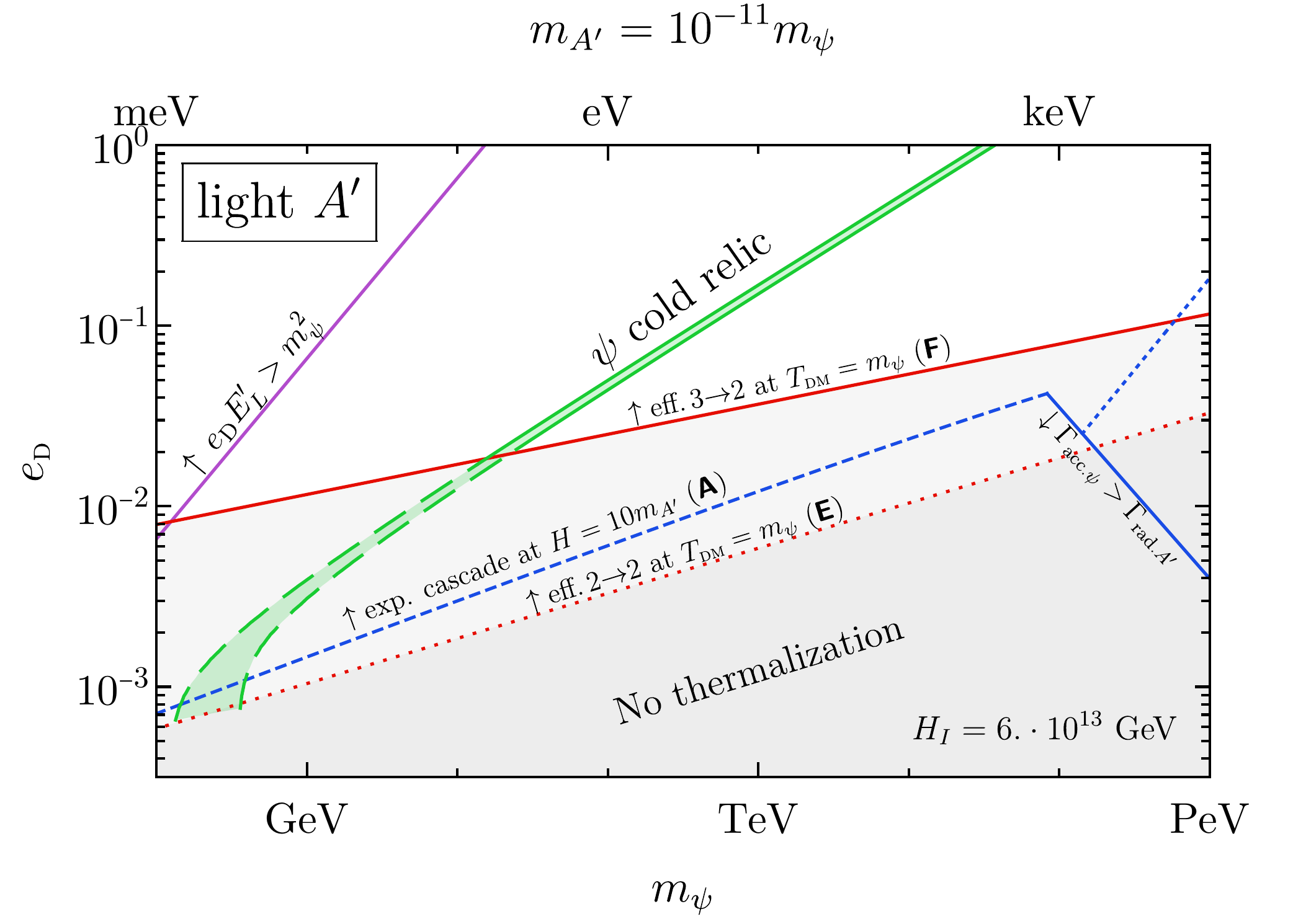}
    \caption{Region where $\psi$ can be efficiently produced in the case of very light $A'$. 
    The blue dashed lines mark where an exponential cascade process can take place before $\astar$ (we choose the time when $H=10\,\mA$), even if the condition $\qd \ELvac >\mps^2$ is not satisfied.
    The dark photon mass $\mA$ can be pushed down to about $10^{-16}\mps$ before the blue line moves above the solid green one, implying that the dark sector cannot reprocess its energy on the relic density line, and that $\rhoA$ would follow the usual prediction of \cite{Graham:2015rva}.}
    \label{fig:veryLightPhotonRelicAbundance}
\end{figure}

The situation changes slightly for $\qd \ELvac < \mps^2$. At $\Hcasc$ the three relevant rates to achieve an exponential cascade are still slower than $H$: (1) $\Gamma_\text{acc. $\psi$}$ to accelerate $\psi$'s to a large enough boost $\gamma_\psi$ so that $A'_T$ radiated around the UV cutoff (i.e.~$\omega_{A'}\sim \gamma_\psi^2 (\qd \ELvac/\mps)$) satisfy $\chi\gtrsim 1$, (2) $\Gamma_\text{rad. $A'$}$ for the emission of a hard enough dark photon from the accelerated $\psi$, (3) $\Gamma_\text{pair}$ for the $\psi\overline \psi$ production rate from an $A'_T$ as given in Eq.~\ref{eq:cascade rate} for $\chi \sim 1$.
The time by which all of these rates are faster than $H$ arrives later than in the region $\qd \ELvac > \mps^2$. 
In particular, we estimate $\Gamma_\text{pair}>\Gamma_\text{rad. $A'$}$, so the exponential cascade has a bottleneck either in the acceleration of $\psi$ to large enough boosts or in radiating the energetic $A'$ from the fermion. 
We skip the description of our estimates for these rates, given that these details can only slightly affect the lower range of $\mA$ and can be ultimately addressed only by a simulation.
The corresponding history of $\nps$ is the following: it saturates temporarily at early times when the last frozen $\psi$ modes that were gravitationally produced during inflation (see e.g.~\cite{Lyth:1996yj,Kuzmin:1998kk,Chung:2011ck}) re-enter the horizon at $H\sim \mps$, then $\nps a^3$ does not vary significantly until $H\sim \min(\Gamma_\text{acc. $\psi$},\Gamma_\text{rad. $A'$})$, when it grows exponentially until it reaches the screening regime. 
We show in Fig.~\ref{fig:veryLightPhotonRelicAbundance} the region of parameter space (for $\mA\sim 10^{-11}\mps$) where this exponential cascade (depicted in diagram \circled{A} in Fig.~\ref{fig:cartoon thermalization} or in Fig.~\ref{fig:qed_cascade}) is achieved before $H=10\,\mA$. 
The blue solid line defines the two regions where either $\Gamma_\text{acc. $\psi$}$ or $\Gamma_\text{rad. $A'$}$ holds the cascade back, and in the region above the blue dashed lines the dark sector reaches the screening regime. 

\subsection{Screening era -- expanded}
\label{app:screening regime}
Once the exponential cascade production of fermions starts, it clearly cannot continue forever, and it eventually gets cut off because it depletes the electric field.
This point can be computed in a few different ways. 
The first estimate is by writing down the charge density that screens a constant DC electric field of strength $E'$ over a physical length $H^{-1}$: this yields a fermions density $\nps \sim \frac{E' H}{\qd}$. 
This simplistic argument actually already captures the broad physics.  
We can refine this picture by considering the time oscillation of the electric field with frequency $H$. Charges are accelerated to a typical energy $\qd E' H^{-1}$, so the fermion bath extracts all available energy from the electric field when their two energy densities are comparable:
\begin{equation}
\label{eq:screening density}
\nps \qd E' H^{-1} \sim E^{\prime 2}
\end{equation}
which again gives $\nps \sim \frac{E' H}{\qd}$. 
Finally, a last argument that we will revisit in App.~\ref{app:E screening plasma} involves the plasma frequency $\omp^2=\frac{\qd^2 \nps}{\omps}$, that by the previous estimates turns out to be $\omp\sim H$ in the screening regime. When $\omp \gtrsim H$, at horizon entry the electric field experienced by the plasma is suppressed, compared to the vacuum value, by $H^2/\omp^2$. Therefore, when the regime of Eq.~\ref{eq:screening density} is reached, the electric field starts to be screened in the plasma.

As the fermion number density approaches this threshold, the dark sector enters the screening regime and the exponential cascade stops.  Each new $A'_L$ mode entering the horizon has an electric field that induces oscillatory motions in the dark fermion plasma at a frequency $\sim H$, making the $\psi$'s radiate during these accelerations.
The moving charges emit predominantly $A'_T$ radiation, which (as already discussed in App.~\ref{app:before Hcasc}) is mostly soft, with $\omega_{A'}\sim H$. 
We can expect that, in the screening regime, an approximate equipartition of energy among the electric field, $\psi$ and $A'_T$ is reached. Correspondingly, we estimate $n_{A'_T}\sim E^{\prime2}/H$ at this stage. 
We do not consider possible energy injections into the $\psi-A'_T$ system from the residual electric field on scales smaller than the Debye length (which is of the order of the Hubble radius in the screening regime), our goal being a conservative estimate of the path towards thermalization in the dark sector.

\subsection{Plasma dominated era -- expanded}
\label{app:first steps thermalization}
Armed with an understanding of the typical energies and number densities for $\psi$ and $A'_T$ in the screening regime, we can estimate which perturbative processes are efficient. This is a necessary step in order to understand how the dark sector moves beyond the screening regime.

The first relevant process for this discussion is shown by process \circled{B} in Fig.~\ref{fig:cartoon thermalization} (Compton scattering), where a $\psi$ scatters off one of the many $A'_T$ soft photons and heats it up. 
The corresponding rate $n_{A'_T} \sigma v \sim \frac{\ELvacSq}{H}\frac{\qd^4}{H (\qd \ELvac /H)}$ becomes faster than Hubble at $H\sim \Hplas \equiv \qd \Hcasc$. 
The final states of this process both have energy $\sim\qd \ELvac/H$. 
The same estimate also shows that the resulting population $\nA^\text{(hot)}\sim \qd^2 E^{\prime 2}_L/H$ of hot photons can re-scatter efficiently off a soft $A'_T$. For the parameter space where $\qd \ELvac>\mps^2$, this process has a large enough center-of-mass energy to produce a $\psi\overline \psi$ pair (process \circled{C}).
Starting from $\Hplas$, the combination \circled{B}+\circled{C} of inverse Compton scattering and pair production
\footnote{On top of the combination of inverse Compton scattering (\circled{B}) followed by pair production (\circled{C}), also other processes contribute to fermion production, as for example the direct pair production (\circled{C}) from a pair of radiated $A'_T$. This process starts to be kinematically disfavored when $H\lesssim \mps$, given that $\omega_{A'_T}$ for radiated $A'_T$ is typically of order $H$ with a red-tilted distribution in $n_{A'_T}(\omega)$.}
increases $\nps$ without relying on further energy injection from the $A'_L$ on top of the work done by the electric field in setting the screening regime. 

The increase of $\nps$ around $\Hplas$ brings in two other ingredients that affect the remainder of this story.
The first one was already mentioned in App.~\ref{app:screening regime}, and will be further investigated in App.~\ref{app:E screening plasma}: as $\omp >H$, the electric field experienced by the plasma is suppressed with respect to what expected without a plasma. 
This reduces the energy density accessible to the $\psi-A'_T$ system from the electric field component of $\rho_{A'_L}$.
The second point is that, given that the process \circled{B} (inverse Compton scattering) is efficient, the mean free path $\nu_{A'_T}^{-1}$ of $\psi$ starts to be $<H^{-1}$ starting from $\Hplas$. 
This reduces the typical energy of the fermions, because the (plasma-suppressed) electric field acts on them for a distance $\nu^{-1}<H^{-1}$.
This second point is further elaborated in App.~\ref{app:towards thermalization}.

\subsection{Towards thermalization -- expanded}
\label{app:towards thermalization}
In order to achieve thermalization in the dark sector, the $\psi-A'_T$ system must access through particle scatterings the bulk of the energy density stored in the mass term of the $A'_L$ field. 
This occurs via Compton scattering $\psi A'_L\to \psi A'_T$ (we recall that the emission of $A'_L$ in the final state is suppressed by $\mA^2/s$), as shown by diagram \circled{D} in Fig.~\ref{fig:cartoon thermalization}.
The cross section for this process is proportional to $s^{-2}\sim H^{-2}\omps^{-2}$. 
As the electric field starts to be screened ($E'< \ELvac$, where we remind the reader that $\ELvac\equiv \mA \HI /(2\pi)$) and the $\psi$ mean free path becomes smaller than the Hubble radius ($\nu>H$), the typical fermion energy $\omps\sim \qd E'/\nu$ decreases from its value in the screening regime, facilitating process \circled{D} (Compton scattering off an $A'_L$).
Its rate compared to Hubble is
\begin{equation}
 \frac{\Gamma_{\psi A'_L \to \psi A'_T}}{H} \sim \frac 1H n_{A'_L} \sigma_{\psi A'_L\to\psi A'_T} v 
   \sim (H \HI^2) \frac{\qd^4 \mA^2}{H^2\omps^2}
   \sim \pare{\qd \frac{\ELvac}{E'} \frac{\nu}{H}}^2
\end{equation}
which becomes $>1$ soon after $\Hplas$. 
This means that each $\psi$ converts at least one $A'_L$ into $A'_T$ in a Hubble time, with the effects of softening the typical $\omps$ (hence making \circled{D} even more efficient) and boosting number-changing processes like \circled{F} ($3\to 2$ scattering), where the large $n_{A'_T}$ is reprocessed in the $\psi-A'_T$ system and moves it closer to being thermal.

The overall picture is that the dark sector energy density is gradually moved from $A'_L$ modes (as they cross the horizon) into the $\psi-A'_T$ bath, without relying on coherent effects through the electric field.
Within this simplified picture, we estimate that the $\psi-A'_T$ system is in thermal equilibrium and has thermalized the whole $\rhoA$ at horizon entry for $\qd > (\mA/\HI)^{1/13}\sim \mathcal O(0.01-0.07)$.

In summary, the history that we have outlined is a conservative estimate of the processes leading to thermalization in the dark sector. 
As discussed in Sec.~\ref{sec:thermalization}, the ultimate requirement is that number-changing processes like \circled{F} ($3\to2$ scattering) are efficient. 

We can compare the sketch of the cosmological history of the dark sector for the case of pure massive $A'$ (Fig.~\ref{fig:massiveVectorEvol} adapted from \cite{Graham:2015rva}) to the analogous scheme for the case of massive dark QED with $\mps\gg \mA$ in Fig.~\ref{fig:plasma_effects}.
Comparing that case to our scenario of massive dark QED, the cosmological history of the dark sector does not change until $\aplas$, when the $\psi-A'_T$ plasma (whose production is discussed in App.~\ref{app:before Hcasc}) starts affecting the evolution of the super-horizon $A'_L$ modes. 
During this screening regime (described in App.~\ref{app:screening regime}) and the first stages of thermalization (App.~\ref{app:first steps thermalization}), two main points change in the evolution of $A'_L$ modes.
They start dissipating their energy into the $\psi-A'_T$ plasma predominantly at horizon crossing (red-shaded band), and do not start oscillating until horizon crossing (see Eq.~\ref{eq:H osc plasma}) even after they have become non-relativistic. 
Therefore $\astar$ does not play a special role in our case.
These plasma effects on $A'_L$ cease when the dark fermions freeze-out at $a_\text{f.o.}$ and $\nps$ drops (see App.~\ref{app:SH_plasma} for more details on this era).
The overall effect of this delayed onset of oscillations for super-horizon $A'_L$ modes is a logarithmic change in $\eta^2$. This parameter, that we defined in Eq.~\ref{eq:TDMa}, counts the number of $e$-folds in $k$ that yield the same energy density $\rhoA(k)\sim\mA^2 \HI^2$ when all super-horizon $A'$ modes start oscillating.
In the pure $A'$ case of \cite{Graham:2015rva}, assuming instantaneous reheating, the epoch between $a_\text{end}$ and $\astar$ is radiation-dominated (hence $k_\text{end}/\kstar=\sqrt{\HI/\mA}$) and from Eq.~\ref{eq:eta} we obtain
\begin{equation}
\label{eq:eta pure A'}
\eta^2=\frac 12 \log\frac{\HI}{\mA} \,.
\end{equation}
In the dark QED case, the range of momenta that display a flat power spectrum is enlarged down to $k_\text{f.o.}$, and $k_\text{end}/k_\text{f.o.} \sim \HI x_f /\mps$ (see also Fig.~\ref{fig:plasma_effects}). The parameter $\eta^2$ changes accordingly into 
\begin{equation}
\label{eq:eta dark QED}
\eta^2=\log\frac{\HI x_f}{\mps}\,.
\end{equation}
The typical value of $\eta$ within the parameter space where we obtain the relic abundance (for maximum $\HI$ and instantaneous reheating) lies in a small range $\sim 5-6$, due to the logarithmic dependence.
We choose an illustrative range $\eta\sim 3-6$ in the relic abundance lines in Figs.~\ref{fig:lightPhotonRelicAbundance} and \ref{fig:heavyPhotonRelicAbundance}, in order to account for the possible variation due to smaller $\HI$ or lower $H_\text{end}$.

\subsection{The evolution of the superhorizon vector modes -- before freeze-out}
\label{app:E screening plasma}
The presence of a $\psi\overline\psi$ plasma affects the evolution of $A'_L$ modes in FRW in an appreciable way as soon as $\omp \gtrsim H$. 
The physics of these systems has already been illustrated in general terms in App.~\ref{app:ProcaPlasmas}.
In this section, we build on the results of App.~\ref{app:ProcaPlasmasGR} and numerically solve the equations governing the evolution of $A'_L$ and plasma for a given cosmological evolution of $\omp$ and the fermion collision rate $\nu$. 
The results are shown in Fig.~\ref{fig:plasma_A,J,E}.
\begin{figure}[t] \centering
    \includegraphics[width=.9\textwidth]{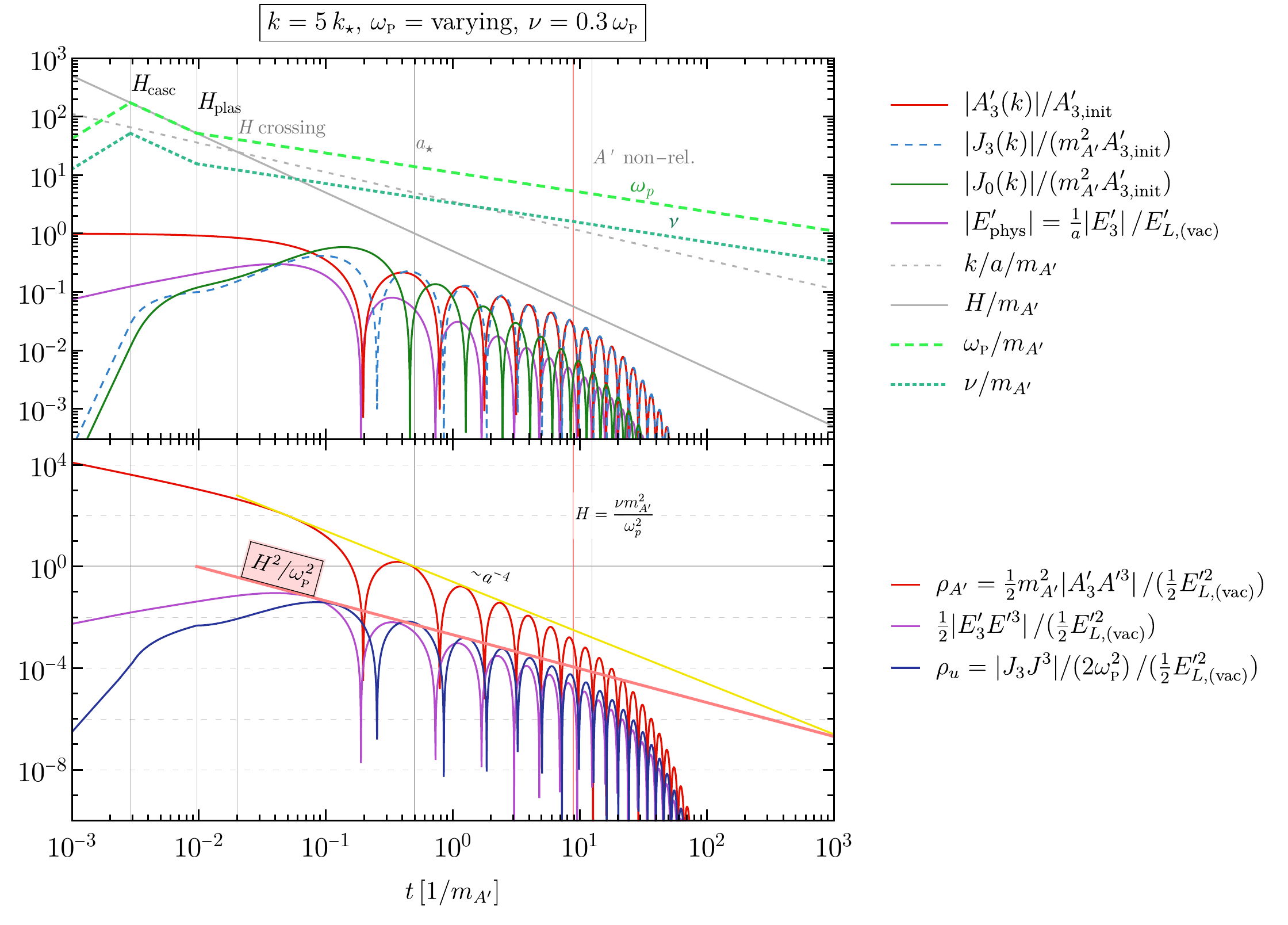}
    \caption{Evolution of the longitudinal field $A'_L$ (denoted by $A'_3$), the vector current $J_3$ and the current density $J_0$, as governed by the Maxwell-Proca equation \ref{eq:A3eomGR}, the continuity equation \ref{eq:J0eomGR} and the Euler equation \ref{eq:J3eomGR}.
    \textbf{Upper panel}: Solution for $A_3,\,J_3,\,J_0$ and electric field $E_3$ with plasma mass $\op$ given in input as shown in the figure, and collision rate $\nu$ set to $0.3\,\op$. 
    The results are normalised to either the initial value $A'_\text{3,init}$ or the electric field at horizon entry in vacuum $\ELvac\equiv\mA\HI/(2\pi)$.
    We choose the mode $k=5\kstar$, but very similar results hold for any $k$ entering the horizon between $\Hplas$ and the decay time of the modes, shown with a vertical red line.
    \textbf{Lower panel:} Corresponding energy densities, as defined in Table~\ref{tab:longWavelengthModes}.
    The most relevant point for our discussion is that the kinetic energy density of the plasma is capped by $\frac{H^2}{\op^2} \ELvacSq$, as shown by the pink line.}
    \label{fig:plasma_A,J,E}
\end{figure}

The three equations that determine the evolution of longitudinal mode $A'_3$, the current density $J_0$ and the vector current $J_3$ are the Maxwell-Proca equation \ref{eq:A3eomGR}, the continuity equation \ref{eq:J0eomGR} and the Euler equation \ref{eq:J3eomGR}. 
We solve for this system of equations with the initial conditions discussed in App.~\ref{app:ProcaPlasmasGR}, and by plugging a fixed time evolution for $\omp$ and $\nu$ as shown in the upper panel of Fig.~\ref{fig:plasma_A,J,E}. The choice of $\omp$ is motivated by the cosmological history discussed until now: it rises up to $\omp \sim H$ at the onset of screening regime $\Hcasc$, and it starts deviating from it at $\Hplas$. From that moment on, we scale it as $a^{-2/3}$: this choice is motivated by an energy scaling $H^2/\omp^2$ that we discuss at the end of this section, but it is important to stress that it does not change the story as long as $\omp$ decreases more slowly than $H$. Finally, we fix a value $\nu/\omp<1$, as a reminder that in a thermalized plasma $\omp\sim\qd T$, $\nu\sim\qd^4 T \sim \qd^3\omp$, but again this choice is not crucial.

The results of the numerical solution of the system is shown in the upper panel of Fig.~\ref{fig:plasma_A,J,E} for a mode $k>\kstar$ entering the horizon after $\Hplas$ and before $\astar$.  For modes $k<\kstar$, the qualitative evolution is mostly the same, with the onset of oscillations around zero delayed until when $k\sim aH$ (see Eq.~\ref{eq:H osc plasma}) even if the mode would be non-relativistic and would have already started oscillating without the presence of the plasma.
As $\omp$ reaches $H$, the currents $J_0$, $J_3$ are excited. At horizon crossing, $A'_L$ starts oscillating around zero%
\footnote{\label{smalloposc}Strictly speaking, $A_3$ and $J_3$ start displaying very small oscillations of frequency $\omp$ around a constant value, as soon as $\omp>H$ (see Sec.~\ref{app:SH_plasma} for details). For the purpose of our discussion, these can be neglected, and when talking about oscillations of $A_3$ and $J_3$ we refer to the large oscillations around zero.} 
together with $J_0,\,J_3$.
The electric field $\tfrac 1a E'_3$ detected by a comoving observer (where $E'_3$ is given by the term in brackets in Eq.~\ref{eq:u3eomGR}) is shown with a purple line, and grows as $a$ until horizon crossing.

The lower panel of Fig.~\ref{fig:plasma_A,J,E} shows the energy densities of the various components of the system, normalized to $\tfrac 12 \ELvacSq$. 
The main take-home messages from this exercise are as follows:
First, as anticipated in App.~\ref{app:ProcaPlasmasGR}, the initial conditions for the vector field and the fermion current select a dispersion relation that resembles very closely the slow mode depicted in Fig.~\ref{fig:cold_dispersions} for flat space. This mode is gapless and resembles an acoustic mode, with a dispersion relation close to $\omega=k$ for large $\omp$. Second, 
the energy associated to the plasma kinetic motion ($\rho_u$, in blue) is of order $\frac{H^2}{\omp^2} \ELvacSq$, as shown by the pink straight line, with a suppression of $(H/\omp)^2$ compared to when the plasma mass is negligible. 
This quantifies the suppression of the kinetic energy of the $\psi-A'_T$ plasma, starting from $\Hplas$.

\subsection{The evolution of the superhorizon vector modes -- after freeze-out}
\label{app:SH_plasma}

Up until the time of fermion freeze-out, we have seen that all $A'$ modes entering during the era of the thermalized subhorizon dark plasma dissipate their energy into the thermal bath, contributing logarithmically to its energy budget. After fermion freeze-out, dissipation drops to zero as collisions are no longer efficient. However, the number density in fermions can still be high enough that it modifies the evolution of modes entering after freeze-out. In this section we quantify this statement and prove that the energy density of these modes is parametrically smaller than that of the fermions. 

The equations of motion for $A_3'$, $J_3$, and $J_0$, given in Eqs.~\ref{eq:A3eomGR}, \ref{eq:J0eomGR} and \ref{eq:J3eomGR}, can be expanded to lowest order in $k^2/(a^2\mA^2)$ for nonrelativistic superhorizon modes, yielding
\begin{gather}
\label{eq:A3newksmall2}
\partial_t^2 A'_3+H\pare{1+\frac{2k^2}{\mA^2a^2}}\partial_t A'_3+\pare{\frac{k^2}{a^2}+\mA^2}A'_3=J_3-2H\frac{ik}{\mA^2}J_0\\
\label{eq:J3newksmall2}
\frac{\di J_3}{\di t}=-\op^2\pare{1-\frac{k^2}{\mA^2a^2}}\partial_t A'_3+\op^2\frac{ik}{\mA^2}J_0+\frac{\dot{n}_\psi}{n_\psi}J_3\\
\label{eq:J0newksmall2}
\partial_t J_0+3HJ_0-\frac{ik}{a^2}J_3=0.
\end{gather}
where we have set $\nu=0$. As discussed in App.~\ref{app:ProcaPlasmasGR}, the initial conditions to these equations set by inflation are $A'_{3,\text{init}}=\frac{k\HI}{2\pi\mA}$ and $\partial_tA'_{3,\text{init}}=-\frac{k^2}{3a^2H}\frac{k\HI}{2\pi\mA}$, which means that initially the mass term $\mA^2 A'_3$ dominates. When the charges appear early on, the $A'$ field sets up superhorizon currents of the mode $k$, and the large coupling between the plasma and $A'$ enforces the condition $\mA^2 A_3'=J_3$. 

To get some analytic intuition of this statement, consider the $k=0$ limit of Eqs.~\ref{eq:A3newksmall2}, \ref{eq:J3newksmall2} \& \ref{eq:J0newksmall2} and a constant $\op$. Then the Lorentz force law Eq.~\ref{eq:J3newksmall2} can be solved exactly to give $J_3(t)=-\op^2\pare{A_3'(t)-A_{3,\text{init}}}$. Plugging this into Eq.~\ref{eq:A3newksmall2} and solving with the aforementioned initial conditions gives
\begin{equation}
    \label{eq:A3_is_J3}
    A'_3(t)=\frac{\op^2}{\mA^2+\op^2}A'_{3,\text{init}}+\mathcal{A} t^{1/4}J_{1/4}\pare{t\sqrt{\op^2+\mA^2}}+ \mathcal{B} t^{1/4}Y_{1/4}\pare{t\sqrt{\op^2+\mA^2}},
\end{equation}
where $\mathcal{A}$ and $\mathcal{B}$ are constants that can be determined analytically, whose key feature is that they are suppressed by $\mA^2/\op^2$ compared to the constant first term. Looking back at $J_3(t)$, this gives $J_3(t)\simeq\mA^2 A_3'(t)$ to zeroth order in $\mA^2/\op^2$. This expression demonstrates that inflationary initial conditions project onto what we would call the ``slow mode'' in the flat spacetime limit (see App.~\ref{app:ProcaPlasmasMinkowski}), which is this configuration of fields and charges unique to Proca plasmas. The Bessel functions are manifestly the projection onto the ``fast mode,'' as they oscillate in $\op$ (these are the small oscillations we alluded to in footnote \ref{smalloposc}). This solution is valid until $k/a\sim H$, at which point we should keep the higher order terms in the momentum $k$. Numerically we find that a time-dependent $\op$ leads to a qualitatively similar behavior for the system. The crucial condition, as we will also see below in more detail, is $\op\gg\mA$, and is largely independent of the details of its time dependence.

We proceed to studying Eqs.~\ref{eq:A3newksmall2}, \ref{eq:J3newksmall2} \& \ref{eq:J0newksmall2} to first order in $k^2/(a^2\mA^2)$, with the additional assumption $J_3(t)=\mA^2 A'_3(t)$ which we motivate from the discussion above and which we also see holding numerically. In this case, the Lorentz force law Eq.~\ref{eq:J3newksmall2} becomes
\begin{equation}
\label{eq:Lorentz_tight}
  \frac{\di J_3}{\di t}=\frac{ik\op^2}{\op^2+\mA^2}J_0+\frac{\mA^2}{\mA^2+\op^2}\frac{\partial_tn_\psi}{n_\psi}J_3,
\end{equation}
where we have approximated $\op^2\pare{1-k^2/(a^2\mA^2)}\simeq\op^2$. The two terms dominate in the different regimes $\op\gg\mA$ and $\op\ll\mA$ respectively, which means that they can be treated separately. 

Notice that $\op$ cannot fall below $\mA$ before freeze-out: the plasma would be hot and $\op=\mA$ would happen at $\TDM=\mA/\qd<\mps$, which is contradictory. Therefore, $\op$ crosses $\mA$ either during or after freeze-out. Using the energy density of fermions after freeze-out from Eq.~\ref{eq:psi_freeze-out}, we calculate the time $H_\text{np}$ that $\op$ crosses $\mA$ after freeze-out to be
\begin{equation}
\label{eq:late_np}
    \frac{H_\text{np}}{H_\text{f.o.}}=\frac{3}{(2\pi)^{7/3}}\pare{\frac{33}{5}}^{2/3}\frac{\xfo^{4/3}\qd^2}{\Gstarinf^{1/3}}\pare{\frac{\mA}{\mps}}^{4/3}\pare{\frac{H_I\eta}{\mps}}^{2/3},
\end{equation}
where $H_\text{f.o.}$ is the Hubble rate at the end of freeze-out. First let us discuss the regime where this ratio is smaller than unity.

In this regime and during the period $\op\gg\mA$, we can drop the $J_3$ term from Eq.~\ref{eq:Lorentz_tight}, and plug the solution for $J_0$ into the continuity Eq.~\ref{eq:J0newksmall2} to get
\begin{equation}
\label{eq:J3_osc}
\partial^2_t J_3+3H\partial_tJ_3+\frac{k^2\op^2}{a^2(m^2+\op^2)}J_3=0,
\end{equation}
where we have dropped corrections of $\mathcal{O}(k^2/(a^2\mA^2))$ and $\mathcal{O}(\mA^2/\op^2)$ in the Hubble term, but kept them in the frequency term to highlight that the oscillation frequency of $J_3$ is manifestly that of the ``slow mode'' derived in App.~\ref{app:ProcaPlasmasMinkowski}, with $k/a\equiv k_\text{phys}$, the physical momentum. In fact, it was not necessary to assume that $\op$ was constant in deriving this result, in this limit. Solving this equation with initial conditions set by the relation $J_3(t)=\mA^2 A'_3(t)$, and approximating the frequency as $\sim k/a$, yields
\begin{equation}
\label{eq:analytic_smallk}
J_3(t)=\mA^2 A'_{3,\text{init}}\parea{\frac{1}{a}\cos\pare{\frac{k}{aH}\parea{1-\frac{1}{a}}}-\pare{\frac{k}{aH}}^{-1}\sin\pare{\frac{k}{aH}\parea{1-\frac{1}{a}}}},
\end{equation}
where we set $a(t_\text{init})=1$ for clarity. This solution tracks the current as the mode enters the horizon, showing that it remains frozen until re-entry and then oscillates with the ``slow-mode'' frequency, while redshifting as $a^{-1}$. The $A'_3$ mode follows the same behavior, by the condition $J_3(t)=\mA^2 A'_3(t)$. Notice that this is self-consistent, in the sense that Eq.~\ref{eq:A3newksmall2} seen with the solution Eq.~\ref{eq:analytic_smallk} as a source is just a driven oscillator with driving $J_3(t)$. A ``de-tuned'' driven oscillator has simply its amplitude and frequency set by the driver. Therefore, $A'_3$ remains frozen until it re-enters the horizon and subsequently its energy density redshifts as radiation. These results are in excellent agreement with numerical simulations of the full system. Naturally, there are corrections of $\mathcal{O}(\mA^2/\op^2)$, which include a part that oscillates with frequency $\sim\op$.

After $\op=\mA$, the current essentially decouples from the $A'_3$ modes and redshifts simply as a free current, $\partial_t J_3\simeq (\partial_t n_\psi/n_\psi)J_3$ (we remind the reader that the covariant velocity $u_3$ does not redshift). The solution to this equation is $J_3=J_3(t_\text{np})n_\psi/n_\psi(t_\text{np})\propto t^{-3/2}$, where $t_\text{np}$ is the time when $\op\simeq\mA$. Note that the initial condition is $J_3(t_\text{np})=\mA^2A_3'(t_\text{np})$ as set from the evolution until this point.

This is then the source term in Eq.~\ref{eq:A3newksmall2}. Dropping all higher order terms, i.e.\ the $k^2/(a^2\mA^2)$ corrections to the Hubble term and the frequency term, as well as the $J_0$ term that redshifts very fast, and redefining $\tau\equiv\mA' t$ and $\tilde{A}_3\equiv A'_3/A'_3(t_\text{np})$, the field $\tilde{A}_3$ obeys
\begin{equation}
    \label{eq:A3_tau}
    \ddot{\tilde{A}}_3+\frac{1}{2\tau}\dot{ \tilde{A}}_3+\tilde{A}_3=\pare{\frac{\tau_\text{np}}{\tau}}^{3/2}, \quad \tau>\tau_\text{np},
\end{equation}
where dots denote derivatives with respect to $\tau$. Since $H_\text{np}\ll \mA$, $\tau\gg 1$, so that this equation has the simple solution
\begin{equation}
    \label{eq:A3sol_np}
    \tilde{A}_3(\tau)=\pare{\frac{\tau_\text{in}}{\tau}}^{3/2}+\frac{3\sin(\tau-\tau_\text{in})}{2\tau^{1/4}\tau_\text{in}^{3/4}}-\frac{9\cos(\tau-\tau_\text{in})}{64\tau^{5/4}\tau_\text{in}^{3/4}}.
\end{equation}
From this we see that $A'_3\propto a^{-3}$, until the $\tau^{-3/2}$ term has diluted enough that the dominant contribution comes from the $\tau^{-1/4}$ term. Note that once $\op=\mA$, all modes will follow this behavior, both superhorizon and those whose energy was redshifting as $a^{-4}$ that we discussed right before. This is saying that $A'_3$ drops the energy stored in the solution $J_3=\mA^2 A_3'$ very fast, and then reverts to a free field, with a significantly reduced amplitude. That this is the expected qualitative behavior can be seen directly from Eq.~\ref{eq:A3_tau}: the source $(\tau_\text{init}/\tau)^{3/2}$ coming from $J_3$ redshifts faster than the Hubble dilution, so eventually it will become subdominant and $A_3'$ will evolve as a free field thereafter. This should be contrasted to the solution in the $\op\gg\mA$ era, Eq.~\ref{eq:analytic_smallk}, when seen as a source for the equation of the $A'$ field. This dilutes \emph{slower} than Hubble, so the free field regime can never be reached. Thus, $\rhoA$ will redshift as $a^{-8}$ once $\op$ crosses $\mA$ and it will redshift as matter after a time
\begin{equation}
    \tau >\pare{\frac{2}{3}}^{4/5}\tau_\text{in}^{9/5}, \quad \text{or} \quad
    H <\frac{2^{11/5}}{3^{4/5}}\mA\pare{\frac{H_\text{np}}{\mA}}^{9/5}\equiv H_\text{matter}
\end{equation}
Now that we know how the modes that are still superhorizon when freeze-out happens behave until late times, we can compute the energy density stored in them and compare it to that of the fermions. Because of the $a^{-4}$ redshift of the energy of these modes, after they enter the horizon, all modes that enter between freeze-out and $\op=\mA$ contribute the same to the subhorizon energy density, leading to a logarithmic enhancement factor $\eta'^2\equiv\frac{1}{2}\log\pare{H_\text{f.o.}/H_\text{np}}$, similar to the $\eta$ of the earlier dark plasma. This stems from the fact that the longer the wavelength the earlier the mode exited during inflation and their energy redshifted as $a^{-2}$, whereas this relative enhancement of the shorter wavelengths cancels exactly from their earlier re-entry and subsequent $a^{-4}$ redshift of their energy during radiation domination. In addition, because the energy in all modes irrespective of $k$ starts redshifting as $a^{-8}$ at $H_\text{np}$, every $k<k_\text{np}$, where $k_\text{np}$ is the mode entering at $H_\text{np}$, will contribute less to the overall energy budge.

Since $\rho_\psi$ redshifts as matter after freeze-out, the requirement is that the contribution from modes $k_\text{np}\lesssim k \lesssim k_\text{f.o.}$ is subdominant to that of the fermions at least at $H_\text{matter}$. Accounting for the $a^{-4}$ redshifting of the energy of modes after horizon re-entry and for the $a^{-8}$ scaling between $H_\text{np}$ and $H_\text{matter}$, the condition becomes
\begin{equation}
\label{eq:final_req}
\left. \frac{\rho_{k<k_\text{f.o.}}}{\rho_\psi}\right|_{H=H_\text{matter}} 
 = \frac{(H_I\eta')^2H_\text{np}^{1/2}H_\text{f.o.}^{3/2}}{4\pi^2\rho_\psi^{\text{f.o.}}} \pare{\frac{H_\text{matter}}{H_\text{np}}}^{5/2}
 \sim \qd^9 \pare{\frac{\mA}{\mps}}^{4/3}\pare{\frac{H_I\eta}{\mps}}^{2/3}.
\end{equation}
In the second line we kept only the parametric dependence of this ratio on the energy scales and the dark electric charge. Notice that the scaling with these parameters is the same as that of Eq.~\ref{eq:late_np} (with additional powers of the dark charge $\qd<1$), which means that the assumption $H_\text{np}<H_\text{f.o.}$ leads necessarily to the conclusion that $\rho_\psi$ will not be overcome by dark photon modes. Including all prefactors and the logarithm $\eta'$ does not change this parametric argument.

In this analysis we have assumed that superhorizon currents remain unchanged during freeze-out, because the $\dot{n}_\psi/n_\psi$ term comes with a suppression $\mA^2/\op^2$ in the equation of motion for the current, Eq.~\ref{eq:Lorentz_tight}. Since cold relic freeze-out happens when the annihilation rate becomes similar to $H$, this period should not alter appreciably the evolution studied above. Numerically we find that this rate can be at most $\mathcal{O}(5)$ larger than the usual number density dilution term $\propto 3H$. However, note that even if the $\dot{n}_\psi/n_\psi$ term was appreciable compared to the $J_0$ one, it would only lead to a further decrease in the energy density of the $A_3'$ modes, which would strengthen the conclusions of this analysis.

Finally, let us address a possible drop of $\op$ below $\mA$ during the freeze-out process, which might happen for heavy enough dark photons (see Eq.~\ref{eq:late_np}). In that case, the energy in all modes will start redshifting as $a^{-8}$ right after freeze-out, until $H_\text{matter}$, given now by
\begin{equation}
    \frac{H_\text{matter}}{H_\text{f.o.}}=\pare{\frac{H_\text{f.o.}}{\mA}}^{4/5}.
\end{equation}
 The mode that contributes the most to the energy density is the one entering right at freeze-out, since all other $k<k_\text{f.o.}$ had redshifted for longer during inflation. The requirement $\rho_{k_\text{f.o.}}/\rho_\psi<1$ is parametrically the condition $\qd^2(\mps/\mA)^2(\mps/H_I)>1$, which has the same scaling as the requirement that $\op=\mA$ during freeze-out, Eq.~\ref{eq:late_np}, but is additionally suppressed by $\qd^{7/2}$. Therefore, if $\op=\mA$ happens during freeze-out, the energy in superhorizon modes is again always parametrically subdominant to that of the fermions.

To conclude, even though plasma effects dramatically alter the evolution of superhorizon modes long after fermion freeze-out, the energy density in those modes is eventually always parametrically subdominant and cannot overclose the universe.

\subsection{Case of heavy \texorpdfstring{$A'$}{A'}: differences with respect to the case of light \texorpdfstring{$A'$}{A'}}
\label{app:thermalization heavy A}
As summarized in Sec.~\ref{sec:heavyPhotonThermalization}, in the case of heavy $A'$ ($\mA \gg 2\mps$) the onset of thermalization of the dark sector at early times does not change significantly with respect to the light $A'$ case that we have discussed until now. The only change is that the fermion mass is the smaller scale in the problem and thus does not constitute a kinematical threshold. 

We first discuss the case of large $\qd>\mathcal O(10^{-2})$, for which the dark sector thermalizes and we can easily compute the DM relic abundance.
The dark electric field is always above the critical value for this range of couplings, and the kinematical restrictions that we have discussed in App.~\ref{app:first steps thermalization} are not an issue.
The only relevant difference concerns the final step of the cosmological evolution of the dark sector.  For heavy $A'$, all the dark photons decay to fermions after a time $\GamA^{-1}$ in their rest frame, so the last step is not $\psi$ freeze-out as for light $A'$, but $A'$ decay. 
Accordingly, we evaluate the $3\to2$ and $2\to2$ rates (that ensure thermalization and define in Fig.~\ref{fig:heavyPhotonRelicAbundance} the region of validity of the relic prediction) at the decay time.
If the photons are relativistic at decay (with a boost factor $\gamA$), then the decay time in the cosmological frame is dilated by $\gamA$. For the region of large $\qd$ that we are considering, the photons have interacted enough that their state is much closer to the thermal one than the initial cold state set up after inflation. For this reason, it is a good approximation to estimate these rates by assuming the thermal number density and $\gamA\sim \TDM/\mA$, to check whether the dark sector is in thermal equilibrium at the last possible stage.
The red lines that we show in Fig.~\ref{fig:heavyPhotonRelicAbundance} are then evaluated at
\begin{equation}
H=\frac{\GamA}{\gamA}=\pare{\frac{2\pi^2 \Gstarinf}{15}}^{1/6} \frac{\GamA^{2/3} \mA^{2/3}}{(\HI\eta)^{1/3}},
\end{equation}
leading to 
\begin{subequations}
\label{eq:heavyAthermalizationRates}
\begin{gather}
\label{eq:heavyA-3to2}
\qd > 3.5 \pare{\frac{\mA}{\HI}\frac{6}{\eta}}^{1/8} \quad \text{(eff. $3\to2$ at $H=\GamA/\gamA$)},
\\
\label{eq:heavyA-2to2}
\qd > 1.8 \pare{\frac{\mA}{\HI}\frac{6}{\eta}}^{1/5} \quad \text{(eff. $2\to2$ at $H=\GamA/\gamA$)}.
\end{gather}
For comparison, we show in Fig.~\ref{fig:heavyPhotonRelicAbundance} with a violet line also the corresponding lines evaluated at $H=\GamA$, conservatively ignoring the aforementioned effect of time dilation:
\begin{gather}
\qd > 3.1 \pare{\frac{\mA}{\HI}\frac{6}{\eta}}^{1/10} \quad \text{(eff. $3\to2$ at $H=\GamA$)},
\\
\qd > 1.5 \pare{\frac{\mA}{\HI}\frac{6}{\eta}}^{1/6} \quad \text{(eff. $2\to2$ at $H=\GamA$)}.
\end{gather}
\end{subequations}

For values of $\qd$ below the value in Eq.~\ref{eq:heavyA-2to2}, and potentially already below the value of \ref{eq:heavyA-3to2}, the dark sector does not achieve full thermalization and we cannot trust the estimate provided in Eq.~\ref{eq:heavyARelicAbundance}. Numerical simulations would be able to provide the value of $\mps$ yielding the correct DM abundance for a given value of $\qd$.
We can bracket the range for the correct $\mps$ as follows. 
The highest possible value is given in Eq.~\ref{eq:heavyARelicAbundance}, corresponding to the maximum conversion of the initial $\rhoA$ into kinetic energy, and therefore lowest achievable $\nA$ before decay.
The lowest possible $\mps$ is achieved when $\qd$ is so tiny that it does not allow the large $\nA$ generated after inflation to get reprocessed before decay. In that case, the final DM abundance is 
\begin{equation}
\label{eq:heavyA-lower-mDM}
\Omega_{\psi,0}^{\text{heavy $A'$, no inter.}} \sim    0.26 \left( \frac{\mps}{250 \eV} \right) \left( \frac{10\TeV}{\mA} \right)^{1/2} \left( \frac{\HI}{6 \cdot 10^{13} \GeV} \frac{\eta}{6}\right)^2
\end{equation}
This regime would allow to reach very low masses for fermion DM. Phenomenologically though, it cannot be realized when we take into account the following two requirements.
The dark photons need to decay early enough so that the fermions have enough time to redshift and become non-relativistic before what required by Large-Scale-Structures constraints on warm DM. This implies a lower value for $\qd$ in the ballpark $10^{-6}-10^{-10}$, depending on $\mps$.
On the other hand, in order to prevent a significant impact of SFQED processes on the dark sector, we must require that $\chi<1$ (see Eq.~\ref{eq:SFQED_condition}) when $H<\Hcasc$. Accounting for the redshift of the dark electric field after $H=\mA$ in absence of a plasma, this gives the condition $\qd < 2\pi \mps^{5/4}/(\mA^{1/4} \HI)$ which points to much smaller values $\qd<10^{-15}-10^{-20}$.

In summary, below the thermalization line shown in solid red in Fig.~\ref{eq:heavyARelicAbundance}, it is not possible to provide an analytical estimate of the value of $\mps$ leading to the correct DM abundance. We can only provide a lower estimate for it, as given in Eq.~\ref{eq:heavyA-lower-mDM}. 
This is an interesting regime in that it provides a minimal cosmological production mechanism for fermion DM well below the GeV scale, but it is not appealing in terms of direct detection because of the larger yield obtained via freeze-in in the experimentally accessible parameter space (see Sec.~\ref{sec:dark_fermion_heavy}).

\section{Freeze-in Abundances}
\label{app:freeze-in}

In this appendix we calculate the freeze-in abundances of dark photons and dark fermions when there is a kinetic mixing of the dark $U(1)$ with the SM photon.

\paragraph{Dark Photon Freeze-in}

Here we briefly state the results of Refs.~\cite{0811.0326, 1407.0993}, which are important for understanding how the constraints from energy injections during BBN/decoupling and photon injections into the IDPB change for our model. We discuss these modifications in App.~\ref{app:darkPhotonAppendix}.

\uline{Region $\mA<1$ MeV}\smallskip\\
For masses below $1$ MeV, dark photons are produced by the resonant conversion of plasmons. The yield $Y\equiv n/s$ (where $n$ is the species number density, and $s$ the entropy density of the thermal bath it interacts with) at the resonance was calculated in Ref.~\cite{0811.0326} and is approximately
\begin{equation}
\label{eq:yield_resonance}
      Y_2\approx 4.1\times 10^{18} \epsilon^2 \pare{\frac{\text{MeV}}{\mA}},
\end{equation}
where we have assumed that at the resonant temperature the electrons are still relativistic, so that $T_r= 3 \mA/(2\alpha\pi)\simeq 8 \mA$. This is a good assumption near the peak of the `Thermal' bound at $\sim 1\text{ MeV}$ that we are considering here. In addition, in this region we are away from particle thresholds, so we have also set $g_\star\simeq 10$. 

Using entropy conservation, the relic energy density today is
\begin{equation}
\label{eq:dpO_today}
\Omega_2 h^2\simeq 2.82\times 10^5\frac{\mA}{\text{MeV}}Y_2.
\end{equation}

\uline{Region $\mA>1$ MeV}\smallskip\\
For masses above $\simeq 2m_e$, the dominant production channel is lepton pair coalescence \cite{0811.0326}. The rate for this process in vacuum is
\begin{equation}
\label{eq:decay_to_leptons}
 \Gamma_{\ell\bar{\ell}\to A'}=\frac{\epsilon^2\alpha}{3}\mA\pare{1+2\frac{m_\ell^2}{\mA^2}}\sqrt{1-4\frac{m_\ell^2}{\mA^2}}.
\end{equation}
Thermal effects (and in particular resonant production) can be ignored for our estimates because their contribution compared to pair-coalescence is parametrically suppressed by $\mathcal{O}(\alpha^{1/2})$  \cite{1407.0993}. By including pair-coalescence of only electrons and positrons obeying a Maxwell-Boltzmann distribution and away from particle thresholds the yield at $T=\mA$ is, to a good approximation \cite{1407.0993}:
\begin{equation}
\label{eq:e_an_yield}
    Y_{\text{pair}}=\frac{9}{4\pi}\frac{\mA^3 \Gamma_{A'\to e^+ e^-}}{\left.(H s)\right|_{T=\mA}}
\end{equation}
As noted in Ref.~\cite{1407.0993}, this yield roughly agrees with that of Eq.~\ref{eq:yield_resonance}, which applies for $\mA\lesssim 1$ MeV, barring the $\mathcal{O}(1)$ kinematic factors of Eq.~\ref{eq:decay_to_leptons}. This is why the freeze-in line in Fig.~\ref{fig:dark photon constraints} is straight across $\mA\simeq1$ MeV, even though the physical process contributing to the dark photon abundance in each side are different.

\paragraph{Dark Fermion Freeze-in}
In this section we estimate the slow-leakage of energy to the DS via freeze-in. This constitutes an upper bound on $\epsilon \qd$ since our mechanism assumes no significant energy exchange between the SM and the DS. We do not attempt a detailed calculation here, but rather do a simple estimate to highlight the relevant energy scales in the region of parameter space we are interested in. We refer the reader to Ref.~\cite{1911.03389,1112.0493,1902.08623} for more thorough calculations. 

\underline{Light dark photon.} Dark fermions can be produced through pair annihilation of SM charged particles and from plasmon or $Z$-boson decays to a fermion-antifermion pair. We estimate the cross section of electron-positron pair annihilations for $\TSM \gg m_e$ to be
\begin{equation}
    \avg{\sigma v}_{e^+e^-\to \bar{\psi}\psi} \approx\frac{1}{8\pi}\epsilon^2 e^2 \qd^2\frac{1}{\TSM^2}=\frac{2\pi\alpha\ald\epsilon^2}{\TSM^2}
\end{equation}

Since each electron carries energy $\sim T_\text{SM}$ while relativistic, we estimate the total energy injected while this annihilation channel is open to be
\begin{equation}
\label{eq:totalinj}
    \rho_\text{Fr.in} \sim\int_{T_\text{f}}^{T_\text{in}}n_e^2\avg{\sigma v}\TSM\frac{\mpl}{\TSM^3}\parea{\frac{a(\TSM)}{a(T_\text{f})}}^4\di \TSM 
    \simeq \alpha\ald\epsilon^2 \mpl T_\text{f}^4\int_{T_\text{in}}^{T_\text{f}}\pare{-\frac{1}{\TSM^2}}\di \TSM,
\end{equation}
where $T_\text{f}= \mps$ is the temperature when the channel shuts off and $T_\text{in}$ is some early time when the process starts. The integral is dominated by late times and the resulting energy density redshifts as matter thereafter for $\mps>m_e$. To get the freeze-in abundance it is enough to ask that this energy density is equal to $\rho_\text{SM}\sim T_\text{mre}^4$ at matter-radiation equality. This gives a bound
    $\epsilon \qd\lesssim(T_\text{mre}/\mpl)^{1/2}/e\, .$
We see that $m_\psi$ drops out of this estimate as anticipated, since the dark fermions are produced cold at $T=\mps$ and their abundance is fixed to be the dark matter.

Including prefactors and the effective degrees of freedom we find
\begin{equation}
\begin{split}
\label{eq:ldp_freezein_bound}
    \epsilon \qd \lesssim \frac{4\pi^{17/4}\gstar(\mps)^{1/4}\gstarS(\mps)^{1/2}}{9\times 5^{3/4}\zeta(3)e}\parea{\frac{\gstar(\text{mre})}{\gstarS(\text{mre})}}^{1/2}\pare{\frac{T_\text{mre}}{\mpl}}^{1/2}\sim 8 \times10^{-12}
    \end{split}
\end{equation}
where ``mre'' denotes the value of the quantity at matter-radiation equality and we have taken $\mps=1$ GeV for the estimate. This bound corresponds to the line above our parameter space, as plotted in Figs.~\ref{fig:Reach-DD_light-A'} and \ref{fig:Fermion-bounds_light-A'}.

We have checked that the contribution to the freeze-in abundance from plasmon decays is a subdominant contribution for the parameter space we are interested in. Here we have not discussed possible production from $Z$ exchange \cite{1112.0493}, but we do account for it in the freeze-in line we show.

In Fig.~\ref{fig:Fermion-bounds_light-A'} we show a larger part of the dark fermion parameter space, far below the masses our mechanism favors. Below an MeV, freeze-in production of fermions is dominated by plasmon decays and not by the annihilation process considered here. We do not estimate this part of the line, but rather use the results of Ref.~\cite{1902.08623}.

It is worth mentioning that whenever both a dark photon and a dark fermion are present in the theory, in the regime $\mA<2\mps$ and $\mA<2m_e$, the dark matter relic could be the dark photon \cite{1911.03389}. This statement depends on $\qd$ and the subsequent evolution of the dark sector. According to Refs.~\cite{0811.0326,1911.03389}, for $\mA<100$ keV, there is no significant production of dark photons compared to dark fermions, but above this value and up to $2m_e$, resonant production of dark photons can overcome the fermions. However, this region is already constrained \cite{0811.0326}, which we take into account in the scatter plot of Fig.~\ref{fig:scatter-plot}.

\underline{Heavy dark photon.} The calculation proceeds similarly, with the difference that the cross-section is given by 
\begin{equation}
\avg{\sigma v}_{e^+e^-\to \bar{\psi}\psi} = 2\pi\alpha\ald\epsilon^2 \cdot 
    \begin{cases} 
    \displaystyle \TSM^{-2}, & \TSM> \mA \\[0.1cm]
    \displaystyle\mA^{-2}, & \TSM< \mA
    \end{cases}
\end{equation}
so that the main contribution comes from the times around $\TSM=\mA$.

Since $\mA>\mps$ the final injected energy density in this case redshifts as radiation until $\mps$ and as matter afterwards. Overall this results in an additional factor of $2\mps/\mA$, the $2$ coming from calculating the integral. The final bound on $\epsilon \qd$ including prefactors is given by:
\begin{equation}
\begin{split}
\label{eq:hdp_freezein_bound}
 \epsilon \qd &\lesssim \frac{4\pi^{17/4}\gstar(\mA)^{1/4}\gstarS(\mA)^{1/2}}{9\times 5^{3/4}\zeta(3)e}\parea{\frac{\gstar(\text{mre})}{\gstarS(\text{mre})}}^{1/2}\pare{\frac{T_\text{mre}}{\mpl}}^{1/2}\pare{\frac{\mA}{2\mps}}^{1/2}\\
  &\simeq 10^{-11}\pare{\frac{10}{\mA/\mps}}^{1/2}
  \end{split}
\end{equation}

The larger the mass ratio between the dark photon and the dark fermion, the weaker the bound, as the point at which the freeze-in abundance is set moves even earlier compared to $T_\text{mre}$.

\section{Dark Photon Bounds} \label{app:darkPhotonAppendix}
\subsection{More details on the dark photon bounds}
In the `light dark photon' case, there is some left-over abundance of dark photons after the dark fermion freezes out. When this abundance overcomes the freeze-in one that the original works on injections during BBN, CMB \cite{1407.0993} and IDPB \cite{0811.0326} assumed, the constraints need to be re-assessed. The results of the following estimates are shown in Fig.~\ref{fig:dark photon constraints} of Sec.~\ref{sec:darkPhotonConstraints}.

\textbf{Bounds from BBN and CMB} As stated in Sec.~\ref{sec:darkPhotonConstraints}, these constraints are `islands' in the dark photon parameter space because they are shaped by the combined requirements of having enough number density to disrupt the standard cosmological history, but to also decay within the appropriate time. Thus, these islands can be thought of as the region between two contours of constant decay times -  corresponding to the start and end times of the relevant era - and above a contour of constant number density - enough so that the energy injections are in conflict with observations.

We do not attempt a careful re-analysis of these constraints for our scenario, but in what follows, we will roughly outline how these ought to change.

The freeze-in yield, Eq.~\ref{eq:e_an_yield}, scales as $Y_\text{pair} \mpl^{-1}\propto \epsilon^2 \mA^{-1}$, while the rate scales as $\Gamma_{A'\to e^+ e^-}\sim \epsilon^2 \mA$. As a result, keeping the rates fixed and going to higher masses will eventually render the yield too small to produce a measurable effect. However, the yield from our mechanism $Y_{A'}$ (see Eq.~\ref{eq:psi_freeze-out}) is independent of $\epsilon$ and depends only logarithmically on $\mA$ through the $\eta$ factor of the dark sector bath temperature. Therefore, $Y_{A'}$ is essentially constant and the bound can in principle extend to arbitrarily low $\epsilon$ such that $\Gamma_{A'\to e^+ e^-}$ is kept constant. 

Below the dashed black line of Fig.~\ref{fig:dark photon constraints} is where the dark photon abundance of our mechanism overwhelms the freeze-in one, transforming the islands into `stripes'. Note that if the black line falls below an island (for smaller $\HI$ for example), then the respective stripe disappears. This is the case because, as mentioned above, below it the abundance is independent of both $\epsilon$ and $\mA$ (up to the $\eta$ factor) and thus constant.

We should point out that since the decay products of the dark photons have energies of order $\mA$, they might be less efficient in altering BBN and decoupling, thus weakening the bounds \cite{1407.0993}. We do not take this effect into account and remain conservative in taking the more stringent bounds.

\textbf{Intergalactic Diffuse Photon Background (IDPB)} If the dark photon is lighter than the electron, it decays primarily to $3\gamma$ through an electron loop, which will contribute to the IDPB. The bound is derived by imposing that the total flux produced from the decay of the dark photon does not exceed the measured one \cite{0811.0326}:
\begin{equation}
\label{eq:IDGBbound}
    \frac{\mA\,\tau}{\text{GeV$\,\cdot\,$s}}\lesssim 10^{27}\pare{\frac{\omega}{\text{GeV}}}^{1.3}\pare{\frac{\Omega_2 h^2}{0.1}},
\end{equation}
where $\omega\simeq \mA/3$ is the peak of the decay spectrum and $\tau\equiv \tau_{A'\to\gamma\gamma\gamma}$ is given by 
\begin{equation}
    \tau_{A'\to\gamma\gamma\gamma}^{-1}=\frac{17 \alpha^4\epsilon^2}{11\,664\,000\,\pi^3}\frac{\mA^9}{m_e^8}.
\end{equation}
In deriving the bound for the light dark photon case of this work, we use Eq.~\ref{eq:dpO_today} for the last factor of Eq.~\ref{eq:IDGBbound}, with the yield $Y_{A'}$ coming from Eq.~\ref{eq:psi_freeze-out}. The resulting bound is plotted in Fig.~\ref{fig:dark photon constraints}. The left bound changes slope when the freeze-in abundance $Y_2$ of Eq.~\ref{eq:yield_resonance} becomes equal to $Y_{A'}$. \medskip\

\subsection{Combined parameter space}
\label{app:scatter plot}
In this section we discuss the parameter space that could be of interest for the combined direct detection of $\psi$ and $A'$.
The detection of a millicharged particle lying in the green region of Fig.~\ref{fig:Reach-DD_light-A'} below the freeze-in line, would already be a significant hint to the production mechanism that we discuss in our paper.
As a smoking-gun signature, it would be significant to look for the dark photon, and identify the $\mA$ range that is more likely to be probed in the near future with the experiments described in Sec.~\ref{sec:darkPhotonConstraints}.
\begin{figure}[ht] \centering
\hspace{-1em}
\includegraphics[width=.52\textwidth]{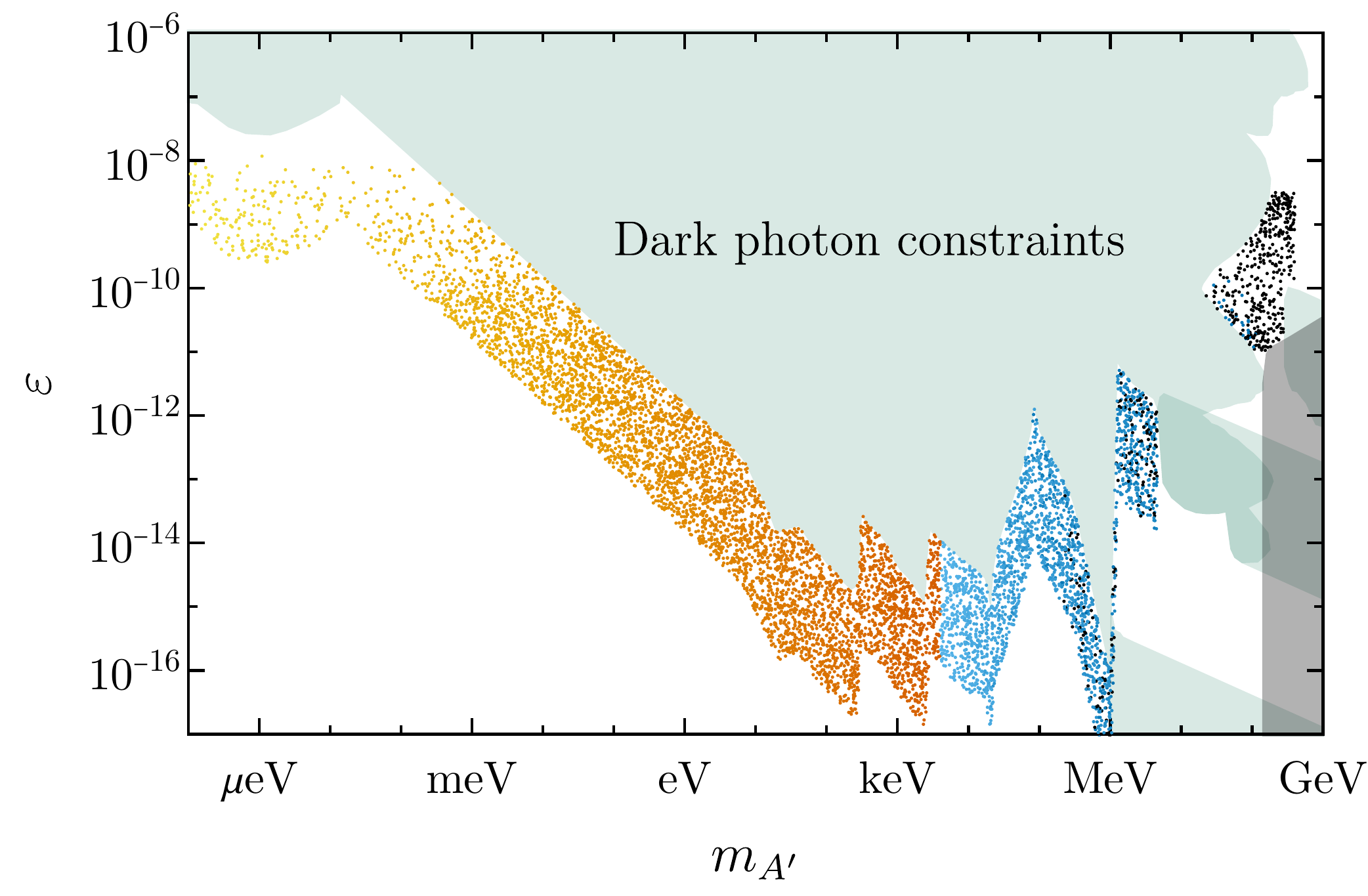}\hspace{-6em} \hfill
\includegraphics[width=.5\textwidth]{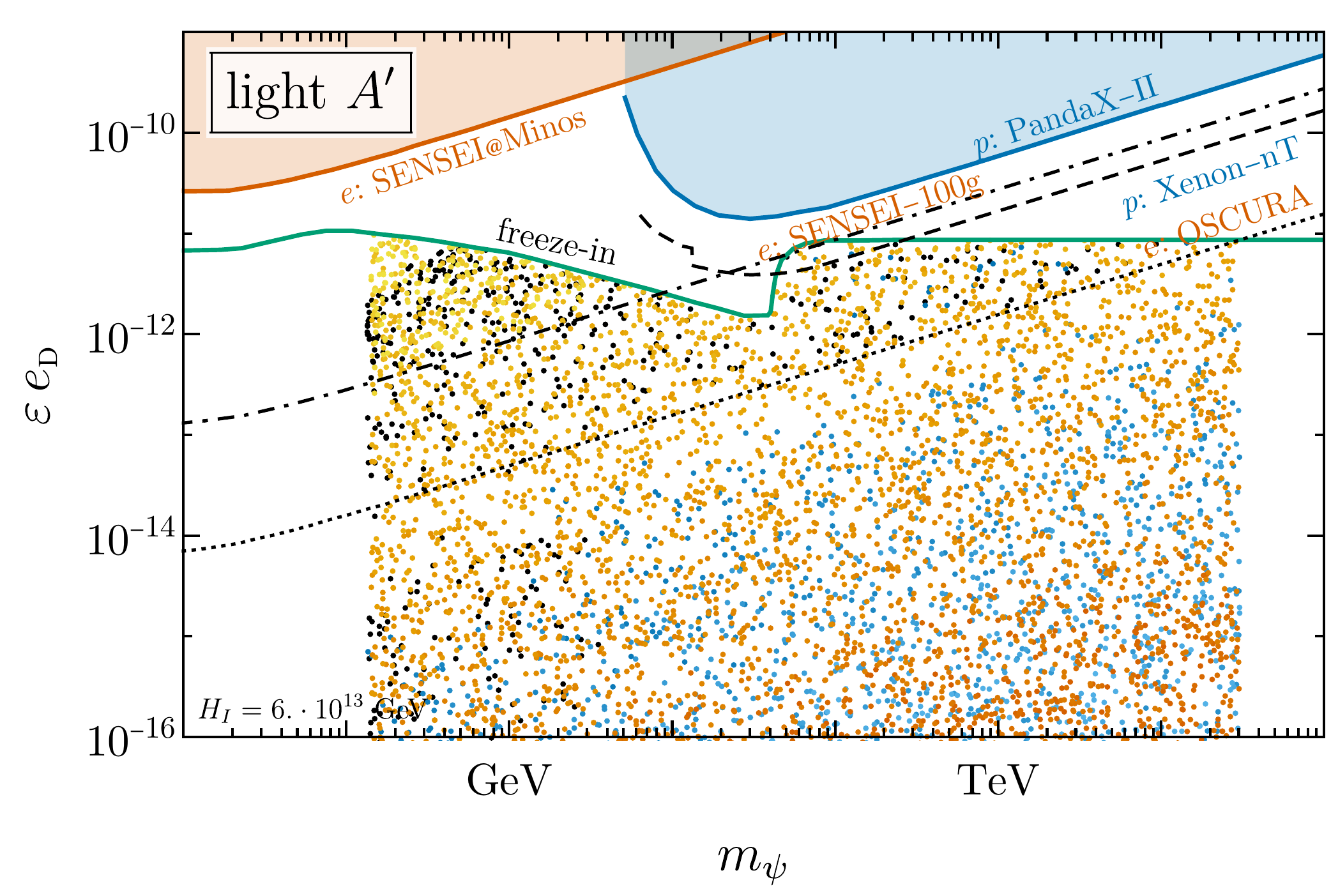}
\caption{Scatter plot sampling our 4-dimensional parameter space $(\mps,\qd,\mA,\epsilon)$, with the goal of identifying points that allow for a successful production of DM and are potentially close to discovery in the dark photon parameter space. The detection of the dark fermion in the favored region (shown in Fig.~\ref{fig:Fermion-bounds_light-A'} and in the right plot) would be suggestive of our mechanism and provide indication on the favoured range for $\mA$ to be successfully probed.
\textbf{Left:} Projection in the $(\mA,\epsilon)$ parameter space. We sample values of $\epsilon$ that are within 2 orders of magnitude from present bounds (for which we refer to Fig.~\ref{fig:dark photon constraints}).
The points are coloured according to whether $A'$ is a light enough mediator for $\psi$ direct detection: orange/red refers to light mediator for electron scattering ($\mA$ smaller than the typical momentum transfer $\alem m_e$) \textit{and} nuclear scattering on Xenon ($\mA<\mu_{\psi N_\text{Xe}}v_\DM$, where $\mu_{\psi N_\text{Xe}}$ is the reduced mass of $\psi$ and the Xe nucleus), blue for only nuclear scattering, and black for the remaining points.
The depletion of points in the upper left corner is due to the requirement of having a dark electromagnetic cascade (see Fig.~\ref{fig:veryLightPhotonRelicAbundance}).  
\textbf{Right:} Projection in the $(\mps,\eps \qd)$ plane which is relevant for direct detection. Orange/red points are within reach of both electron (SENSEI, OSCURA) and nuclear recoil (PandaX, Xenon-nT) searches, whereas blue points are only within reach of nuclear recoil searches.
}
\label{fig:scatter-plot}
\end{figure}

With this goal in mind, we scan the 4-dimensional parameter space $(\mps,\qd,\mA,\epsilon)$ for dark QED in the presence of kinetic mixing with the following prescription:
We begin by sampling points in the $(\mps,\qd,\mA)$ space that yield the right relic density and reach kinetic equilibrium at $\TDM= \mps$ (as the green lines in Fig.~\ref{fig:lightPhotonRelicAbundance}) and can reach the cascade regime even for very light $A'$ (see Fig.~\ref{fig:veryLightPhotonRelicAbundance}).
Then, for each triplet we sample the kinetic mixing parameter $\epsilon$ which lies within at most two orders of magnitude below the present constraint at that value of $\mA$, and provided it does not fall in the gray shaded region in Fig.~\ref{fig:dark photon constraints}.

We project the resulting sample onto the two different planes that are relevant for direct detection of $A'$ and $\psi$ (left and right plots respectively in Fig.~\ref{fig:scatter-plot}).
For the direct detection of $\psi$, we focus on the parameter space in which $A'$ is light enough to rely on the more promising projections for the reach on the detection of electron and nuclear recoils assuming the form factor $F(q^2)=\alem m_e/q^2$ (appropriate for light mediator).
In order to establish whether the corresponding point of our parameter space qualifies for this criterion for either electron or nuclear recoils, we color the point respectively in orange/red if $\mA$ is below both thresholds $\alem m_e$ (the typical momentum exchange in electron recoils) and $\mu_{\psi,\text{Xe}} v_\textsc{dm}$ (the typical momentum transfer for nuclear recoils, where $\mu_{\psi,\text{Xe}}$ is the reduced mass of $\psi$ and a Xe nucleus and $v_\textsc{dm}\sim 10^{-3}$ is the DM typical velocity), and in blue if only the latter is satisfied.
In other words, orange/red points are within reach of all the experiments shown with black lines, whereas for blue points the projections of SENSEI and OSCURA do not apply, and only the line of Xenon-nT should be considered.
Finally, black points correspond to the case of $\mA$ comparable to $\mps$, whose form factor would not fall in either the heavy ($F(q^2)=1$) or light ($F(q^2)=\alem m_e/q^2$) cases.

A lesson of this exercise is that, in the case of  detection of a millicharged $\psi$ lying below (but quite close to) the freeze-in line, and considering the allowed dark photon parameter space, experimental $A'$ searches should focus on $\mA$ corresponding to the color code of Fig.~\ref{fig:scatter-plot} or lighter.

\section{Reheating}
\label{app:reheating}
In this section, we address considerations about the robustness of our scenario against variations in the history of reheating after inflation. We first discuss how dark matter produced by inflation remains relatively unaffected by the history of reheating, provided it completes before $H$ is equal to the mass of the DM. We then show how this is in contrast with another well-studied production mechanism for DM, namely the freeze-in production of dark states from the SM mediated by gravity. If reheating is instantaneous, gravitational freeze-in production dominates. However, in simple models of non-instantaneous reheating, we find our mechanism is dominant for reasonable values of the reheating parameters.
\subsection{Effects of non-instantaneous reheating on the energy density in the dark sector at late times}
Ref.~\cite{Graham:2015rva} considers explicitly the case of instantaneous reheating immediately following the end of inflation. Refs.~\cite{Ahmed:2020fhc,Kolb:2020fwh} carefully study the impact of the details of more general reheating scenarios. The broad conclusion is that the late-time dark matter density of dark photons produced by inflation is largely independent of the details of reheating, as long as reheating completes before $H = \mA$.  We briefly recapitulate why that is.\par

\begin{figure}
 \centering
 \begin{subfigure}[b]{.49\textwidth}
     \centering
     \includegraphics[width=\textwidth]{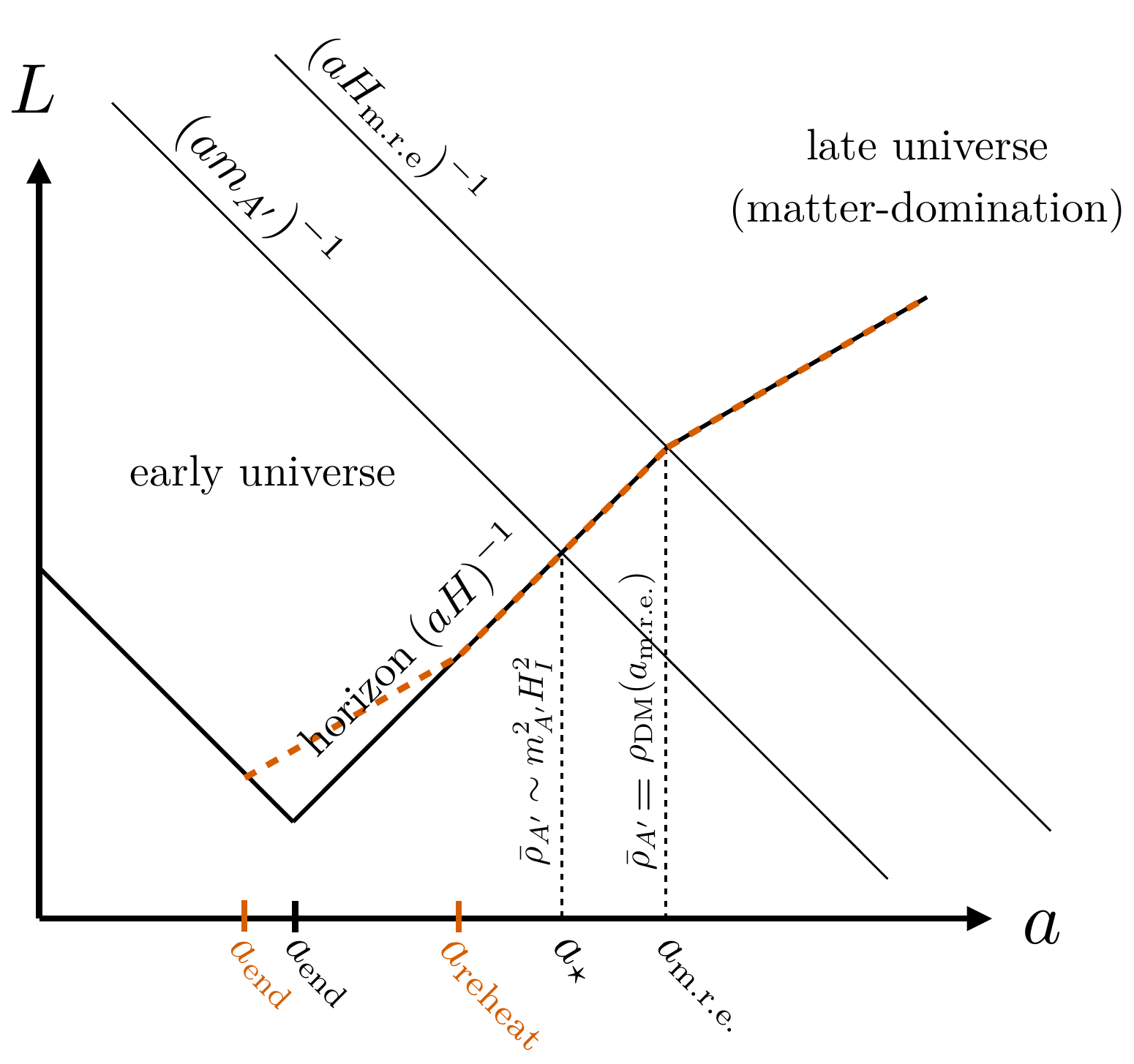}
     \label{fig:reheating1}
 \end{subfigure}
 \hfill
 \begin{subfigure}[b]{.49\textwidth}
     \centering
     \includegraphics[width=\textwidth]{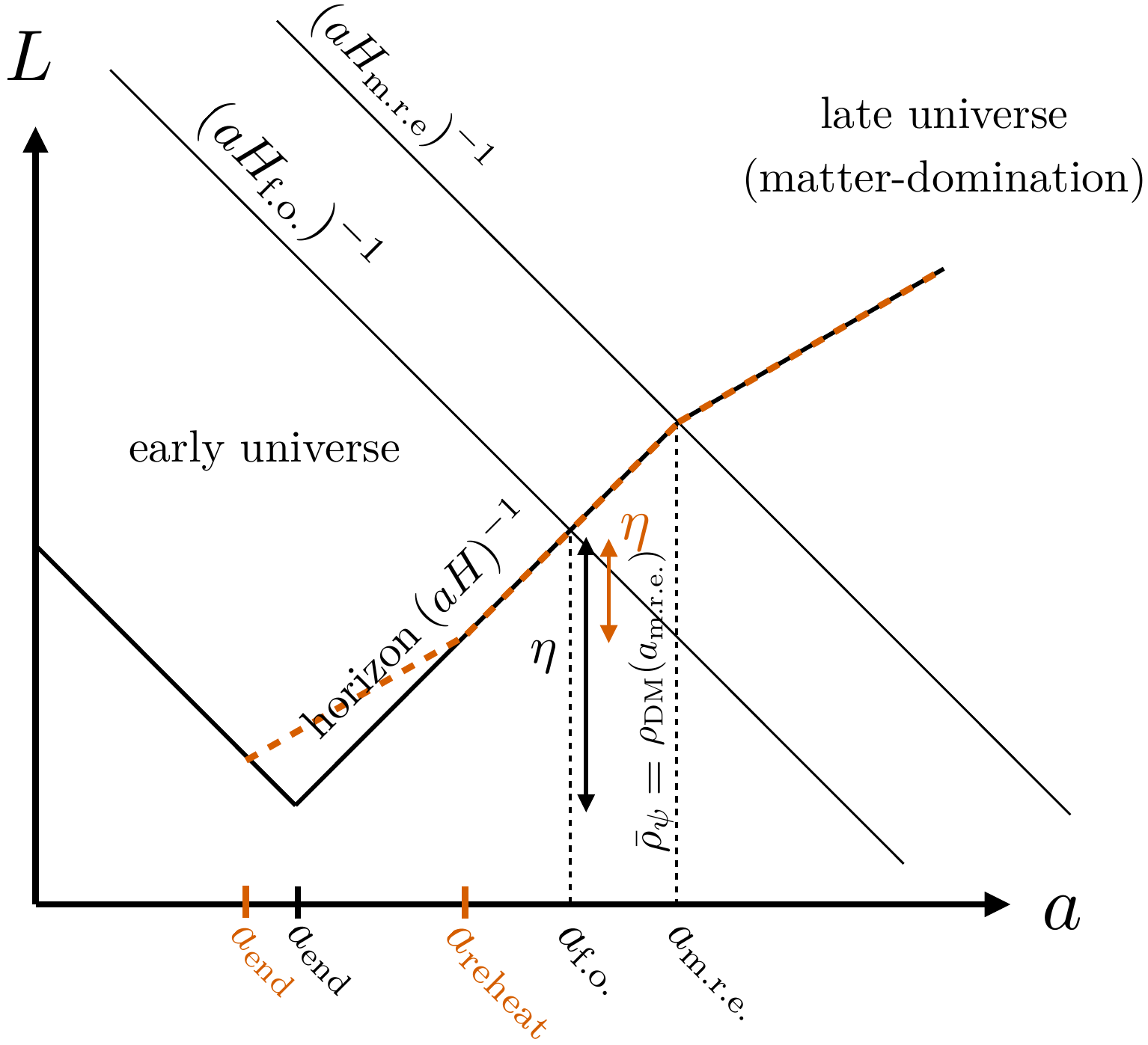}
     \label{fig:reheating2}
 \end{subfigure}
 \caption{Illustration of the impact of non-instantaneous reheating. We compare the evolution of the important scales in the scenario of instantaneous reheating immediately following the end of inflation, and when inflation is followed by a period of matter-domination (dashed orange lines). \emph{Left}: In the non-interacting theory, the dominant component of the late-time energy density is from the set of modes entering at $a_\star$ (i.e. when $H = \mA$). This set contributes energy density $\sim \mA^2 \HI^2$ independently of the history of reheating, and subsequently redshifts like matter. One can see that only the subdominant UV tail of the power spectrum will suffer additional redshift. \emph{Right:} In dark QED, the history of reheating can reduce the number $\eta$ of modes that re-enter the horizon after the universe has entered RD and contribute to the energy density of the dark thermal radiation bath at late times before $\psi$ freeze-out. The effect on $\eta$, however, is only logarithmic.}
    \label{fig:reheating}
\end{figure}

The phase space distribution \ref{eq:occupation_number} of superhorizon modes only depends on the fact that they are superhorizon (i.e. $k/a > H$), and is independent of how $H$ itself depends on $a$.
The late time abundance is dominated by the scale $k_\star$, which enters the horizon at $H=\mA$. This remains true even if reheating is preceded by a period of matter-domination (Fig.~\ref{fig:reheating}). No matter \emph{when} this happens, the occupation number of that mode at that time is $\sim \HI^2/\mA^2$, and the energy density is $\rho_{A'}(k_\star,a_\star)\simeq \HI^2 \mA^2$. This energy density subsequently redshifts like matter. The vector particle will have the correct energy density to be DM if it matches the known DM density at matter-radiation equality. As long as the universe has entered radiation-domination before $H=\mA$, the condition for the vector particle to have to correct abundance to be DM reads
\begin{equation}
\label{eq:requirement for DM density}
\HI^2 \mA^2\left(\frac{H_\text{m.r.e.}}{\mA}\right)^{3/2} \approx \rho_\text{DM}(a_\text{m.r.e.}),
\end{equation}
where $H_\text{m.r.e.}$ and $\rho_\text{DM}(a_\text{m.r.e.})$ are the known value of the Hubble parameter and the density of CDM at the time of matter-radiation equality. Note that the requirement \ref{eq:requirement for DM density} is independent of the evolution of $H$ prior to $H = \mA$ since it only cares about the dependence of $H$ on $a$ between $H=\mA$ and $H = H_\text{m.r.e.}$, which is assumed to occur in RD. One can see that the effect of a period of matter-domination preceding reheating is to suppress the UV tail of the vector power spectrum, which is already a subdominant component (Fig.~\ref{fig:reheating}). \par
In the dark QED scenario, the time $H=\mA$ is no longer special if a dark plasma can form before that time, and the DM spectrum is no longer peaked at the special scale $k_\star$. If thermalization is allowed to complete, the special time $H=\mA$ is in a sense replaced by the time $H=H_\text{f.o.}$ (compare Figs.~\ref{fig:massiveVectorEvol}, \ref{fig:plasma_effects}). Modes that enter the horizon after the end of reheating contribute their energy to the late-time energy of the dark bath. The history of reheating only affects the number $\eta$ of such modes logarithmically, and thus only has a mild effect on the temperature of the dark bath relative to that of the SM (Fig.~\ref{fig:reheating}). The dark thermal relic abundance is therefore essentially unaffected by the history of reheating, as long as it ends before $H_\text{f.o.}$.

\subsection{Redshift-Temperature Relation}
We now turn our attention to the gravitational freeze-in production of dark states. Our goal is to contrast the sensitivity of this mechanism to the relative robustness of the one proposed in this work.\par
We first follow Ref.~\cite{Giudice:2000ex} in deriving how the SM radiation bath redshifts during the reheating period.
If the inflaton decays to SM with a constant rate $\Gamma_\phi$ the Boltzmann equations for the system are
\begin{align}
\frac{\di \rho_\phi}{\di t}+3H \rho_\phi&=-\Gamma_\phi\rho_\phi\\
\frac{\di \rho_R}{\di t}+4H \rho_R&=\Gamma_\phi\rho_\phi\label{eq:reheat_rad}
\end{align}
where $\phi$ is the inflaton, $\Gamma_\phi$ its constant decay rate and $\rho_R$ the energy density of the resulting radiation bath, i.e. of the SM. We assume that the SM thermalizes quickly, so that we can define its temperature $T$ via the usual relation
$\rho_R=\pi^2g_\star(T) T^4/30$.
The effective relativistic degrees of freedom don't change during the reheating period and so $g_\star(T)\equiv g_\star\simeq 106.75$. 

We define $\Phi\equiv \rho_\phi a^3$ and $R\equiv \rho_R a^4$. The inflaton dominates the energy density so that $H^2\simeq 8\pi G\rho_\phi/3$. Approximating $\Phi\simeq \Phi_I$ (its value at the end of inflation) so that $\rho_\phi\propto a^{-3}$, the solution to Eq.~\ref{eq:reheat_rad} is
\begin{equation}
\rho_R=\frac{2}{5}\Gamma_\phi\sqrt{\frac{3\Phi_I}{8\pi G}}\pare{a^{-3/2}-a^{-4}},
\end{equation}
where we have defined that at the end of inflation $a=a_I=1$. The temperature is thus given by 
\begin{equation}
\label{eq:reheat_temp}
T=\pare{\frac{6\Gamma_\phi}{\pi^2 g_\star}}^{1/4}\pare{\frac{3\Phi_I}{8\pi G}}^{1/8}\pare{a^{-3/2}-a^{-4}}^{1/4}
\end{equation}
Note that this is \emph{not} the temperature of the dominant energy density (as in RD) since here most of the energy is in the non-thermalized and decaying $\phi$.

From Eq.~\ref{eq:reheat_temp} we see that the temperature rises to a maximum value $T_{\max}$ and then redshifts as $T\propto a^{-3/8}$, i.e. slower than in RD. We define reheating to be the time when the SM radiation dominates the energy density of the universe, namely when the inflaton decays completely at $\Gamma_\phi=H$. The reheating temperature is then
\begin{equation}
\label{eq:reheat_trh}
\TRH=\pare{\frac{45}{4\pi^3 g_\star}}^{1/4}\pare{\Gamma_\phi \mpl}^{1/2}
\end{equation}
where $\mpl=G^{-1/2}$. At the beginning of reheating the energy density in $\phi$ is just $\rho_\phi=\Phi_I=3 \HI^2 \mpl^2/(8\pi)$, where the second equality follows from the definition $a_I=1$ at the beginning of reheating. The maximum temperature then is
\begin{equation}
\label{eq:reheat_tmax}
T_{\max} = \frac{3^{3/20} \sqrt[4]{5}}{2 \sqrt[5]{2}}\pare{\frac{6\Gamma_\phi}{\pi^2 g_\star}}^{1/4}\pare{\frac{3\Phi_I}{8\pi G}}^{1/8}
     \simeq 0.12\pare{\sqrt{\frac{\Gamma_\phi}{\HI}}\HI \mpl}^{1/2}
\end{equation}
for $g_\star= 106.75$. The ratio of the two temperatures is
\begin{equation}
\label{eq:reheat_tratio}
\frac{T_\text{max}}{\TRH}=\frac{3^{3/20}}{2^{6/5}}\pare{\frac{\HI}{\Gamma_\phi}}^{1/4}\simeq 0.51\pare{\frac{\HI}{\Gamma_\phi}}^{1/4}
\end{equation}
After $T_\text{max}$ is attained, $T\propto a^{-3/8}$ so that the scale factor is related to the temperature by 
\begin{equation}
\label{eq:reheat_scalef}
a =\frac{3}{2\pi^2}\pare{\frac{3}{2 g_\star^2}}^{1/3}\pare{\sqrt{\frac{\Gamma_\phi}{\HI}}\HI \mpl}^{4/3}T^{-8/3}
= \pare{\frac{9\HI^{2} \mpl^{2} \TRH^{4}}{20\pi^3g_\star}}^{1/3} T^{-8/3}
\end{equation}
which means that $H$ evolves as $\propto T^4$ or more precisely as
\begin{equation}
\label{eq:reheat_hubble}
H^2 =\frac{8\pi G}{3}\rho_\phi 
 =\HI^2 a^{-3}\\
 =\frac{20\pi^3g_\star}{9}\frac{T^{8}}{\TRH^4 \mpl^2}.
\end{equation}

\subsection{SM Freeze-in Estimate During Reheating}
Now we can turn to the pair-annihilation of SM states into DM $\bar{\psi}\psi$ pairs. For $T\gg m_\text{SM},\mps$ (where $m_\text{SM}$ is the mass of the SM states), the thermally averaged cross-section is 
\begin{equation}
\avg{\sigma v}\simeq C\frac{T^2}{\mpl^4}
\end{equation}
the coefficient $C$ in front depends on the degrees of freedom of the initial and final states. For annihilation to fermion DM, SM fermions have $C=C_{1/2}=24\pi/5$, SM vectors $C=C_1=C_{1/2}$ and SM scalars $C=C_0=2\pi/5$. Our calculations agree with the results of Refs.~\cite{1709.09688,1708.05138}. In this section $T$ denotes the temperature of the SM radiation bath given by Eq.~\ref{eq:reheat_temp}.

Each annihilation of SM states transfers energy $T$ into the DM sector. The energy transferred per unit time per unit volume from a fermion $f$ of the SM thermal bath is just $\mathcal{W}\sim n_f^2 \avg{\sigma v} T$. These resulting DM fermions will be relativistic during the reheating period (assuming $\TRH \gg \mps$), so their energy density will redshift as radiation afterwards and until the end of reheating. After reheating is over at $H=\Gamma_\phi$ we enter RD where $H\propto T^2/\mpl$ and the injected energy into the DS will continue to redshift as radiation until $T=\mps\ll \TRH$. 

The total energy density injected into the DS is then
\begin{equation}
\mathcal{\rho}_\text{inj}=\int n_f^2 \avg{\sigma v} T \parea{\frac{a(t)}{a_\text{end}}}^4\di t=\int n_f^2 \avg{\sigma v} T \frac{\di a}{a H}
\end{equation}
where $a_\text{end}$ is some final time that corresponds to the upper limit of the integral.

At the end of reheating, taking $a_\text{end}=a(\TRH)$, the injected energy density into the dark sector is
\begin{equation}
\label{eq:reheat_injRH}
\mathcal{\rho}_\text{inj,RH} = \pare{\frac{3\zeta(3)g_f}{4\pi^2}}^2 \frac{C}{\mpl^4} \pare{-\frac{4 \mpl \TRH^2}{\sqrt{5\pi^3 g_\star}}}\int_{T_\text{max}}^{\TRH} T^6 T^2 T\pare{\frac{\TRH}{T}}^{32/3}\frac{\di T}{T^5}
\simeq\frac{27 Cg_f^2\zeta(3)^2}{68\sqrt{5\pi^{11}g_\star}}\frac{\TRH^7}{\mpl^3}
\end{equation}
where in the last line we used that the integral is dominated by \emph{late} times. 

A similar power-counting of the analogous integral for RD using $T\propto \gstar^{-1/3}a^{-1}$ and $H\propto T^2/\mpl$, gives an integrand that scales as $\sim T^6 T^2 T (T_\text{end}/T)^4 (1/T^3)=T^2$, so that now the earliest times contribute the most, giving at $T_\text{end}$ (neglecting the change is $g_{\star,s}$)
\begin{equation}
\label{eq:reheat_injRD}
\mathcal{\rho}_\text{inj,RD} \simeq\frac{9\sqrt{5} Cg_f^2\zeta(3)^2}{32\sqrt{\pi^{11}g_\star}}\frac{T_\text{end}^4 \TRH^3}{\mpl^3}\parea{\frac{g_{\star,s}(T_\text{end})}{g_{\star,s}(\TRH)}}^{4/3}
\end{equation}
This is the contribution from the annihilations of a single SM fermion species. There are $N_{1/2}=45$ such possible fermion annihilations and $g_f=4$. For the $N_1=12$ possible massless gauge boson contributions, $g_f\to g_V=2$ and one needs to change $\zeta(3)\to 4\zeta(3)/3$ to account for the different statistics in the number densities used (we used the fermionic one in the calculation above). Finally for the $N_0=4$ possible scalar contributions $g_f\to g_S=4$ and again $\zeta(3)\to 4\zeta(3)/3$. Factoring out $C$ and $g_f$ from Eq.~\ref{eq:reheat_injRD}, overall the SM amounts to a factor of 
\begin{equation}
D\equiv45C_{1/2}g_f^2+12C_{1}g_V^2\pare{\frac{4}{3}}^2+4C_{0} g_S^2\pare{\frac{4}{3}}^2=\frac{1598336\pi}{405}\simeq 10^{5}
\end{equation}
We wish to compare this energy density to that of our mechanism at $H=\mA$. The dark photons of Eqs.~\ref{eq:reheat_injRH} and \ref{eq:reheat_injRD} redshift as relativistic matter so compared to $\rho_\text{DS}=\mA^2 \HI^2\eta^2/(16\pi^2)$ at $H=\mA$ we find
\begin{equation}
\frac{\rho_\text{inj,gr}}{\rho_\text{DS}} =\frac{8829\sqrt{5}\zeta^2(3)D}{136\pi^{13/2}\parea{g_\star({\TRH})}^{3/2}}\frac{\TRH^3}{\HI^2\mpl\eta^2}\parea{\frac{g_{\star,s}(T_\text{end})}{g_{\star,s}(\TRH)}}^{4/3} 
 \simeq0.24 \frac{\TRH^3}{\HI^2\mpl\eta^2}
\end{equation}
where we took $g_\star(\TRH)=g_{\star,s}(\TRH)= 106.75$ and $g_{\star,s}(H=\mA)\simeq 80$. The graviton-exchange contribution is subdominant for
\begin{equation}
\TRH \lesssim 1.6 \HI^{2/3} \mpl^{1/3}\eta^{2/3}
\simeq 5.7\times 10^{15}\eta^{2/3}\text{ GeV}\pare{\frac{\HI}{6\times 10^{13}\text{ GeV}}}^{2/3},
\end{equation}
or equivalently
\begin{equation}
\frac{\Gamma_\phi}{\HI} \lesssim 18 \eta^{4/3}\pare{\frac{\HI}{\mpl}}^{1/3} 
\simeq 0.8 4^{2/3}\eta^{4/3}\pare{\frac{\HI}{6\times 10^{13}\text{ GeV}}}^{1/3}.
\end{equation}
For $\eta=7$ this gives $\Gamma_\phi\lesssim 2.4\times 10^{14}\text{ GeV}$, for the highest inflationary scale. Note that this is actually higher than its upper limit, which is $\HI$ itself.

In terms of temperatures, we can define the temperature of an instantaneous reheating at $H=\HI$ as
\begin{equation}
T_\text{inst}\equiv\pare{\frac{45}{4\pi^3g_\star}}^{1/4}\pare{\HI \mpl}^{1/2}
\end{equation}
Defining as $T_\text{rh,th}$ the temperature at which the ratio of Eq.~\ref{eq:reheat_tratio} is equal to $1$ and keeping the $g_\star(\TRH)$ we find
\begin{equation}
\frac{T_\text{rh,th}}{T_\text{inst}} \simeq 0.12 \parea{g_\star({\TRH})}^{3/4}\eta^{2/3}\pare{\frac{\HI}{\mpl}}^{1/6}
\simeq 0.54\,\eta^{2/3} \pare{\frac{\HI}{6\times 10^{13}\text{ GeV}}}^{1/6}
\end{equation}
for $g_\star(\TRH)\simeq 106.75$. Setting $\eta=7$ makes this ratio $\simeq 2$ for the highest inflationary scale. As a result, for the highest inflationary scale we consider, this process turns out to give a similar contribution to ours. Therefore, we assume a few e-folds of reheating (a smaller $\TRH$) so that this contribution becomes subdominant. Our mechanism is insensitive to lower reheating temperatures, as long as we reheat above the lowest temperature needed to thermalize the dark sector.

\subsection{Contributions from direct inflaton annihilations}

If the inflaton annihilates directly to DM particles (via a graviton exchange) we need to look at the Boltzmann equation for the DS. The resulting bath will be relativistic as each inflaton annihilation event injects energy $m_\phi\gg m_X$, where $X$ is a DS state. We assume that the DS does not have any other couplings to the Standard Model.

The number density of $X$ evolves as \cite{1708.05138} 
\begin{equation}
\label{eq:reheat_infl_dec}
\frac{\di (n_X a^3)}{a^3\di t}=\frac{n_\phi^2}{m_\phi^2}\mathcal{F}g_i^2\sigma
\end{equation}
where $\mathcal{F}=\parea{(p_1\cdot p_2)^2-m_1^2m_2^2}^{1/2}=|\mathbf{p}_i|\sqrt{s}$, $g_i$ are the spin degrees of freedom of the initial state. In the CM frame $\mathcal{F}=|\mathbf{p}_i|\sqrt{s}$, where $\mathbf{p}_i$ is the 3-momentum of the initial state. The cross-section for a scalar initial state is
\begin{equation}
\label{eq:appHcross-section}
 \sigma= \frac{1}{32\pi s S }\frac{|\mathbf{p}_f|}{|\mathbf{p}_i|}\int\cos\theta|\mathcal{M}|^2
 \end{equation}
where $S$ is a symmetry factor equal to 2 if the final state consists of neutral scalars or gauge bosons (otherwise it's 1). The squared amplitude includes a \textit{sum} over polarizations.

Using Eq.~\ref{eq:appHcross-section} we find for the production of fermions and vectors respectively:
\begin{align}
\mathcal{F}\sigma_{1/2}&=\frac{\pi m_\phi^2 M_X^2}{4\mpl^4}\pare{1-\frac{M_X^2}{m_\phi^2}}^{3/2}\\
\mathcal{F}\sigma_{1}&=\frac{\pi m_\phi^4}{4\mpl^4}\pare{1+\frac{M_X^2}{m_\phi^2}+\frac{19 M_X^4}{4m_\phi^2}}\sqrt{1-\frac{M_X^2}{m_\phi^2}}
\end{align}

As previously noted in \cite{2102.06214}, production of fermions suffers a helicity suppression as the rate is $\propto M_X^2$.

Integrating Eq.~\ref{eq:reheat_infl_dec} we find that it is dominated by \emph{early} times, in contrast to SM particle annihilations. Comparing the resulting energy density $\rho_{\text{inj,}\phi}$ at $H=\mA$ to $\rho_\text{DS}=\mA^2\HI^2\eta^2/(16\pi^2)$ of our mechanism we find
\begin{equation}
\begin{split}
\frac{\rho_{\text{inj,}\phi}}{\rho_\text{DS}}&=\frac{75\times  3^{8/15} 5^{2/3} \pi^2 g_\star^{1/3}(\TRH)}{32\times 2^{14/15}\eta^2}\parea{\frac{g_{\star,s}(T_\text{end})}{g_{\star,s}(\TRH)}}^{4/3}\pare{\frac{m_\phi^3\TRH^4}{\HI^5 \mpl^2}}^{1/3}\parea{\frac{M_X^2}{m_\phi^2}}_{1/2}\\
&=\frac{1125\times 3^{1/5}}{16\times 2^{13/5}\eta^2}\parea{\frac{g_{\star,s}(T_\text{end})}{g_{\star,s}(\TRH)}}^{4/3}\frac{m_\phi}{\HI}\pare{\frac{\Gamma_\phi}{\HI}}^{2/3}\parea{\frac{M_X^2}{m_\phi^2}}_{1/2}\\
&\simeq 33 \eta^{-2}\frac{m_\phi}{\HI}\pare{\frac{\Gamma_\phi}{\HI}}^{2/3}\parea{\frac{M_X^2}{m_\phi^2}}_{1/2}\\
\end{split}
\end{equation}
where the last factor inside the brackets applies only in the case of fermionic DS states. In terms of the reheating temperature:
\begin{equation}
\frac{\rho_{\text{inj},\phi}}{\rho_\text{DS}}\simeq 2.8\times 10^2 \eta^{-2}\frac{m_\phi}{\HI}\pare{\frac{\TRH}{\HI}}^{2/3}\pare{\frac{\TRH}{\mpl}}^{2/3}\parea{\frac{M_X^2}{m_\phi^2}}_{1/2}
\end{equation}
For fermions, the helicity suppression renders inflaton decays irrelevant for reasonable choices of reheating parameters. For vectors, in order for this contribution to be subdominant at that time we need
\begin{equation}
\label{eq:infl_bound1}
\frac{m_\phi}{\HI}\lesssim 3\times 10^{-2}\eta^2\pare{\frac{\HI}{\Gamma_\phi}}^{2/3}
\end{equation}

Assuming a Yukawa coupling to the SM, the inflaton decay rate is $\Gamma_\phi=y^2 m_\phi/(8\pi)$ and we can recast the bound of Eq.~\ref{eq:infl_bound1} into a bound for $y$:
\begin{equation}
y\lesssim 7.1\pare{\frac{\HI}{m_\phi}}^{5/4}.
\end{equation}

\end{document}